\documentclass{aa}  

\usepackage{graphicx}
\usepackage{txfonts}
\usepackage{natbib}
\usepackage{booktabs}
\usepackage{amsmath}
\usepackage{siunitx} % Optional for number formatting
\usepackage{enumitem}

\usepackage{float}

\usepackage{tikz}
\usepackage{graphicx}
\usepackage{adjustbox}
\usepackage{caption}
\usetikzlibrary{arrows.meta, positioning}

\tikzstyle{startstop} = [rectangle, rounded corners, minimum width=3.2cm, minimum height=1cm, text centered, draw=black, fill=gray!20]
\tikzstyle{process} = [rectangle, minimum width=3.2cm, minimum height=1cm, text centered, draw=black] %fill=blue!10
\tikzstyle{decision} = [rectangle, minimum width=3.2cm, minimum height=1cm, text centered, draw=black, fill=green!10]
\tikzstyle{arrow} = [thick,->,>=stealth]
\tikzstyle{final} = [rectangle, minimum width=3.2cm, minimum height=1cm, text centered, draw=black, fill=gray!40] %magenta!20

\usepackage[bookmarks=false,colorlinks=true,linkcolor=black,citecolor=cyan,filecolor=black,urlcolor=cyan]{hyperref}
\usepackage[symbol]{footmisc}

\begin{document}

   \title{The incidence of eROSITA X-ray AGN in the local Universe:\\from dwarf to massive galaxies}

   \author{Z. Igo
          \inst{1,2,3}, 
          A. Merloni\inst{1},
          A. Georgakakis\inst{4},
          J. Buchner\inst{1},
          R. Arcodia\inst{5},
          M. Salvato\inst{1,2},
          J. Aird\inst{6},
          K. Nandra\inst{1},
          B. Trakhtenbrot\inst{7},
          P. G. Boorman\inst{1},
          J. Comparat\inst{1},
          G. Lamer\inst{1},
          B. Laloux\inst{1},
          M. Kluge\inst{1},
          W. Roster\inst{1},
          E. Bulbul\inst{1},
          F. Balzer\inst{1},
          T. Dwelly\inst{1},
          W. N. Brandt\inst{8,9,10},
          R. Seppi\inst{11},
          S. Morrison\inst{12},
          E. Kyritsis\inst{1},
          J. Gelfand\inst{13},
          S. F. Anderson\inst{14},
          D. P. Schneider\inst{8, 10}
          }

   \authorrunning{Z. Igo et al.}
   
   \institute{Max-Planck-Institut für Extraterrestrische Physik (MPE), Giessenbachstrasse 1, 85748 Garching bei München, Germany
              \email{zigo@mpe.mpg.de}
         \and
             Exzellenzcluster ORIGINS, Boltzmannstr. 2, 85748, Garching, Germany
        \and 
        European Space Agency (ESA), European Space Astronomy Centre (ESAC), Camino Bajo del Castillo s/n, E-28692 Villanueva de la Ca\~{n}ada, Madrid, Spain
        \and 
        Institute for Astronomy and Astrophysics, National Observatory of Athens, V. Paulou and I. Metaxa, 11532, Greece
        \and 
         MIT Kavli Institute for Astrophysics and Space Research, Massachusetts Institute of Technology, Cambridge, MA 02139, USA
        \and 
        Institute for Astronomy, University of Edinburgh, Royal Observatory, Edinburgh EH9 3HJ, UK
        \and
        School of Physics and Astronomy, Tel Aviv University, Tel Aviv 69978, Israel
        \and
        Department of Astronomy and Astrophysics, 525 Davey Lab, The Pennsylvania State University, University Park, PA 16802, USA
        \and
        Department of Physics, 104 Davey Laboratory, The Pennsylvania State University, University Park, PA 16802, USA
        \and
        Institute for Gravitation and the Cosmos, The Pennsylvania State University, University Park, PA 16802
        \and
        Department of Astronomy, University of Geneva, Ch. d'Ecogia 16, CH-1290 Versoix, Switzerland
        \and
        Department of Astronomy, University of Illinois at Urbana-Champaign, Urbana, IL 61801, USA
        \and 
        New York University Abu Dhabi, PO Box 129188, Abu Dhabi, UAE
        \and
        Department of Astronomy, University of Washington, Box 351580, Seattle, WA 98195, USA
         }

   \date{Received XX; accepted YY}

% \abstract{}{}{}{}{} 
% 5 {} token are mandatory
 
  \abstract
   {Combining deep, wide-area X-ray surveys with multi-wavelength catalogues provides insights into rare, highly-accreting active galactic nuclei (AGN) and low-mass galaxies at low redshift, the latter potentially representing local analogues of the first galaxies in the early Universe.}
   {We use eROSITA, aboard the \textit{Spectrum Roentgen Gamma} satellite (SRG), and its four-pass All Sky Survey (eRASS:4), to select the largest catalogue of X-ray AGN in a highly complete sample of low-redshift galaxies, including low-mass ($\log M_*/M_{\odot}\leq10$) ones. We probe their distribution of specific accretion rates, $\lambda_{\rm SAR}\propto L_{\rm X}/M_*$, and the cumulative AGN fraction above varying $\lambda_{\rm SAR}$ thresholds.}
   {Our parent sample consists of $\sim 5.35$ million galaxies selected from the tenth Data Release of the Legacy Imaging Survey with $z$-band fluxes brighter than 20~mag and redshifts of $0.03<z<0.2$ ($\sim 17$\% of which are spectroscopic, with the rest being good-quality photometric redshifts). We place particular emphasis on the detailed characterisation of our sample, including: (i) estimating unbiased physical galaxy properties through SED fitting; (ii) rigorous cleaning and validation of the X-ray aperture photometry and associations with optical host galaxy counterparts; and (iii) building a stellar mass- and luminosity- complete sample.}
   {We identify 874 X-ray AGN in low-mass galaxies, most of them newly discovered as X-ray emitters, with some reaching $2$–$10$~keV luminosities above $10^{43}$~erg~s$^{-1}$. Thanks to a Bayesian framework that makes use of the X-ray information from all parent sample galaxies, we constrain the specific accretion rate distribution, $p(\log \lambda_{\rm SAR} | M_*, z)$, across a wide range of $\lambda_{\rm SAR}$ and uncover second-order mass-dependent effects. We detect a break at high $\lambda_{\rm SAR}$, possibly indicating Eddington-limited, self-regulated black hole growth. Integrating $p(\log \lambda_{\rm SAR} | M_*, z)$ above $\lambda_{\rm SAR}\geq10^{-3}$, we find a cumulative AGN fraction of $\sim 1\%$ for low-mass galaxies, placing a firm lower limit on the black hole occupation fraction in this regime. We also observe a suppression in the efficiency of fuelling AGN beyond $\lambda_{\rm SAR} \geq 10^{-2}$ at both low and high masses, in comparison to those located in galaxies with $\log M_*/M_{\odot}\sim 10-10.5$.
   }
   {Overall, our specific accretion rate distributions, sampling down to the as-of-yet unexplored low-mass regime, highlight a more nuanced, mass-dependent view of AGN growth and accretion history that must be taken into account in future modelling.}

   \maketitle

%
%-------------------------------------------------------------------

\section{Introduction}
\label{sec:intro}

Supermassive black holes (SMBHs) in the centres of massive galaxies are now deemed ubiquitous and thought to co-evolve with their host galaxies \citep[e.g.][]{Ferrarese&Merritt2000, Kormendy&Ho2013, HeckmanandBest2014}. However, at the time of writing, the jury is still out regarding the incidence of massive ($M_{\rm BH} \sim 10^{4-7}~M_{\odot}$; MBHs) or intermediate-mass black holes ($M_{\rm BH} \sim 10^{2-5}~M_{\odot}$; IMBHs) in the centres of smaller galaxies \citep[`dwarfs'; see reviews by e.g.][]{Mezcua2017, Greene2020}. 

The answer to this question may hold the keys to our understanding of black hole seeding in the early Universe and their subsequent growth across cosmic time. Depending on the seeding mechanism\footnote{Three commonly discussed mechanisms for black hole seeding are: (i) the death of Population III stars at $z>15$ leading to $\sim 10^2-10^3~M_{\odot}$ black holes (`light' seeds); (ii) the collapse of pristine primordial gas clouds directly into $\sim 10^4-10^6~M_{\odot}$ black holes (`heavy seeds'); or (iii) hierarchical growth within high-concentration nuclear star clusters, operating even at later cosmic epochs, where stellar-mass black holes undergo unstable runaway growth through tidal captures \citep[see reviews by e.g.][]{Greene2020, Inayoshi2020, Volonteri2021, Natarajan2021}.}, different black hole occupation fractions (BHOFs) in low-mass galaxies, defined in this work as $M_* \leq 10^{10} M_{\odot}$, are expected. Roughly speaking, light seeding predicts a BHOF close to 100\%, while heavy seeding predicts a steeply falling BHOF towards lower masses with an occupation fraction of around 50\% at $\log (M_*/M_{\odot}) \sim 8-9$ \citep[e.g.][but are also degenerate to the various post-seeding growth channels, see e.g. \citealt{Chadayammuri2023}]{Ricarte2018, Burke2025, Miller2015, Zou2025}.

Observationally, there are two main approaches to address this problem: 1) search for high-redshift ($z>10-20$) black hole `seeds'; and 2) search for the remnants of such seeds in the local Universe ($z<0.2$) that did not grow. The former has seen great advancements thanks to recent observations with the \textit{James Webb Space Telescope} \citep[\textit{JWST}; ][]{Gardner2023} which are pushing the limits of detecting more MBHs earlier in cosmic time \citep[e.g.][]{Harikane2023, Ubler2023, Pacucci2023, Maiolino2024nature, Maiolino2024jades, Juodzbalis2024, Geris2026}.

The second approach, and the one adopted for this work, hinges on the assumption that low-mass galaxies, particularly dwarf galaxies (defined as having $M_*< 10^{9.5} M_{\odot}$), in the local Universe, may be analogous to the first galaxies that formed in the early Universe and can thus be used to test high-redshift black hole growth and seeding models \citep[e.g.][]{Mezcua2017}. In recent years, this field has evolved into a multi-wavelength search and characterisation of black holes in low-mass galaxies, thanks to ever-deeper and ever-wider multi-wavelength surveys \citep[e.g.][]{Greene2004, Greene2007, Nyland2012, Reines2013, Sartori2015, Chilingarian2018, Kaviraj2019, Mezcua2018, Mezcua2019, Reines2020, Zou2023, Eberhard2024, Pucha2025, Zou2025}. However, as \citet{Wasleske2024} demonstrate, different selection methods recover only subsets of the dwarf galaxy population, potentially leading to biased estimates of the BHOF and its lower bound, the `active fraction', defined as the fraction of low-mass galaxies that contain an accreting active galactic nucleus (AGN). 

One wavelength regime that provides a highly complete view of the AGN population is X-rays, as it is less affected by dust obscuration and the nuclear AGN-related emission can be readily distinguished from other contaminating galactic processes, given sufficient ancillary data. This is useful, as signatures of black hole seeding and growth should be imprinted in population-level predictions that trace the luminosity and stellar mass distribution of black holes. To investigate this, we choose to study the fraction, or `incidence', of X-ray AGN as a function of specific black hole accretion rate\footnote{This approach avoids the known degeneracies of flux-limited surveys \citep[e.g.][]{Aird2012} being unable to discriminate between high accretion rate, small mass black holes and low accretion rate, large mass black holes.}, $\lambda_{\rm SAR}$, a quantity proportional to the X-ray luminosity over the stellar mass, which can be considered a proxy for the Eddington ratio. The $\lambda_{\rm SAR}$ distribution has been extensively studied in high-mass galaxies, defined in this work to be $M_* > 10^{10} M_{\odot}$, and found to have a constant decreasing power-law slope that is, to first order, independent of stellar mass \citep[e.g.][]{Aird2012, Bongiorno2012, Georgakakis2017, Aird2018, Igo2024, Zou2024}.

However, computing the AGN incidence as a function of mass-scaled radiative power in the low-mass regime has thus far been greatly hindered by low sample sizes. Some key works in this context are by \citet{Aird2018}, \citet{Birchall2020}, and \citet{Birchall2022}, who compile statistical samples of tens of X-ray AGN in dwarf galaxies to probe the distribution of specific accretion rates. They derive an AGN active fraction of $\sim0.1-1\%$, similar to other X-ray works using different methods \citep[e.g.][]{Pardo2016, Mezcua2018, Pacucci2021, Zou2023}, but struggle to directly connect these results to the high-redshift black hole seeding for reasons discussed in detail in Section \ref{sec:imbh_discussion}. Further notable samples of tens of AGN identified in dwarf galaxies include: early work from pointed X-ray observations targeting optically-selected AGN in low-mass galaxies with \textit{Chandra}, \textit{XMM-Newton} and \textit{Swift} or existing (deep) surveys from these instruments \citep[e.g.][]{Schramm2013, Lemons2015, Miller2015, Mezcua2016, Baldassare2017, Chilingarian2018, Mezcua2018}; combining the NASA-Sloan Atlas catalogue \citep[NSA; e.g.][]{Blanton2011} with the early eROSITA data releases \citep[e.g.][]{Latimer2021, Sacchi2024, Eberhard2025}; and combining the MPA-JHU SDSS DR8 catalogue \citep{Brichmann2004, Kauffmann2003, Tremonti2004} with the deepest eRASS data in the eastern Galactic hemisphere \citep{Bykov2024}.

In this paper, we use the deepest western Galactic hemisphere SRG/eROSITA All Sky Survey data \citep[eRASS:4; the stacked four consecutive surveys observed between 2019-12-11 and 2021-12-19, belonging to the German, eROSITA-DE, Consortium;][]{Predehl2021, Sunyaev2021, Merloni2024} and combine it with the tenth Data Release of the DESI Legacy Imaging Survey \citep[hereafter: LS10;][]{Dey2019}. These datasets allow us to build the largest statistical sample of X-ray AGN in low-mass galaxies in the local Universe ($z<0.2$) to date, across an area of more than 13,000~deg$^2$. This information is used to compute the incidence of X-ray AGN in high- and low-mass galaxies over a wide range of $\lambda_{\rm SAR}$, including the rare highly accreting sources that are only observable in very wide-area surveys. Even though we cannot reliably constrain the BHOF with eRASS:4 data alone, this work provides valuable constraints on the properties of accretion across the mass scale, encoded in specific accretion rate distributions covering an as-of-yet unexplored region of parameter space in galaxy mass, redshift and specific accretion rate.

The outline of the paper is as follows. Section 2 describes the building of the parent galaxy sample using LS10. Section 3 presents the eROSITA X-ray-detected sample of sources within this parent sample and the various cleaning procedures taken to ensure the X-ray emission is produced by nuclear AGN emission. Section 4 describes the Bayesian methodology used to infer the incidence of X-ray AGN as a function of $\lambda_{\rm SAR}$. Section 5 presents the results regarding this specific black hole accretion rate distribution and the cumulative AGN fraction as a function of stellar mass. Finally, Section 6 and 7 discuss and summarise the results in the context of black hole accretion mechanisms across the mass scale. A standard flat cosmology with $H_0 = 70\rm{~km~s^{-1}~Mpc^{-1}}$, $\Omega_M=0.3$, and $\Omega_{\Lambda}=0.7$ is used throughout and all magnitudes are AB magnitudes corrected for galactic extinction.

\section{Building the parent galaxy sample}
\label{sec:building_parent_sample}

This section discusses the construction of the parent sample of galaxies including: (i) a description of the optical selection criteria and removal of  contaminants; (ii) the creation of a compilation of good-quality, extragalactic spectroscopic redshifts to supplement the photometric redshifts available for the optical survey used; and (iii) the calculation of galaxy properties, such as stellar masses and star formation rates (SFRs), using two different methods. Further details regarding points (ii) and (iii) are given in Appendices \ref{sec:exgal_specz_compil} and \ref{appendix:details_ml_grahsp}.

We selected our parent sample of galaxies from LS10, which includes photometry in the {\it g, r, i}, and {\it z} bands and WISE forced photometry at the optical source co-ordinates, following \citet{Lang2014, Lang2016}, at 3.4~$\mu$m, 4.6~$\mu$m, 12~$\mu$m, and 22~$\mu$m. The novel features of LS10, compared to previous releases, are its extended footprint, deeper coverage (including data from NEOWISE-Reactivation) and added \textit{i}-band observations. In addition to the observations completed by the Beijing-Arizona Sky Survey (BASS), the DECam Legacy Survey (DECaLS), and the Mayall z-band Legacy Survey (MzLS), the DECam eROSITA Survey \citep[DeROSITAS;][]{Zenteno2025} ensures 5$\sigma$ depths across (almost) the entire eROSITA-DE footprint in the western Galactic hemisphere of 22.7, 23.2, 23.3, and 22.5 mag in the {\it g, r, i}, and {\it z}  bands, respectively. For an overview of the depth and coverage of LS10 in the eROSITA-DE footprint, excluding the Galactic plane (Galactic latitude |b| $< 20^{\circ}$), that is the area covering the entire extragalactic sky in the southern equatorial hemisphere with a declination of $<32.375^{\circ}$, see Figure 1 in \citet{Saxena2024}.

\begin{figure}
    \centering
    \includegraphics[width=0.9\linewidth]{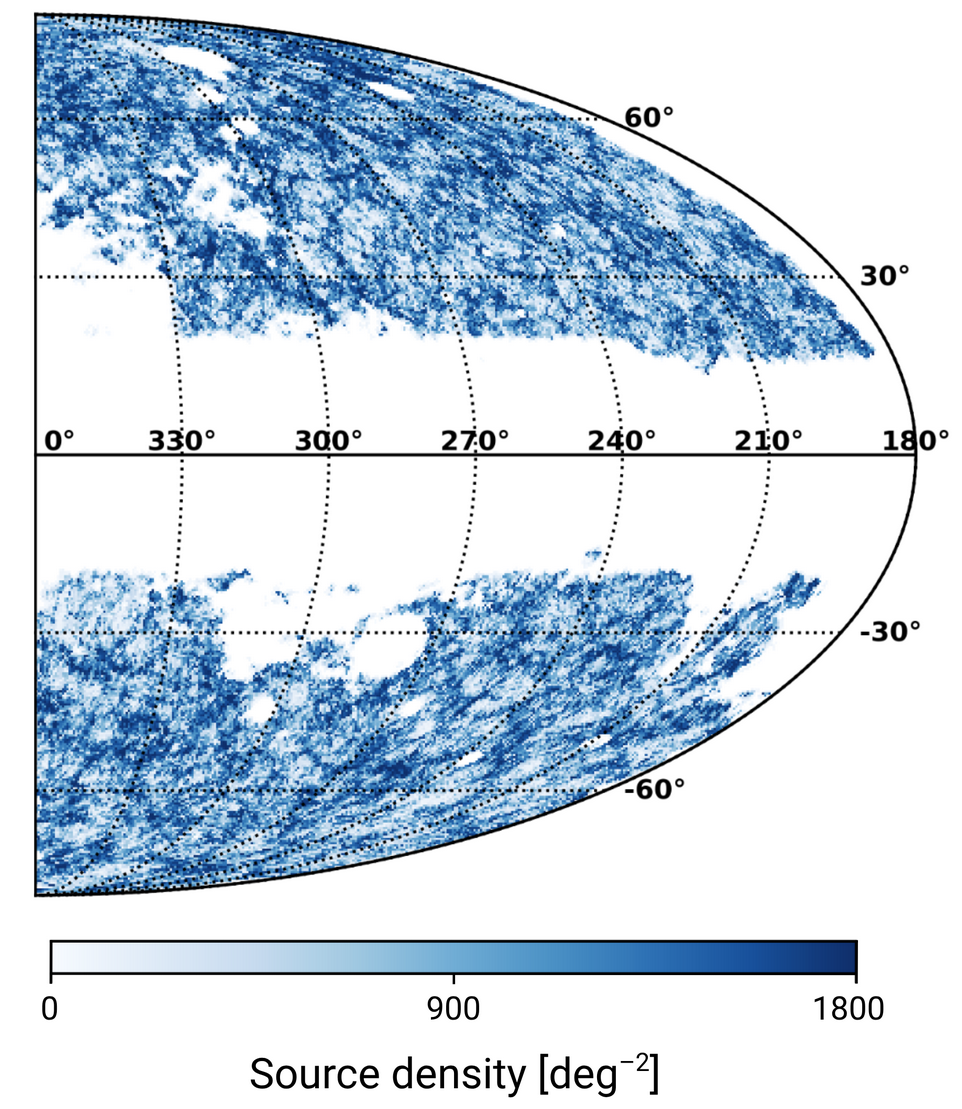}
    \caption{Sky map showing the source density of the parent galaxy sample, as defined in Sect.~\ref{sec:building_parent_sample}, in Galactic coordinates and Mollweide projection.} 
    \label{fig:skymap_parent}
\end{figure}

To build our parent galaxy sample, we applied the following selection criteria to the LS10 sources. Each source must:
\begin{enumerate}
     \item Be located in the western Galactic hemisphere (179.9442$^{\circ}$ $\leq$ galactic longitude $\leq$ 359.9442$^{\circ}$), to overlap with eROSITA-DE footprint \citep{Merloni2024}. 

    \item Have been observed with the {\it g, r, z}, and {\it W1} bands (i.e. \texttt{NOBS\_\{g,r,z,W1\}>0}). 

    \item Not be associated with known problematic photometry\footnote{DR10 bitmasks: \url{https://www.legacysurvey.org/dr10/bitmasks/}}. This includes removing objects with: \texttt{MASKBIT} 0 (secondary detections), \texttt{MASKBIT} 1 (objects touching Tycho sources with \texttt{MAG\_VT}$<$13 and Gaia stars with G$<$13), \texttt{MASKBIT} 12 (objects touching a pixel in a \textit{Siena Galaxy Atlas} large galaxy), \texttt{MASKBIT} 13 (objects touching a pixel in a globular cluster), as well as \texttt{FITBITS} 1, 6, 8, 9, 10, 11, 12, and 13.

    \item Not be a GAIA duplicate source (\texttt{TYPE!=DUP}).
    
    \item Have a high signal-to-noise ratio (S/N) in four bands\footnote{We denote the square root by \texttt{sqrt}.}: \texttt{FLUX\_\{g,r,z,W1\}*sqrt(FLUX\_IVAR\_\{g,r,z,W1\})}>3.
    
    \item Not have a large parallax or a large proper motion (PM), i.e. a stellar astrometry cut: \texttt{PARALLAX*sqrt(PARALLAX\_IVAR) < 5  \& sqrt[(PMRA*sqrt(PMRA\_IVAR))$^2$+ (PMDEC*sqrt(PMDEC\_IVAR))$^2$] < 5}.

    \item Not have stellar colours according to \citet{Mara_ctp_efeds}: \textit{z}$-$\textit{W1} > (0.8 $\times$ (\textit{g}$-$\textit{r})$-$1.2), where colours have been corrected for Galactic extinction.

    \item Have a low Galactic extinction: \texttt{E(B-V)}$<0.1$.

    \item Satisfy the \textit{z}-band magnitude cut: 12 $<$ z\_mag $\leq$ 20.

    \item Satisfy the redshift cut: $0.03\leq z\leq0.2$ (spectroscopic, if it exists; otherwise, photometric).

    \item Not lie within the mask defined by large foreground galaxies from the Heraklion Extragalactic Catalogue \citep[HECATE v2.0; ][]{Kyritsis2025, Kyritsis2026}.

    \item Not lie within a mask defined by the $R_{500}$ of eRASS1 X-ray clusters nor in the known spurious over-dense regions flagged in the eRASS1 catalogue (see text below for details).

\end{enumerate}
Figure \ref{fig:skymap_parent} shows the sky density distribution of the parent sample that consists of 5,352,526 galaxies obtained applying the above cuts, in Galactic co-ordinates\footnote{Approximately 400 sources are not shown via the Mollweide projection as the eROSITA `western Galactic hemisphere' cuts through the location of Sgr A*, $(l, b) = (359.9442^{\circ},-0.04616^{\circ})$, and not the origin of the Galactic co-ordinate system $(l,b)=(0,0)$.}.

We selected our sources in the {\it z}-band, as per criterion 9 above, as it offers more uniform sky coverage \citep[see Fig.~1 in][]{Saxena2024}. The bright- and faint-end limits were chosen to avoid saturation effects and to remain competitive with Dark Energy Spectroscopic Instrument (DESI) Bright Galaxy Survey \citep[BGS;][]{Hahn2023}, which selects objects with an {\it r}-band limit of 19.5 mag, respectively. To illustrate the effect of selection criteria $1-9$, we took four representative extragalactic (eROSITA-DE) LS10 sweep files (each 0.25$^{\circ}~ \times$ 0.25$^{\circ}$). An initial {\it z}-band magnitude cut retained around 1.1 million sources ($\sim$12\%). Subsequent photometric and observational quality cuts (criteria 2, 3, 4, and 5) removed around 65\% of these, leaving roughly 385,000 sources. The non-stellar selection criteria (cuts 6 and 7) further reduced the sample to around 207,000 sources. After additionally applying criterion 8, only $\sim$2\% of the total LS10 sources within these regions contribute to the parent sample. This value is sensitive to Galactic latitude, with higher losses in areas of high extinction; in this example, 87\% of the sources in the selected sweep files fell within low-extinction regions (\texttt{E(B–V)}$<0.1$).

We focused on the low-redshift Universe of $z\leq0.2$, where we can build complete samples of low-mass galaxies (see Sect. \ref{sec:mstar_completeness_paper3}), but did not explore $z<0.03$, as the photometric redshifts (photo-zs) are known to be problematic in the very local Universe \citep[e.g.][]{Hearin2010, Dahlen2013}. We started by curating an extragalactic ($z>0.002$) spectroscopic redshift (spec-z) compilation that includes most of the largest catalogues in the literature and aims towards high completeness. Full details of this extragalactic spec-z compilation are presented in Appendix \ref{sec:exgal_specz_compil}. The photo-zs are computed for all LS10 galaxies with good quality photometric information by \citet{Zhou2021, Zhou2023}. In Appendix \ref{sec:exgal_specz_compil}, we also validate the LS10 photo-zs explicitly for the case of our parent sample. \citet{Zhou2021, Zhou2023} show that these photo-zs are in excellent agreement with spectroscopic samples; for example, comparing photo-zs without {\it i}-band to spec-z galaxies from the Galaxy And Mass Assembly survey \citep[GAMA;][]{Driver2022} there is a normalised bias of 0.017 and an outlier fraction of 1.2\% (see Appendix \ref{sec:exgal_specz_compil} for formal definitions of these statistical metrics). Finally, for each galaxy in our parent sample, we assigned the best possible redshift (\texttt{BEST\_Z}) that exists in the following decreasing priority order: spec-z, photo-z with i-band or photo-z without i-band. Overall, $\sim 17$\% of our parent galaxy sample has spectroscopic redshifts.

Large, extended foreground galaxies can cause fragmentation of LS10 photometry. This is an effect whereby the Legacy Survey Tractor pipeline \citep{Dey2019} breaks up the galaxy into smaller `fragments', to which it then assigns an entry in the catalogue with its own (erroneous) properties. To limit the effect of fragmentation in our parent galaxy sample, we masked out the $D_{25}$ region around HECATE v2.0 galaxies, using the \texttt{HealSparse} Python library. The $D_{25}$ region is defined as the two-dimensional ellipse fitted to where the B-band brightness profile drops below 25~mag~arcsec$^{-2}$ \citep{Kovlakas2021, Kyritsis2025}. The size of this ellipse is encoded via the minor and major axis, along with the positional angle showing the projected orientation of each galaxy on the sky. This mask removes $<1\%$ of the LS10 galaxies selected with criteria $1-9$ above. Similarly, the masking of sources associated with \textit{Siena Galaxy Atlas} \citep[SGA;][]{Moustakas2023} galaxies, often suffering from fragmented optical data, aims to remove the matches to bright off-nuclear X-ray sources \citep[e.g. `ultra-luminous X-ray sources';][]{Fabbiano1989, Walton2022}, which are not the focus of this study (but may be important in getting a complete understanding of the BHOF; see discussion in Sect.~\ref{sec:imbh_discussion}).

Lastly, in Section \ref{apetool_methods} we describe how we compute X-ray fluxes via aperture photometry extracted at the locations of the optical galaxies. Given the relatively large point spread function (PSF) of eROSITA, nearby bright X-ray point-like and/or diffuse, extended sources, such as clusters of galaxies, can leak X-ray flux into our target apertures and cause biases. Therefore, we also created a \texttt{HealSparse} map of the $R_{\rm 500}$ region around the 12,247 eRASS1 clusters \citep{Bulbul2024, Kluge2024}. $R_{\rm 500}$ defines the radius at which the local density equals 500 times the critical density of the Universe and for clusters at a low redshift of $z<0.2$ it ranges from $\sim500$~kpc to 1~Mpc. For clusters with no $R_{\rm 500}$ information we set a value of 500~kpc. We masked all 12,247 clusters regardless of their redshift as even background sources ($z>0.2$) could bias the X-ray aperture photometry. In addition, we removed galaxies that lie in spurious X-ray regions based on the eRASS1 source over-density analysis described in Section 5.2. of \citet{Merloni2024}, including supernova remnants and PSF wings of bright point sources. Overall, these effects flagged a non-negligible $\sim$10\% of the parent galaxy sample and are visible in Figure \ref{fig:skymap_parent} (white ellipse mask regions with zero sources).

\subsection{Calculating host galaxy properties: stellar masses and star formation rates}

\subsubsection{Spectral energy distribution (SED) fitting with LePHARE}
\label{sec:lephare}

Using the accurate photo-zs described in the previous section, we computed host galaxy properties, such as the stellar mass and SFR using the Photometric Analysis for Redshift Estimation code \citep[LePHARE;][]{Arnouts1999, Ilbert2006, Ilbert2009, Shirley2026} with our six photometric bands\footnote{In favour of having a uniform and homogenised set of photometry available in LS10 for all galaxies in our parent sample, we did not include photometry from other multi-wavelength surveys that may only cover a fraction of the western Galactic hemisphere.} \textit{g, r, i, z, W1}, and \textit{W2} (when available). We used the \citet{BruzalandCharlot2003} stellar evolution models, taking a \citet{Chabrier2003} initial mass function (IMF) and \citet{Calzetti2000} dust extinction curves to introduce a reddening through varied \texttt{E(B-V)}. The modelled parameter grid included two metallicity values ($Z=0.008, 0.02$), two star formation histories, exponential decline [${\rm SFR}(t) \propto \exp (-t/\tau)$] and delayed exponential [${\rm SFR}(t) \propto t \times \exp (-t/\tau)$], with star formation timescales ($\tau$) equal to 0.1, 0.3, 1, 3, 5, 30 Gyr and 1, 3 Gyr, respectively. Importantly, we built galaxy-only SED models, which did not include models to explain the mid-infrared (MIR) emission from galactic and nuclear dust heated by star-formation or the AGN \citep[e.g.][]{Mullaney2011, Mor2012, Dale2014, Lyu2017}, or the big blue bump of the AGN accretion disk \citep[e.g.][]{Richards2006} (see Sect.~\ref{sec:grahsp_mstar} for how we treated sources that may have required these components). The $k$-correction for each input photometric band (used in Sect.~\ref{sec:mstar_completeness_paper3} to calculate absolute magnitudes) was derived from the apparent magnitude at the nearest redshifted band. Lastly, we did not apply emission line templates for the SED fitting with LePHARE, but did account for this in later re-fits with a more sophisticated Bayesian SED code (see Sect.~\ref{sec:grahsp_mstar}) as their inclusion can be important in deriving unbiased physical galaxy parameters \citep[e.g.][]{Mobasher2015, Santini2015}.

\begin{figure}
    \centering
    \includegraphics[width=\linewidth]{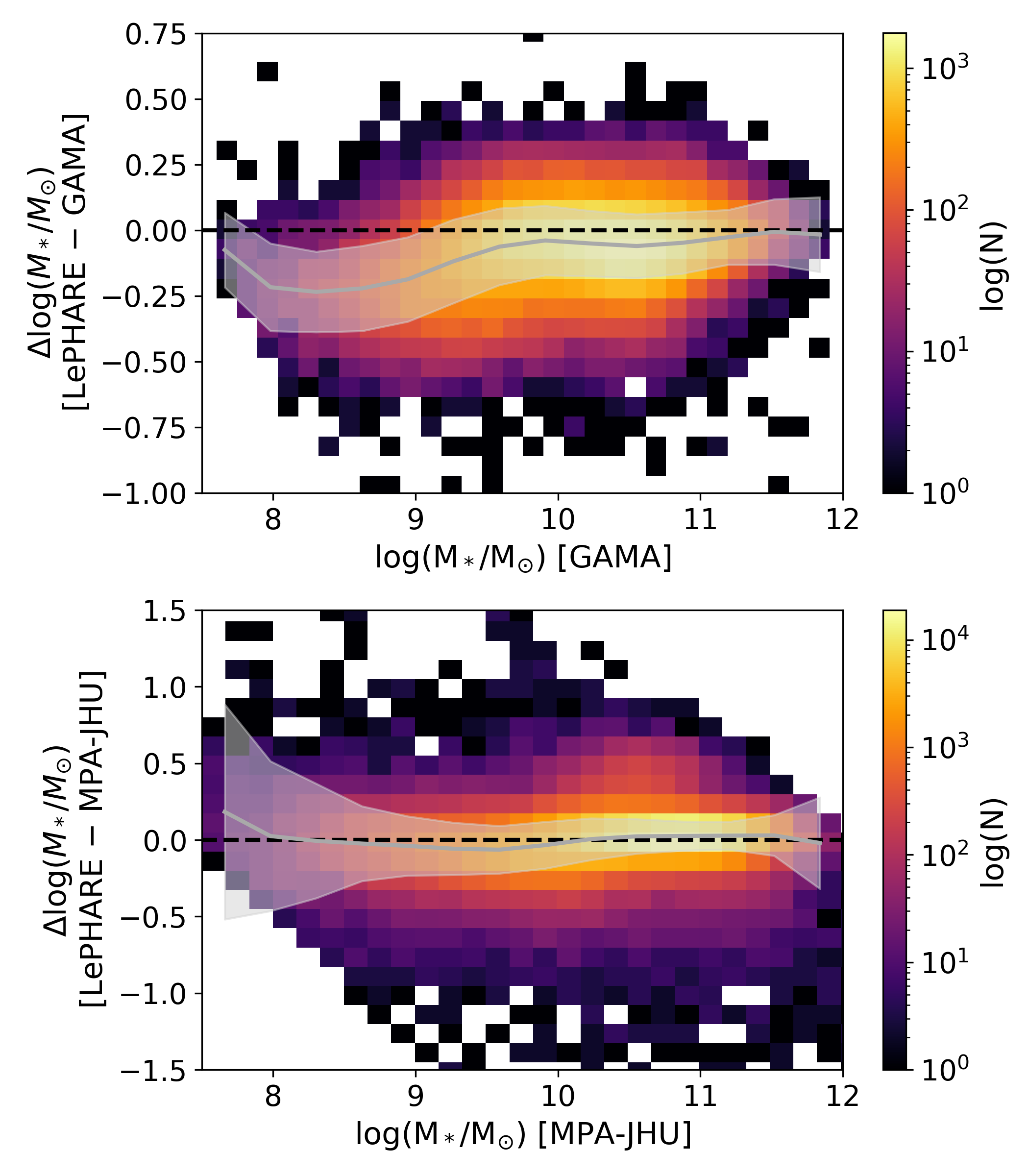}
    \caption{Stellar mass versus difference in stellar mass between measurements from the GAMA and MPA-JHU surveys and those computed with LePHARE for sources in common from our parent sample (with redshifts in agreement). The grey solid line marks the running median with the shaded region indicating the standard deviation of the difference in mass.}
    \label{fig:mstar_comparison}
\end{figure}

In order to validate the stellar masses computed with \mbox{LePHARE}, we matched (using optical co-ordinates) to the GAMA DR4 catalogue \citep{Driver2022}, which provides robust stellar mass estimates derived via SED fitting using extensive photometric coverage from the UV to the far-IR. We found 71,962 sources with $|(\texttt{BEST\_Z}-z_{\rm GAMA})|/(1+z_{\rm GAMA})<0.01$ and both LePHARE- and GAMA-computed $\log M_*/M_{\odot} > 7.5$. The top panel of Figure \ref{fig:mstar_comparison} shows the difference in stellar masses derived from GAMA and from LePHARE, as a function of GAMA stellar mass. The solid grey line marks the running median (offset) with the shaded region indicating the standard deviation of the difference in mass. It is clear that down to around $\log M_*/M_{\odot} \sim 9$ the stellar masses have a small downward offset of $\sim -0.05$~dex, however, this increases to $\sim -0.18$~dex between $\log M_*/M_{\odot} \sim 7.5-9$. The overall scatter (average standard deviation of the offset) is 0.14~dex, in agreement with the findings of \citet{Zou2019}. Similarly, in the bottom panel of Figure \ref{fig:mstar_comparison}, we show the comparison with the MPA-JHU DR8 catalogue, with 258,516 matched galaxies. We find a small positive offset of $\sim 0.005$~dex and scatter of $\sim0.24$~dex across the mass range. Overall, the stellar masses computed with LePHARE are (closely) consistent with previous work, albeit with significant scatter. 

We repeated the analysis to compare the LePHARE-derived SFRs to GAMA and MPA-JHU. We concluded that the SFR computed with only six photometric filters were highly biased and scattered, so we decided to not use them further in this study \citep[see Sect.~\ref{sec:non_agn_processes} for how we set a conservative upper limit on the global star formation properties of the sample using the star-formation main sequence;][]{Speagle2014}.

\subsubsection{Stellar mass estimates including an AGN and mid-IR emission component with \texttt{GRAHSP}}
\label{sec:grahsp_mstar}

In order to further reduce the stellar mass bias and scatter of our final sample, we built a machine-learning-based classifier to identify sources whose LePHARE-derived stellar masses are unreliable (given the criterion mentioned below), due to the lacking MIR emission and AGN models. For these sources, we then re-compute their physical galaxy properties with a more sophisticated (but more computationally intensive) approach using the Bayesian algorithm called \texttt{GRAHSP} \citep[][]{Buchner2024_grahsp}. We used the same photometry, stellar population models, and IMF as the LePHARE run, but now included continuum and line emission from ionised gas \citep{Boquien2013, Boquien2019}, dust attenuation using the Small Magellanic Cloud (SMC) attenuation curve \citep{Prevot1984}, reprocessing in the IR using the \citet{Dale2014} templates and several AGN components. The training sample for the classifier was constructed by randomly sampling the parent galaxies in stellar mass and redshift space, with an additional emphasis on X-ray detected sources, which are more likely to host AGN and therefore represent potentially outlying stellar-mass systems. Further details on the building of this training sample, the SED-fitting using \texttt{GRAHSP} and the performance of the classifier are given in Appendix \ref{appendix:details_ml_grahsp}. We note that computing stellar masses using \texttt{GRAHSP} for the whole parent sample of over five million sources is computationally costly and largely unwarranted, given that AGN are relatively rare events and Figure \ref{fig:mstar_comparison} already shows good agreement with previous stellar mass estimates across the mass range \citep[see also][for a similar approach]{Aird2018}. 

We trained our eXtreme Gradient Boosting machine learning classifier \citep[XGBClassifier;][]{Chen2016} to distinguish galaxies for which the LePHARE- and \texttt{GRAHSP}-derived stellar masses are consistent (inliers) or inconsistent within 0.4\,dex (outliers), using numerous features including optical properties, X-ray luminosity and LePHARE SED fit statistics ($\chi^2$). Overall, our classifier achieves an accuracy of 92\% (see Figs.~\ref{fig:mass_outliers_training} and \ref{fig:confusion_matrix_and_PRcurve}, left) and high recall for both inliers (93\%) and outliers (86\%). Applying the trained classifier to the full parent sample results in 32,548 galaxies (about 0.6\%) deemed to have unreliable stellar masses, likely due to unaccounted-for AGN emission. For these objects, we re-computed stellar masses using \texttt{GRAHSP}, replacing the LePHARE-derived median values with those obtained from \texttt{GRAHSP}. An exception was made for galaxies with unconstrained \texttt{GRAHSP}-derived stellar masses, defined as those with uncertainties exceeding 2.5\,dex at the 2$\sigma$ level, for which we instead adopted the 2$\sigma$ upper-limit value. The unconstrained cases account for 9,423 of the 32,548 sources. These are the final stellar masses used for the rest of this work, including deriving the mass completeness limits (Sect.~\ref{sec:mstar_completeness_paper3}) and the incidence distributions in Sect.~\ref{sec:dwarf_incidence_results}. 

%%A&A requested layout makes the figure too small
% \begin{figure*}
% \sidecaption
% \includegraphics[width=12cm]{FIGURES/0_zvsMstar_parent.png}
% \caption{Stellar mass versus redshift of the parent sample of LS10 galaxies (background blue shaded density grid, with darker colours indicating higher number of sources as in the colour bar), overlaid with the X-ray detected low-mass galaxies (light red circles) and high-mass galaxies (small gray dots). $2\sigma$ upper limits in stellar mass are indicated with downward arrows. The 90\% mass-completeness curve for the high- and low-mass galaxies is plotted in yellow (dashed) and dark green (dot-dashed), respectively. The 70\% mass-completeness curve, used for computing the incidence in this work, is plotted with a solid light green curve.}
% \label{fig:sample_zm_distrib}
% \end{figure*}

\begin{figure*}
    \centering
    \includegraphics[width=0.7\linewidth]{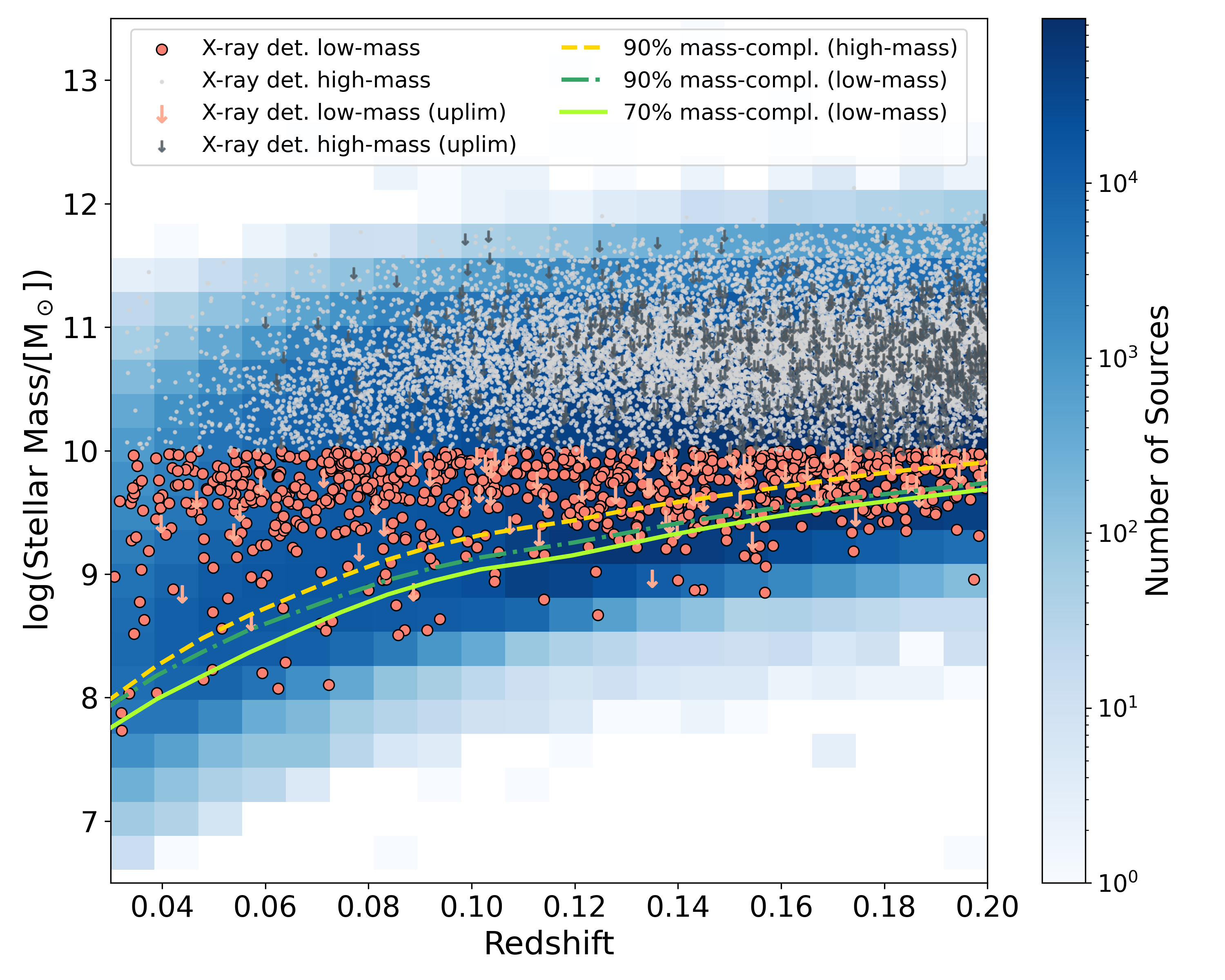}
    \caption{Stellar mass versus redshift of the parent sample of LS10 galaxies (background blue shaded density grid, with darker colours indicating higher number of sources as in the colour bar), overlaid with the X-ray detected low-mass galaxies (light red circles) and high-mass galaxies (small grey dots). $2\sigma$ upper limits in stellar mass are indicated with downward arrows. The 90\% mass-completeness curve for the high- and low-mass galaxies is plotted in yellow (dashed) and dark green (dash-dotted), respectively. The 70\% mass-completeness curve, used for computing the incidence in this work, is plotted with a solid light green curve. }
    \label{fig:sample_zm_distrib}
\end{figure*}

\subsection{Stellar mass completeness}
\label{sec:mstar_completeness_paper3}

Given our magnitude-limited sample, we derived a redshift-dependent stellar mass completeness cut, to ensure that our later results would not be biased by our optical selection. To do so, we first $k$-corrected the $z$-band magnitudes of our parent galaxy sample to derive an absolute $z$-band magnitude ($M_z$) and computed their mass-to-light ratios, $M_*/L$, where
\begin{equation}
    L/L_{\odot}=10^{-0.4(M_z - M_{\odot,z})},
    \label{eq:lumin_from_mag}
\end{equation}
and the absolute magnitude of the Sun in the $z$-band is $M_{\odot,z}=4.50$ \citep{Willmer2018}. Given the variance in $M_*/L$ with mass, whereby more massive galaxies tend to be quiescent, red and less luminous compared to lower mass, bluer, star-forming galaxies, we split our sample into low-mass and high-mass galaxies to derive more physically motivated completeness limits. 

Taking first the low-mass galaxy subsample, we binned our sample into small redshift bins of $\Delta z=0.01$ and computed the corresponding absolute $z$-band magnitude limit and luminosity (Eq.~\ref{eq:lumin_from_mag}), given by the survey selection: $z\leq$ 20~mag. Then we took the upper 50th percentile of the $M_*/L$ ratios in each $\Delta z$ to signify the galaxies that are as luminous as, or less luminous than, the average for that redshift range. Multiplying this limiting luminosity and $M_*/L$ ratio together and taking the upper 70th or 90th percentile gives the limiting mass, at a level of 70\% and 90\%, respectively, that a galaxy can have and still enter into the sample selection. The same process was repeated to derive the mass-completeness limit for the high-mass galaxy subset. Similar methods have been employed, for example, by \citet{Pozzetti2010}, \citet{Moustakas2013}, and \citet{Guetzoyan2025}. 

Figure \ref{fig:sample_zm_distrib} shows the distribution in stellar mass and redshift of the parent galaxy sample (background blue shaded density grid) and the different mass-completeness curves for the low- and high-mass galaxy subsample. Out of the total $\sim$5.35 million parent galaxies, there are $\sim$2.77 million low-mass and $\sim$2.58 million high-mass galaxies. Since the low-mass galaxy mass-completeness function is rather steep compared to the change in stellar mass value, we adopted the 70\% limit (solid light green curve) to maximise source statistics. In comparison, all high-mass galaxies lie above the 90\% mass-completeness curve (dashed yellow) and so are considered complete in the redshift range probed here. The X-ray detected low-mass (light red) and high-mass (grey) galaxies are also shown in Figure \ref{fig:sample_zm_distrib} and will be described in the following section. The sources with unconstrained \texttt{GRAHSP}-derived stellar masses are plotted with downward arrows at the upper $2\sigma$ confidence level.

\section{eROSITA X-ray detected sample}

In this section we discuss the homogeneous determination of X-ray fluxes using aperture photometry centred on the location of the optical parent galaxy sample co-ordinates. This differs from past work dealing with low-mass galaxies which has typically cross-matched pre-existing optical and X-ray catalogues to find X-ray detected sources, after consideration of chance associations and spurious detections \citep[e.g.][]{Latimer2021, Birchall2020, Sacchi2024, Bykov2024}. Although the methods are different, the problems are similar and we elaborate here on the extensive cleaning and validation procedures we took to make sure that: (i) the X-ray detection is real; (ii) the optical host is the statistically favoured counterpart; and (iii) the X-ray flux is associated with the central AGN and not with other galaxy processes. 

\subsection{X-ray aperture photometry using \texttt{apetool}}
\label{apetool_methods}

We used aperture photometry using the \texttt{apetool} task (\texttt{v1.28 eSASSusers\_240410.0.4}) from the eROSITA Science Analysis Software System \citep[eSASS;][]{Brunner2022} centred on the optical co-ordinates of our galaxy sample to compute eRASS:4 X-ray counts in the main eROSITA band ($0.2-2.3$~keV). \texttt{apetool} computes the total counts ($N$) within the specified aperture, fixed here at the encircled energy fraction (EEF) of 75\% (i.e. the radius of the PSF at which 75\% of the energy is contained), corresponding approximately to a radius of $\sim$~30\arcsec.

The background counts ($C_{\rm B}$) were computed from the source-subtracted background map. We chose to set the radius within which such source-subtraction takes place to 37.5\% of the EEF, which corresponds to roughly three times the 50th percentile of the eRASS:4 positional error (i.e. $\sim$ 10\arcsec; shown in Fig.~\ref{fig:cutout} with the dashed yellow circle). In practice, this means that X-ray sources with centroids at radii between 37.5\% and 75\% of the EEF are not being removed from the source map, and thus they will contribute to an increased local background level as depicted in the schematic in Figure \ref{fig:apetool_explanation}\footnote{We note that due to this inner radius, we are not sensitive to off-centre X-ray emission from large, spatially extended galaxies.}. 

The \texttt{apetool} algorithm computes a per-source Poisson tail probability ($P_{\rm thresh}$) defined as the probability of observing $N\geq N_{\rm min}$ counts given a Poisson distributed background with expected counts, $C_{\rm B}$, where $N_{\rm min}$ denotes the minimum number of counts required for detection at the chosen false-positive probability threshold \citep{Georgakakis2008}: 
\begin{equation}
P_{\rm thresh} = P(N \geq N_{\rm min} \mid C_{\rm B}) = \sum_{N = N_{\rm min}}^{\infty} \frac{C_{\rm B}^{N} e^{-C_{\rm B}}}{N!}.
    \label{eq:poiss_null_prob}
\end{equation}
The smaller the value of $P_{\rm thresh}$, the less likely it is that the total observed counts arise from a background fluctuation, and the more significant the detection of an astrophysical source. This probability can be used to define a minimum threshold for classifying a source as `X-ray detected'. We did so by firstly calculating the running median of $P_{\rm thresh}$ as a function of the one-band (1B; $0.2-2.3$~keV) detection likelihood, \texttt{DET\_LIKE\_0}, of the eRASS:4 source catalogues processed with pipeline version c030 \citep{Ramos-Ceja2026}. Then, using the eRASS1\footnote{Although eRASS:4 is deeper than eRASS1 by a factor of $\sim4$, the spurious fraction results from the eRASS1 `digital twin' still apply at a fixed \texttt{DET\_LIKE\_0} (Seppi et al., in prep.). This is consistent with the fact that, for a given source, the increased exposure time in eRASS:4 can lead to a correspondingly higher \texttt{DET\_LIKE\_0}.} `digital twin' simulations by \citet{Seppi2022}, we can infer the spurious detection fraction when opting for a \texttt{DET\_LIKE\_0} cut of $\geq10$ (D10) or $\geq15$ (D15). For D10, corresponding to a median value of $P_{\rm thresh}\leq 1 \times 10^{-4}$, \citet{Seppi2022} compute a spurious fraction of $\sim$1\% (see their Table 3). However, as there is a relatively large scatter on the running median, we chose to adopt a more stringent cut at D15, which corresponds to a median threshold of $P_{\rm thresh}\leq 4 \times 10^{-6}$, similar to \citet{Georgakakis2008}, with a spurious detection fraction of $\sim0.042$\% (but still potentially reaching up to 1\%, given the scatter). This threshold\footnote{We note that the full threshold is defined as $0\leq P_{\rm thresh}\leq 4 \times 10^{-6}$, as \texttt{apetool} sets sources with zero aperture counts to $P_{\rm thresh}=-9.99$, which would otherwise be considered detections.} thereby sets a minimum number of source photons required, relative to the local background, to be considered a statistically significant X-ray detection.

We converted the aperture source counts (total minus background counts) to a net source count rate in the soft $0.2-2.3$~keV band by dividing by the EEF and the mean vignetted exposure time ($t_{\rm exp}$) at the galaxy position (computed by \texttt{apetool} from the exposure maps). Then, to calculate the soft X-ray flux, we divided the count rate by a constant energy correction factor (ECF\footnote{\url{https://erosita.mpe.mpg.de/dr1/eSASS4DR1/eSASS4DR1_arfrmf/eROSITA_ECF_tutorial.pdf}}) of $1.074 \times 10^{12}$~counts~cm$^{2}$~erg$^{-1}$ \citep[Table D.1 in][]{Brunner2022}, which is derived from an absorbed power law fit with photon index, $\Gamma=2.0$, and a hydrogen (foreground) absorption column density of $N_H=3\times10^{20}$~cm$^{-2}$, and encodes information about the eROSITA effective area: 
\begin{equation}
    F_{0.2-2.3keV}=(N-C_{\rm B})/(t_{\rm exp} \cdot EEF ~\cdot ECF).
    \label{eq:counts_to_flux}
\end{equation}
Finally, we converted the soft X-ray flux to (rest-frame) $2-10$~keV luminosity, using the same photon index as above (i.e. we assume all detected sources are not affected by intrinsic obscuration) and the source redshift. The assumption about a single photon index for our statistical study is justified given the relatively narrow Gaussian distributed photon indices for the AGN population \citep[e.g.][]{Nandra1994, Brandt2015}. The assumption about no intrinsic obscuration is more complex, but using the eROSITA Final Equatorial Depth Survey \citep[eFEDS; ][]{Brunner2022,Mara_ctp_efeds,Teng_eFEDS} as a proxy for the eRASS:4 selection, we find that $\sim$90\% of eFEDS X-ray sources with LS host galaxies matching our parent sample (i.e. $0.03<z<0.2$ and z\_mag $\leq20$) have $N_H <10^{21.5}~\rm{cm}^{-2}$, where $N_H$ is the neutral hydrogen column density. 
Additionally, \citet{Igo2024} explicitly demonstrate that when using only an unobscured selection function for eROSITA, the effects of obscuration appear confined to the lower specific accretion rate regime and so have limited implications on the main conclusions on this work. We discuss the role of obscured AGN for our incidence measurements further in Section~\ref{sec:duty_cycle}.

Running \texttt{apetool} on the parent galaxy sample results in 20,830 X-ray detections (i.e. $0\leq P_{\rm thresh}\leq 4 \times 10^{-6}$). This is split into 4,121 X-ray detected low-mass, $\log M_*/M_{\odot}\leq10$, galaxies and 16,709 X-ray detected high-mass, $\log M_*/M_{\odot}>10$, galaxies.

\subsection{Cleaning spurious associations using the eROSITA X-ray and counterpart catalogues}
\label{sec:xrayvalidation_catalogues}

The $P_{\rm thresh}$ value computed for each galaxy gives an indication of the significance of the X-ray detection; however, we must still assess the reliability of the association of the X-ray photons with the host galaxy. In principle, the advantage of extracting X-ray photometry at the location of each optical galaxy is that it can homogeneously identify the X-ray emission originating from the source without requiring additional multi-wavelength catalogues or cross-matching. However, for the low-mass (and smaller physical size) galaxies in particular, this is complicated by the broad PSF of eROSITA, the rather large eROSITA positional uncertainty and the relatively high X-ray source density, dominated by more massive host galaxies, background AGN, stars, and galaxy clusters. 

Quantitatively, we can show that we expect a high number of spurious associations in our parent sample apertures, especially for the low-mass sample. Taking the $\sim$2.77 million low-mass parent galaxy apertures, each with radius 10\arcsec\ (recall that the apertures themselves are $\sim$30\arcsec\ but sources outside $\sim$10\arcsec\ are not subtracted from the source map and so they contribute to the local background emission), gives a total area of 67~deg$^{2}$. Then, we find that the average sky density of eRASS:4 \texttt{DET\_LIKE} $>10$ extragalactic sources (excluding the south ecliptic pole and Galactic plane at $b\pm20^{\circ}$) is $\sim60$~deg$^{-2}$. Therefore, if these $\sim$2.77 million apertures were randomly placed on the sky, one would expect 4,020 chance associations, a very high number considering that the actual number of detections at the positions of low-mass galaxies is 4,121\footnote{An alternative test is to first match the $\sim$2.77 million low-mass parent sample to the $\sim$1.39 million eRASS:4 \texttt{DET\_LIKE} $>10$ extragalactic X-ray sources within a radius of 10\arcsec, finding 5,582 `real' matches. Then we shift the low-mass parent sample apertures by 60\arcsec\ in their declinations, remove real low-mass galaxies within 10\arcsec\ of the shifted positions ($\sim21,000$ removed), and rematch this catalogue to the same $\sim$1.39 million X-ray sources, providing us with 4,883 chance associations.}. These crude estimates indicate that more than $\sim$90\% of our `detections' may be spurious associations.

\begin{figure*}[h!]
\centering
\adjustbox{max height=0.7\textheight}{
\begin{tikzpicture}[node distance=1.5cm and 2cm]

\node (start) [startstop,  align=center] {Parent galaxy sample: 5,352,526\\ \textcolor{red}{Low-mass ($\log M_*/M_{\odot}\leq 10$): 2,774,970} \\ \textcolor{blue}{High-mass ($\log M_*/M_{\odot} > 10$): 2,577,556}};
\node (xray) [startstop, below=0.5cm of start, align=center] {X-ray sub-sample (\texttt{O $\leq$ APE\_POIS $\leq$ 4e-6}): \\ 
\textcolor{red}{Low-mass: 4,121} \\ 
\textcolor{blue}{High-mass: 16,709}};

\node (q1) [decision, below=0.5cm of xray] {Q1: Has match within 30\arcsec\ to eRASS:4 X-ray source?};

\node (res1a) [process, below right=1cm and 0.5cm of q1, xshift=-3cm, align=center] {\textcolor{red}{3,979 matches} \\ \textcolor{blue}{16,490 matches} \\(including duplicate X-ray sources)};

\node (res1b) [process, below left=1cm and 0.5cm of q1, xshift=0cm,align=center] {\textcolor{red}{142 sources} \\ \textcolor{blue}{219 sources}};

\node (res1b_outcome) [final, below=0.5cm of res1b, align=center] {\textcolor{red}{Keep 24/142 X-ray sources (vis. inspect.)} \\ \textcolor{blue}{Discard all 219 sources (no vis. inspect)}
};

\node (q2) [decision, below= 0.5cm of res1a, align=center] {Q2: Has match within 30\arcsec\ to NWAY CTP in eRASS:4-CTP catalogue?};

\node (res2a) [process, below right=1cm and 0.5cm of q2, xshift=-4cm, align=center] {\textcolor{red}{3,945 matches} \\ \textcolor{blue}{16,350 matches}};

\node (res2b) [process, below left=1cm and 0.5cm of q2, xshift=2cm, align=center] {\textcolor{red}{34 sources} \\ \textcolor{blue}{140 sources}};

\node (res2b_outcome) [final, below=0.5cm of res2b, align=center] {\\ \textcolor{red}{Keep 9/34 X-ray sources (vis. inspect)} \\ \textcolor{blue}{Keep* 2/140, discard rest (no vis. inspect)}};

\node (q3) [decision, below=2.2cm of res2a, xshift=-5cm,align=center] {Q3: Unique optical galaxy -- X-ray source -- NWAY CTP triplet \\and the same X-ray match?};
\node (res3a) [process, below right=1cm and 0.5cm of q3,xshift=-4cm, align=center] {\textcolor{red}{3,839 sources} \\ \textcolor{blue}{15,021 sources} };

\node (res3b) [process, below left=1cm and 0.5cm of q3,align=center] {\textcolor{red}{106 sources} \\ \textcolor{blue}{1,329 sources}};

\node (res3bQ4) [decision, below=0.5cm of res3b, align=center] {Q4: Optical galaxy in agreement \\with best NWAY CTP?};

\node (Q4repeat_b) [process, below left=1cm and 1cm of res3bQ4, xshift=2cm, align=center] {\textcolor{red}{86 sources} \\ \textcolor{blue}{705 sources} \\ → Discard};
\node (Q4repeat_a) [final, below right=1cm and 1cm of res3bQ4, xshift=-2cm, align=center] {\textcolor{red}{Keep 20 X-ray sources} \\
\textcolor{blue}{Keep 624 X-ray sources} };

\node (q5) [decision, below=3cm of q3,xshift=3cm, align=center] {Q4: Optical galaxy in agreement \\with best NWAY CTP?};

\node (res5b) [process, below left=2.5cm and 1.5cm of q5, xshift=2cm, align=center] {\textcolor{red}{3,018 sources} \\ \textcolor{blue}{3,029 sources} \\ → Discard};

\node (res5a) [final, below right=1.5cm and 0.5cm of q5, xshift=-3.5cm, align=center] {\textcolor{red}{Keep 821 X-ray sources} \\ \textcolor{blue}{Keep 11,992 X-ray sources}};

\draw [arrow] (start) -- (xray);
\draw [arrow] (xray) -- (q1);

\draw [arrow] (q1) -- node[midway, above right] {Yes} (res1a);
\draw [arrow] (q1) -- node[midway, above left] {No} (res1b);

\draw [arrow] (res1b) -- (res1b_outcome);
\draw [arrow] (res2b) -- (res2b_outcome);

\draw [arrow] (res1a) -- (q2);

\draw [arrow] (q2) -- node[midway, above right] {Yes} (res2a);
\draw [arrow] (q2) -- node[midway, above left] {No} (res2b);

\draw [arrow] (res2a) -- (q3);

\draw [arrow] (q3) -- node[midway, above right] {Yes} (res3a);

\draw [arrow] (res3a) -- (q5);

\draw [arrow] (q3) -- node[midway, above left] {No} (res3b);

\draw [arrow] (res3b) -- (res3bQ4);

\draw [arrow] (q5) -- node[midway, above left] {No} (res5b);

\draw [arrow] (q5) -- node[midway, above right] {Yes} (res5a);

\draw [arrow] (res3bQ4) -- node[midway, above left] {No} (Q4repeat_b);

\draw [arrow] (res3bQ4) -- node[midway, above right] {Yes} (Q4repeat_a);

\end{tikzpicture}
}
\caption{Flowchart showing the decision tree used to validate the X-ray detections and the associations with their host galaxy counterparts (CTPs). *Two sources in the `Q2:no' branch that were kept in the original visual inspection iteration using only LePHARE-derived stellar masses (see Sect.~\ref{sec:grahsp_mstar}) changed from low- to high-mass, and thus are kept as part of the final high-mass sample.}
\label{fig:flowchart}
\end{figure*}
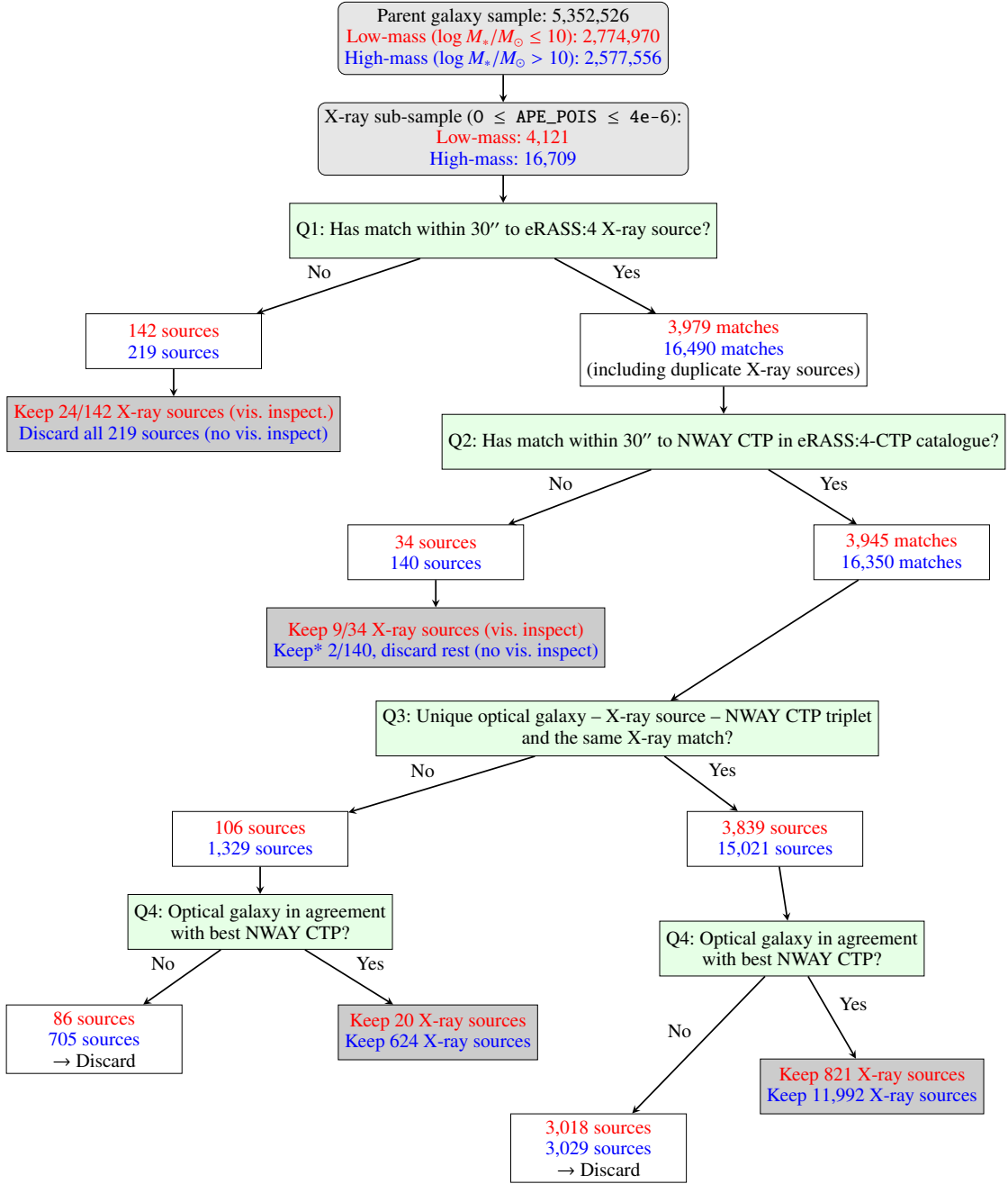

In Appendix \ref{appendix:spur_flowchart} we provide full details of the steps taken to validate the X-ray detections (found via \texttt{apetool}) and their optical host associations (found via NWAY; \citealt{maraNWAY2018}), along with the visual inspection procedure carried out for the low-mass galaxies in order to clean our sample from potentially spurious contaminants. The flowchart in Figure \ref{fig:flowchart} describes these steps in a concise manner.

\begin{figure*}
    \centering
    \includegraphics[width=\linewidth]{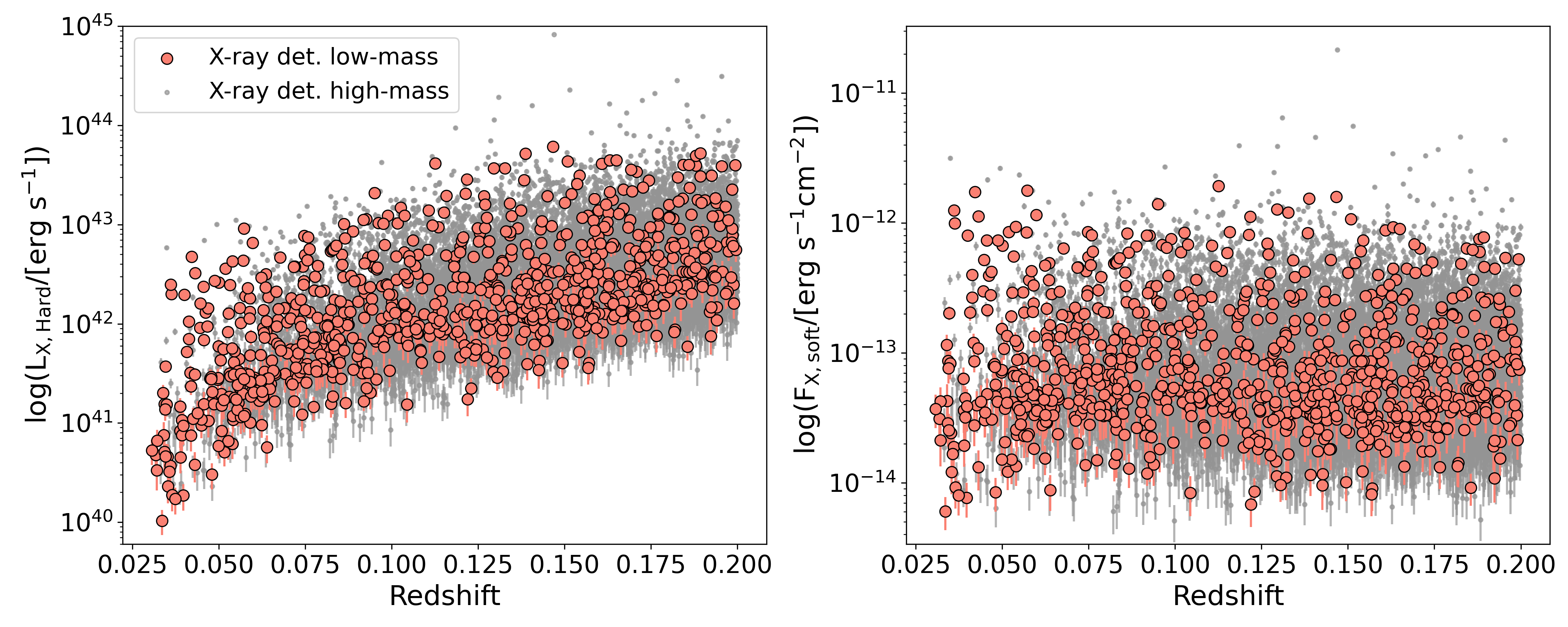}
    \caption{Left: Distribution of rest-frame $2-10$~keV luminosity versus redshift of the X-ray detected low-mass (light red) and high-mass (grey) galaxies from the parent sample.  Right: Distribution of observed $0.2-2.3$~keV flux versus redshift of the same X-ray detected sources. Uncertainties were calculated as described in the text and are often too small to be seen.} 
    \label{fig:z_vs_lumin}
\end{figure*}

Overall, we can confirm 874 X-ray detected low-mass galaxies out of 4,121, all of which have been visually inspected. This means that only 21\% of the initial detections from the aperture photometry method are found to be secure associations, consistent with the high level of spurious contamination predicted from our simple estimates above \citep[see also the extensive cleaning from spurious sources and counterpart associations done in e.g.][]{Birchall2020, Latimer2021, Sacchi2024, Bykov2024}. For the high-mass galaxies, instead, we confirm 12,618 X-ray candidates, using the same method (see Fig.~\ref{fig:flowchart}). This corresponds to 76\% of the original X-ray detections, showing that the high-mass galaxies suffer much less from contamination compared to the low-mass sample. Figure \ref{fig:z_vs_lumin} shows the distribution of rest-frame $2-10$~keV luminosities (left panel) and observed $0.2-2.3$~keV fluxes (right panel) for the sample of X-ray detected low-mass (light red) and high-mass (grey) galaxies as a function of redshift. The 1$\sigma$ uncertainty on the net counts was obtained by adding the Poisson errors on the total and background counts, both independent Poisson variables, in quadrature (i.e. $\sqrt{N+C_{\rm B}}$). This was used in Eq.~\ref{eq:counts_to_flux} to derive the uncertainty on the soft flux and luminosity as shown in Figure \ref{fig:z_vs_lumin}.

\subsection{Determining the origin of the X-ray emission}
\label{sec:non_agn_processes}

While nuclear X-ray emission is one of the most effective indicators of a central accreting black hole, the galaxy can also produce X-ray emission from other processes that are unresolved at the spatial resolution of eROSITA. For example, galactic X-ray emission can come from low- and high-mass X-ray binaries (XRBs). The X-ray emission from the former scales with the stellar mass of the galaxy (tracing the long-lived, older stellar population) and that of the latter with the SFR \citep[tracing the short-lived luminous stellar population; see review by][]{Fabbiano2006}. \citet{Lehmer2016} were the first to parametrise this scaling in the form of:
\begin{equation}
    L_{\rm X, G} = \alpha(1+z)^\gamma M_* + \beta(1+z)^\delta SFR^\theta.
    \label{eq:aird_lx_gal}
\end{equation}
This has since been refined by \citet{Aird2017}, who used a similar method to the one described in Sect.~\ref{sec:plambda_methods} to find a peaked probability distribution as a function of luminosity at $L_X<10^{42}$~erg~s$^{-1}$. This peak, when analysed as a function of stellar mass, is attributed to the `X-ray main sequence of star-formation'. We used the best-fit parameters to  Eq.~\ref{eq:aird_lx_gal} found by \citet{Aird2017}, which are the following: $\log \alpha=28.81 \pm 0.08$, $\gamma= 3.90\pm0.36$, $\log \beta = 39.50\pm0.06$, $\delta = 0.67\pm0.31$ and $\theta = 0.86\pm0.05$.

Hot gas can also emit a faint and diffuse X-ray background via thermal bremsstrahlung processes. This is expected to be around an order of magnitude fainter than the XRB component, as the soft X-ray emission scales only as $L_{\mathrm{Gas}}=(8.3\pm0.1) \times 10^{38} ~ {\rm SFR}/[M_{\odot}~yr^{-1}]$ \citep[][see also Fig. 8 of \citealt{Lehmer2016} for a comparison and recent work by \citealt{Kyritsis2025}]{Mineo2012}. The relation from \citet{Aird2017} inherently includes this component as the authors do not attempt to separate XRB and hot-gas related X-ray emission.

\begin{figure}
    \centering
    \includegraphics[width=\linewidth]{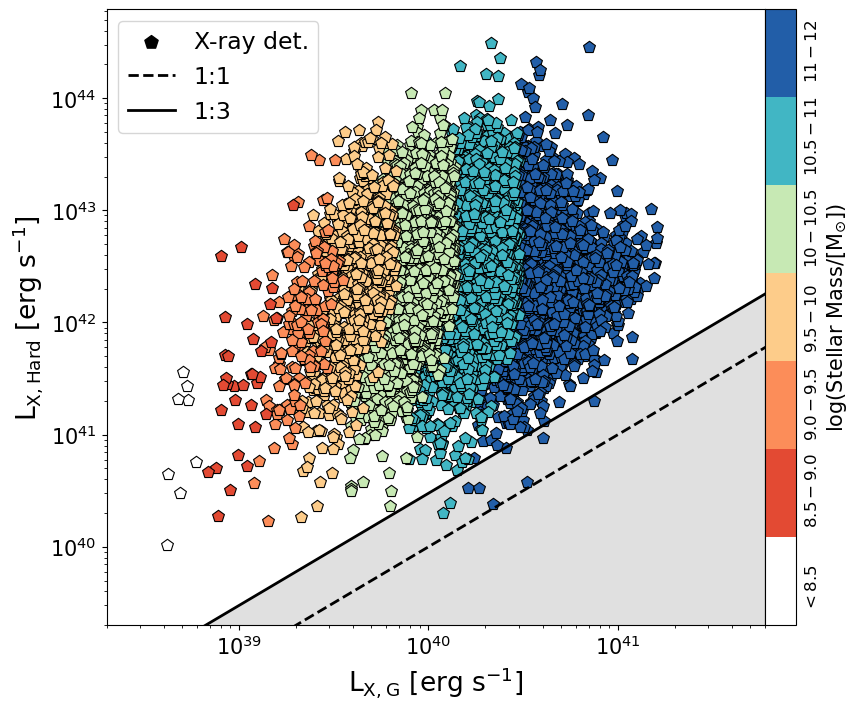}
    \caption{Comparison of the $2-10$~keV luminosity for the X-ray detected sources (colours denote a given stellar mass range as shown by the colour bar) on the y axis, versus the expected X-ray emission from galactic processes, derived using Eq.~\ref{eq:aird_lx_gal} with a SFR equal to the main-sequence value for a given stellar mass, plotted on the x axis. The dashed and solid black lines represent the 1:1 and 3:1 relations, respectively.}
    \label{fig:xrb_hot_gas_removal_IMBH}
\end{figure}

As discussed in Section \ref{sec:lephare}, the SFRs derived with only six photometric bands using LePHARE are not reliable. Therefore, we chose to estimate the SFR using the stellar mass of the galaxies, which we have also validated for AGN-dominated sources using \texttt{GRAHSP}. Given the bluer, star-forming nature of our low-mass galaxy sample \citep[\textit{g-r} $\lesssim0.75$; see also e.g.][]{Kauffmann2003, Baldry2004, Papaderos2008, Kyritsis2025}, we chose to adopt a SFR that is on the main sequence of star formation for a given stellar mass and redshift \citep[Eq. 28 of][]{Speagle2014} and used it in Eq. \ref{eq:aird_lx_gal} to compute the X-ray emission from galactic processes for all galaxies in our sample\footnote{For high-mass galaxies, this estimate will be a conservative upper limit as they tend to be more quiescent systems.}. Figure \ref{fig:xrb_hot_gas_removal_IMBH} shows the comparison of the $2-10$~keV X-ray emission from the parent sample of galaxies versus the expected emission from galactic processes. All X-ray detected low-mass galaxies (and the vast majority of the high-mass ones) lie above the 3:1 solid black line, meaning that their X-ray emission is more than a factor three greater than that expected from galactic processes, and thus AGN-dominated. Therefore, we do not expect significant effects on the AGN incidence distributions, especially for the higher-X-ray luminosity sources (but see Sect.~\ref{sec:plambda_methods} on how we account for this contribution explicitly in the Bayesian formalism).

\section{Methodology}
\label{sec:plambda_methods}

This section presents the Bayesian methodology to compute the distribution of specific black hole accretion rates across the mass range covered by the parent galaxy sample. From here on, we consider only mass-complete samples of galaxies (up to 70\% for low-mass galaxies and $\gg$90\% complete for high-mass galaxies; recall Sect.~\ref{sec:mstar_completeness_paper3}).

Following \citet{Aird2012}, we can define a proxy for the Eddington ratio that we denote as the specific black hole accretion rate, $\lambda_{\rm SAR}$: 
\begin{equation}
     \lambda_{\rm SAR}=\frac{L_{\rm Bol}}{L_{\rm Edd}}=\frac{k_{\rm bol}(L_{\rm 2-10~keV})\cdot~L_{\rm 2-10~keV}}{1.26 \times 10^{38}~\rm{erg~s^{-1}}~M_{\rm BH}/M_{\odot}}, 
    \label{eq:lambda_edd}
\end{equation}
where a bolometric correction factor, $k_{\rm bol}(L_{\rm 2-10~keV})$, converts between hard ($2-10$~keV) X-ray luminosity and bolometric luminosity ($L_{\rm Bol}$). We used the luminosity-dependent bolometric correction from \citet{Duras2020}, implicitly assuming that it remains valid across the mass range considered here \citep[but see discussion in e.g.][]{Zou2023}. We also adopted a mean scaling relation between black hole and stellar mass: $M_{\rm BH} \sim 0.002~M_*$, assuming that the mass of the bulge is equal to $M_*$ \citep{Marconi2003}. To account for uncertainties in this scaling relation, especially in the low-mass regime \citep[e.g.][]{Graham2015, Reines2015, Greene2020}, we included an intrinsic scatter of 0.3~dex \citep{Kormendy&Ho2013}.

We adopted the Bayesian framework described in \citet{Aird2017, Georgakakis2017, Aird2018} where our knowledge of $\lambda_{\rm SAR}$ can be described by a probability distribution:
\begin{equation}
p(\lambda_{\rm SAR} | D_i) d\lambda_{\rm SAR} \propto \mathcal{L}(N_i | \lambda_{\rm SAR},  b_i, t_i, z_i) \; \pi_{\rm AGN}(\lambda_{\rm SAR} | M_{*,i}, z_i) \; d\lambda_{\rm SAR}
\label{eq:pi}
\end{equation}
where $D_i$ is the observed data from source $i$, $\mathcal{L}(N_i | \lambda_{\rm SAR}, b_i, t_i, z_i)$ is the likelihood of observing $N_i$ counts from a source with specific accretion rate $\lambda_{\rm SAR}$, and $\pi_{\rm AGN}(\lambda_{\rm SAR} | M_{*,i}, z_i)$ acts as a prior, describing the true underlying distribution of specific accretion rates of AGN in galaxies with stellar mass, $M_{*,i}$, and redshift, $z_i$.

The likelihood of observing $N_i$ X-ray counts can be described by a Poisson process; thus,
\begin{equation}
\mathcal{L}(N_i | \lambda_{\rm SAR},  b_i, t_i, z_i) = \frac{\mu_i^{N_i} ~~e ^{-\mu_i}}{N_i!},
\label{eq:poisslik}
\end{equation}
where the underlying (non-integer) total expected counts, $\mu_i$, are defined as 
\begin{equation}
\mu_i = k_i(z_i, M_{\rm BH}) \cdot \lambda_{\rm SAR} \cdot M_{*,i} \cdot t_i + b_i,
\label{eq:exp_val}
\end{equation}
where $b_i$ are the (non-integer) expected background counts in the aperture (as in Fig.~\ref{fig:apetool_explanation}), $t_i$ is the exposure time and $k_i(z_i, M_{\rm BH})$ is a source-dependent conversion factor that maps the specific accretion rate to the expected net X-ray count rate, accounting for the ECF, EEF, luminosity distance at redshift $z_i$, and the dependence on black hole mass. We introduced a prior on the black hole mass at fixed stellar mass, $\pi_{\rm M}(M_{\rm BH} | M_{*,i})$, assuming that $\log M_{\rm BH}$ follows a normal distribution centred on $\log M_{\rm BH} \sim \mathcal{N}(\log (0.002\, M_*), \sigma_{\rm BH})$. We fixed $\sigma_{\rm BH}=0.3$~dex, thereby accounting for the intrinsic scatter in the  $M_{\rm BH}$--$M_*$ relation. We marginalised over this distribution when evaluating the likelihood (see Eqs.~\ref{eq:modelA} and \ref{eq:modelB}).

We extended Eq.~\ref{eq:exp_val} to explicitly include an additional contribution arising from galactic processes, given by
\begin{equation}
\mu_i^\prime= [k_i(z_i, M_{\rm BH}) \cdot \lambda_{\rm SAR} \cdot M_{*,i} + l_i \cdot L_{\rm X, G} ] \cdot t_i +b_i,
\end{equation}
where now $l_i \cdot L_{\rm X, G}$ represents the expected count rate arising from galactic processes. We used Eq.~\ref{eq:aird_lx_gal} to derive a prior, $\pi_{\rm GAL}(L_{\rm X,G} | M_{*,i}, \tilde{\mathrm{SFR}}_i)$, on the galactic contribution to the X-ray luminosity. This prior was evaluated using the stellar mass of the galaxy and a SFR equal to the value of the main sequence of star formation at the galaxy's stellar mass, as discussed in Sect.~\ref{sec:non_agn_processes}. We assumed that the logarithm of $L_{\rm X, G}$ follows a normal distribution with an intrinsic scatter of $\sigma_G=0.2$~dex, meaning $\log L_{\rm X,G} \sim \mathcal{N}(\log \overline{L}_{\rm X,G}, \sigma_G)$, and marginalised over this quantity. 

We can thus write the overall likelihood function for all the galaxies in a given stellar mass -- redshift bin ($\mathbf{D}_\mathrm{bin}$) as
\begin{align}
\mathcal{L}_A(\mathbf{D}_\mathrm{bin}) 
&= \prod_{i=1}^{n_\mathrm{source}} \int_0^\infty p(\lambda_{\rm SAR} | D_i) \; d\lambda_{\rm SAR}  \label{eq:modelA} \\
&= \prod_{i=1}^{n_\mathrm{source}} \int_0^\infty \Bigg\{ \int_0^\infty \int_0^\infty
\mathcal{L}_A(N_i | \lambda_{\rm SAR}, M_{\rm BH}, L_{\rm X, G}, b_i, t_i, z_i) \; \nonumber \\
& \cdot \pi_{\rm GAL}(L_{\rm X,G} | M_{*,i}, \tilde{\mathrm{SFR}}_i) \pi_{\rm M}(M_{\rm BH} | M_{*,i}) \; dL_{\rm X, G} \; dM_{\rm BH}\Bigg\} \; \nonumber \\ 
&  \cdot \pi_{\rm AGN}(\lambda_{\rm SAR} | M_\mathrm{*,bin}, z_\mathrm{bin}) \; d\lambda_{\rm SAR}. \nonumber
\end{align}
For notational simplicity, we write the marginalisations as integrals over $M_{\rm BH}$ and $L_{\rm X,G}$, although both $\pi_{\rm M}$ and $\pi_{\rm GAL}$ are specified as normal distributions in logarithmic space. Eq.~\ref{eq:modelA} denotes the likelihood function including components from AGN, galactic and background processes, which we henceforth denote as Model A. However, given the high levels of contamination found in  Sect.~\ref{sec:xrayvalidation_catalogues}, we defined an additional model, Model B, which retained the form of Model A but now explicitly included a contribution from the spurious emission. This was vital to define appropriate confidence intervals at low $\lambda_{\rm SAR}$ (see Sect.~\ref{sec:dwarf_incidence_results}). 

Model B required a component describing the distribution of net count rate from spurious detections, $r_{\rm spur}$. We obtained this by offsetting the parent galaxy apertures by 60\arcsec\ in their declinations (i.e. to an arbitrary location on the sky) and repeating the X-ray aperture photometry in the same way as described in Sect.~\ref{apetool_methods}, while maintaining the same redshift and stellar mass distributions as the real galaxies. Formally X-ray-detected apertures were removed from the offset sample (as if cleaned following Fig.~\ref{fig:flowchart}), and the remaining objects were used to empirically estimate the probability mass distribution (PMF) of the spurious net count rate
\begin{equation}
   r_{\rm spur}=(N_{\rm spur}-b_{\rm spur})/t_{\rm spur}, 
\end{equation}
where $N_{\rm spur}$, $b_{\rm spur}$ and $t_{\rm spur}$ are the extracted counts, local background counts, and exposure times for the offset aperture. As $r_{\rm spur}$ was typically small, we discretised its distribution into bins, $s$, of $\log r_{\rm spur}$ and normalised it such that $\sum_s p(\log r_{\rm spur, s}) = 1$. Therefore, for a given source, $i$, and discrete spurious-rate, $s$, the expected number of counts is
\begin{equation}
\mu_{is}= [k_i(z_i, M_{\rm BH}) \cdot \lambda_{\rm SAR} \cdot M_{*,i} + l_i \cdot L_{\rm X, G} ] \cdot t_i +b_i + r_{\rm spur, s} \cdot t_i.
\end{equation}
Since the spurious rate associated with an individual galaxy was unknown, we marginalised over the empirically measured spurious net count rate distribution, to account for this additional stochastic contribution to the observed counts. This resulted in an overall likelihood function for Model B in a given stellar mass -- redshift bin ($\mathbf{D}_\mathrm{bin}$) of
\begin{align}
&\mathcal{L}_B(\mathbf{D}_\mathrm{bin})=
\nonumber\\
&\quad\prod_{i=1}^{n_\mathrm{source}}
\int_0^\infty
\Bigg\{
\int_0^\infty \int_0^\infty
\sum_s
\mathcal{L}_B(N_i \mid \lambda_{\rm SAR}, M_{\rm BH}, L_{\rm X,G}, r_{\rm spur,s}, b_i, t_i, z_i)
\nonumber\\
&\quad\quad \cdot\;
p(\log r_{\rm spur,s})\,
\pi_{\rm GAL}(L_{\rm X,G} \mid M_{*,i}, 
\tilde{\mathrm{SFR}}_i)\, 
\nonumber\\
&\quad\quad \cdot \pi_{\rm M}(M_{\rm BH} | M_{*,i}) \,
dL_{\rm X,G} \, dM_{\rm BH}
\Bigg\}
\nonumber\\
& \quad\quad \cdot\;
\pi_{\rm AGN}(\lambda_{\rm SAR} \mid M_{\mathrm{*,bin}}, z_{\mathrm{bin}})
\, d\lambda_{\rm SAR}.
\label{eq:modelB}
\end{align}
Recall that the goal of both Model A and B is to derive the intrinsic probability distribution function of $\lambda_{\rm SAR}$, which we now rewrite as 
\begin{equation}
    \pi_{\rm AGN}(\lambda_{\rm SAR} | M_\mathrm{*,bin}, z_\mathrm{bin}) ~ d\lambda_{\rm SAR} = p(\log \lambda_{\rm SAR} | M_{*,\rm bin}, z_{\rm bin}) ~ d \log\lambda_{\rm SAR},
\end{equation}
to indicate a probability density per unit $\log \lambda_{\rm SAR}$ in a given stellar mass ($M_{*,\rm bin}$; sampled in log-space) and redshift ($z_{\rm bin}$) bin. $p(\log \lambda_{\rm SAR} | M_*, z)$ represents the intrinsic distribution of $\lambda_{\rm SAR}$ that gives rise to the observed AGN-associated X-ray counts and we synonymously refer to it as the incidence of X-ray AGN as a function of specific black hole accretion rate. We modelled this distribution as a step function in discrete $\log \lambda_{\rm SAR}$ bins \citep[which is effectively equivalent to a series of Gamma functions used in][for a small enough $\log \lambda_{\rm SAR}$ bins]{Aird2017,Aird2018}. This model is flexible as it does not assume any functional form for $p(\log \lambda_{\rm SAR} | M_{*,\rm bin}, z_{\rm bin})$ and is only constrained by a prior that prefers a smooth variation\footnote{Smoothing was applied by uniformly sampling the first two $\log \lambda_{\rm SAR}$ bins ($\theta_1, \theta_2$) and then recursively defining the following $\log \lambda_{\rm SAR}$ bins via: $\theta_{j} = \theta_{j-1} + \Delta\theta_{j-2}$. Each increment $\Delta \theta$ was drawn from a standard normal prior $\Delta\theta \sim \mathcal{N}(0,1)$, meaning that the subsequent $\theta$ bin can scatter 1~dex around the previous bin value (with no preference on the direction). The smoothing prior exerts its strongest influence at the very low and very high ends of $\log \lambda_{\rm SAR}$, where the data provide little information to constrain the posterior, while in the well-sampled intermediate range the posterior is primarily determined by the likelihood.} across $\log \lambda_{\rm SAR}$ bins and the requirement that this probability distribution function integrates to unity.

Using this parametrisation, we can reduce the overall likelihood functions to a series of likelihood terms from each source $i$, $w_{ij}$ (from the terms inside curly brackets in Eqs.~\ref{eq:modelA} and \ref{eq:modelB}), multiplied by the value of the step function, $\theta_j$, in a given $\log \lambda_{\rm SAR}$ bin, $j$:
\begin{equation} 
\mathcal{L}_{A,B}(\mathbf{D}_\mathrm{bin})= \prod_{i=1}^{n_\mathrm{source}} \sum_j \theta_j \cdot w_{ij}. 
\label{eq:mixturelikelihoods}
\end{equation}
This expression represents a mixture likelihood, where the likelihood of the data for each source is modelled as a weighted sum over all $\log \lambda_{\rm SAR}$ bins, with the mixture coefficients given by the population fractions $\theta_j$. We fitted this hierarchical Bayesian model using \texttt{CmdStan}, the command-line interface to the \texttt{Stan} statistical modelling language \citep{Carpenter2017}, which samples the population parameter space via Markov chain Monte Carlo (MCMC) techniques. Convergence of the Markov chains was assessed using the rank-normalised split-$\hat{R}$ statistic and effective sample sizes \citep{Vehtari2021}. All parameters satisfied $\hat{R}<1.01$, with effective sample sizes exceeding 1000 and there were no divergent transitions. Lastly, we verified with simulations that our models accurately recovered a known input $p(\log \lambda_{\rm SAR} | M_*, z)$ distribution.

\begin{figure}
    \centering
    \includegraphics[width=\linewidth]{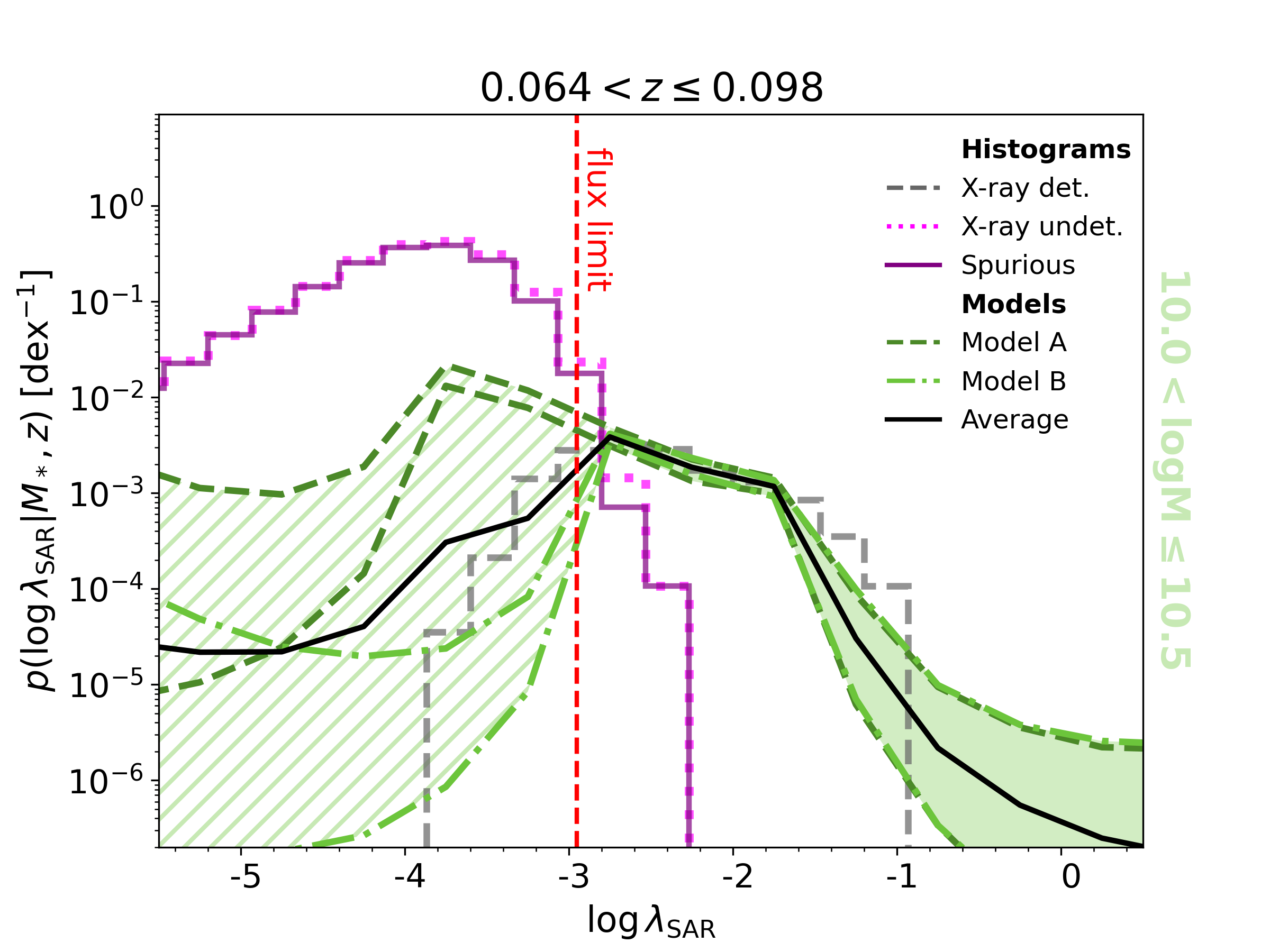}
    \caption{Best-fit $p(\log \lambda_{\rm SAR} | M_*, z)$ distributions from Model A and B for an example stellar mass and redshift bin showing how the average model and the corresponding uncertainty bands were derived in the low- and high-$\lambda_{\rm SAR}$ regimes (see text for details).}
    \label{fig:onebin}
\end{figure}

\begin{figure*}[p]
    \centering
    \includegraphics[width=0.95\linewidth]{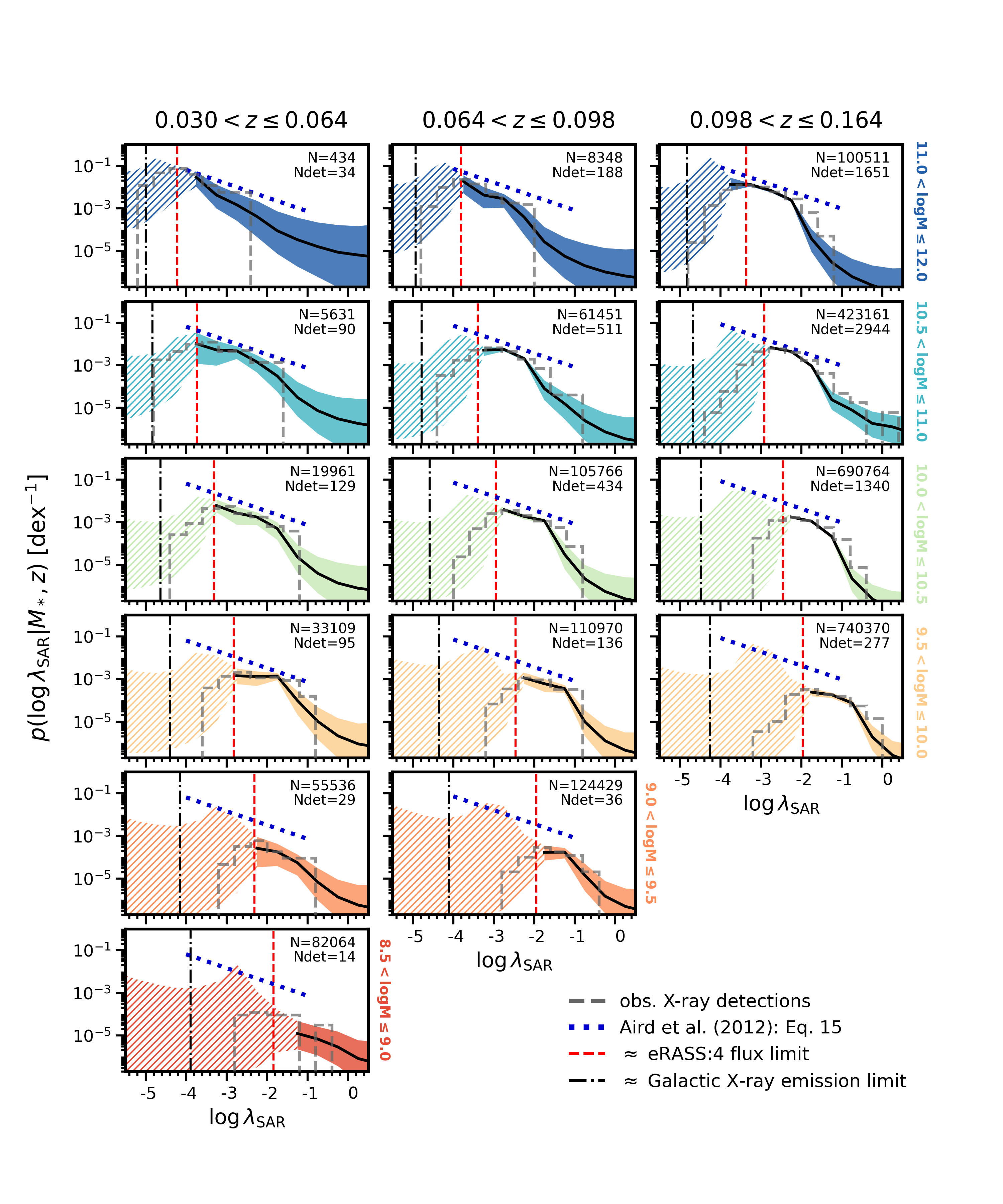}
    \caption[The incidence of eRASS:4 X-ray AGN as a function of $\lambda_{\rm SAR}$, in different mass and redshift bins.]{Incidence of X-ray AGN as a function of $\lambda_{\rm SAR}$, in different mass (rows; in units of $M_{\odot}$) and redshift (columns) bins: $p(\log \lambda_{\rm SAR} | M_*, z)$. The black curves and solid shaded coloured regions mark the best estimate and $1\sigma$ confidence interval, respectively, as found by our Bayesian methodology. The dashed grey histogram represents the observed distribution of X-ray detected sources and the extrapolated results from \citet{Aird2012} are shown in blue dotted lines. Vertical dashed red and dash-dotted black lines indicate the approximate eRASS:4 flux limit and galactic X-ray emission contamination limit, respectively; hatched coloured regions mark the low-$\lambda_{\rm SAR}$ regime where spurious contamination is dominant (see text for details).}
    \label{fig:plambda_curves}
\end{figure*}

\begin{figure*}[ht!]
    \centering
    \includegraphics[width=\linewidth]{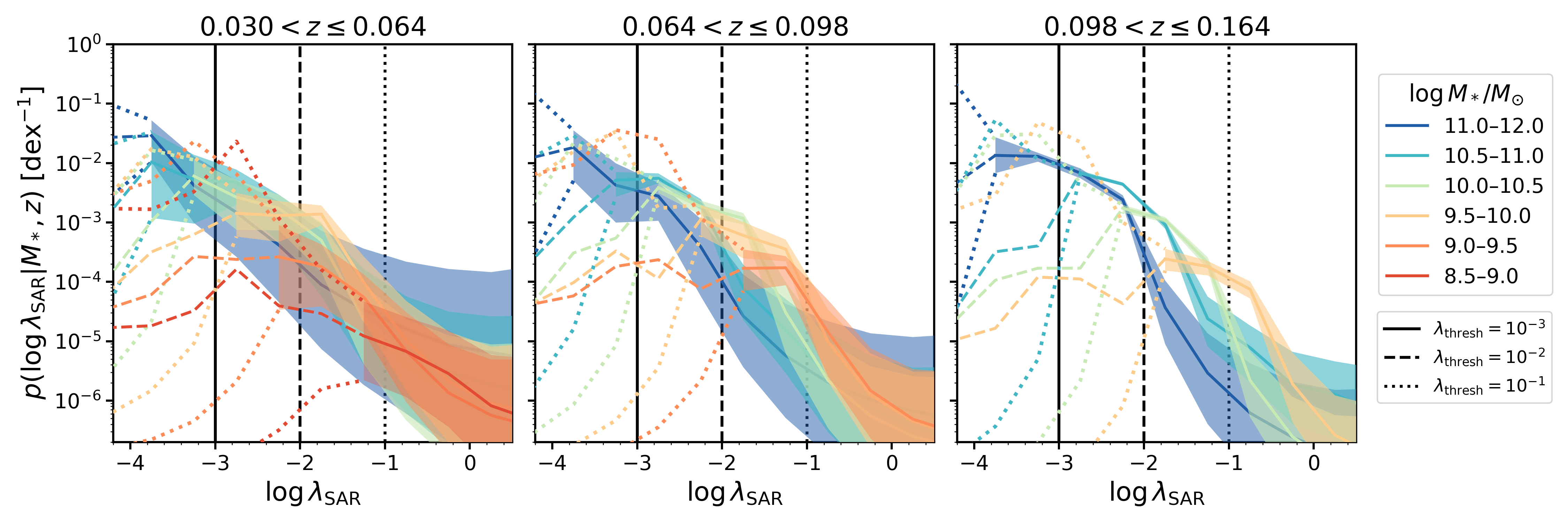}
    \caption[Same $p(\log \lambda_{\rm SAR} | M_*, z)$ distributions as in Figure \ref{fig:plambda_curves}, but overlaying the different stellar mass bins for a given redshift bin.]{Same $p(\log \lambda_{\rm SAR} | M_*, z)$ distributions as in Figure \ref{fig:plambda_curves}, but overlaying the different stellar mass bins (colours) for a given redshift bin (panels). The median and uncertainty envelope of $p(\log \lambda_{\rm SAR} | M_*, z)$ in the low-$\lambda_{\rm SAR}$ regime affected by spurious contamination (hatched in Fig.~\ref{fig:plambda_curves}) are now marked with a dashed and dotted coloured curve, respectively, for clarity. The vertical solid, dashed, and dotted black lines correspond to $\lambda_{\rm thresh} = 10^{-3}, 10^{-2}$ and $10^{-1}$, respectively. These are the thresholds above which the different cumulative AGN fractions were computed (see text).}
    \label{fig:plambda_zpanels}
\end{figure*}

\section{Results}

\subsection{Incidence of X-ray AGN as a function of $\lambda_{\rm SAR}$}
\label{sec:dwarf_incidence_results}

As an illustrative example, Figure \ref{fig:onebin} shows the X-ray AGN incidence distributions and fitted models for a stellar mass bin ($10 < \log M_*/M_{\odot} \leq 10.5$) and redshift bin ($0.064 < z \leq 0.098$). The dashed grey and dotted pink normalised histograms show the observed $\log \lambda_{\rm SAR}$ distribution for the X-ray detected sources and X-ray undetected sources, respectively. Meanwhile, the solid purple normalised histogram shows the contamination from spurious sources, calculated by converting the net spurious counts in the shifted apertures to $\log \lambda_{\rm SAR}$ using Eqs.~\ref{eq:counts_to_flux} and \ref{eq:lambda_edd}. It is clear that the spurious distribution is statistically almost indistinguishable from the undetected population. Fig.~\ref{fig:onebin} also presents the best-fit $p(\log \lambda_{\rm SAR} | M_*, z)$ distributions as inferred from Model A (dashed dark green curves) and Model B (dash-dotted light green curves) described in Sect.~\ref{sec:plambda_methods}. Their average is shown by the solid black curve. The shaded and hatched regions indicate the uncertainty in the inferred distribution. The shaded region corresponds to the $1\sigma$ confidence interval shared by both models at high $\lambda_{\rm SAR}$, while the hatched region shows the envelope defined by the upper and lower $1\sigma$ confidence intervals of Model A and Model B, respectively\footnote{We note that the $\lambda_{\rm SAR}$-histograms are derived using the fixed stellar mass of each galaxy, whereas the $p(\log \lambda_{\rm SAR} | M_*, z)$ curves incorporate an intrinsic scatter in the $M_*-M_{\rm BH}$ relation. This causes the apparent discrepancy between the two at high $\lambda_{\rm SAR}$, which can be reconciled by using $\sigma_{\rm BH}=0$ in Models A and B, as expected.}. The transition between these regimes occurs where the difference between the upper and lower envelope of Model A and Model B, respectively, drops below 1.5~dex when moving from low- to high-$\lambda_{\rm SAR}$. In practice, this transition typically occurs near the approximate $0.2-2.3$\,keV eRASS:4 \texttt{DET\_LIKE\_0} $> 10$ flux limit of $\sim 2\times10^{-14}$\,erg\,s$^{-1}$\,cm$^{-2}$, converted to an estimate of $\lambda_{\rm SAR}$ using Eq.~\ref{eq:lambda_edd} and the median stellar mass and upper redshift bound of each bin (vertical dashed red line). The hatched envelope captures the uncertainty in $\lambda_{\rm SAR}$ associated with the ambiguity between weak AGN emission and spurious signals in the aperture counts.

Figure \ref{fig:plambda_curves} shows the $p(\log \lambda_{\rm SAR} | M_*, z)$ distributions in all six stellar mass bins (rows) in the range of $\log M_*/M_{\odot}=8.5-12$ and all three redshift bins (columns), with the same figure style as explained in Fig.~\ref{fig:onebin}. The upper bound of each of the three redshift bins is defined such that the galaxies are mass-complete (to the 70\% level) at $\log M_*/M_{\odot}= 8.5,\, 9$ and $9.5$. The number of X-ray detections (Ndet) and total number of galaxies (N) within each bin are stated in the top right corner of each panel. Vertical black dash-dotted lines indicate the 90th percentile of $L_{\rm X,G}/M_*$ of all sources in a given bin, converted to $\lambda_{\rm SAR}$ with Eq.~\ref{eq:lambda_edd}, as an estimate of the galactic X-ray contamination. Finally, as a comparison, the $p(\log \lambda_{\rm SAR} | M_*, z)$ curves found by \citet{Aird2012} (for $0.2 < z < 1.0 $ and $9.5\leq\log M_*/M_{\odot}\leq12$), extrapolated to the median of each redshift bin, are plotted with dotted dark blue lines. Figure \ref{fig:plambda_zpanels} shows the same $p(\log \lambda_{\rm SAR} | M_*, z)$ distributions from Figure \ref{fig:plambda_curves} divided into three redshift bins (panels), with stellar mass bins overlaid to highlight their differences.

In general, we observe a broad distribution of $p(\log \lambda_{\rm SAR} | M_*, z)$ with a non-trivial shape, alluding to a complex interplay between the evolution of black hole growth and accretion physics across the mass scale. For high-mass galaxies, the shape of the incidence distribution follows well the power-law scaling with index around $-0.65$, found by \citet{Aird2012} in the regime $-4\leq \log \lambda_{\rm SAR} \leq -2$. Whereas in the low-mass regime, particularly for $\log M_*/M_{\odot}\leq9.5$, we observe a significant departure from the \citet{Aird2012} relation (lower normalisation). We discuss below the observed behaviour of $p(\log \lambda_{\rm SAR} | M_*, z)$ in the high-$\lambda_{\rm SAR}$ regime across the mass scale, as well as the reasons for our relatively large uncertainty envelope at low $\lambda_{\rm SAR}$. Detailed comparison of our results to the literature is presented in Sect.~\ref{discussion:lit_comparison}.

Firstly, there is a noticeable break present between $\log \lambda_{\rm SAR}\sim -2$ and $-1$ across the mass scale, revealed thanks to the large statistical power of the eRASS:4 sample. The presence of this high-$\lambda_{\rm SAR}$ break indicates that AGN do not simply undergo stochastic fuelling from the available gas supply \citep[e.g.][]{Hickox2014}, as deduced from early studies finding a power-law trend \citep[e.g.][]{Aird2012}, but that there is a $\lambda_{\rm SAR}$-dependent change in accretion mechanisms in this regime. This would make physical sense as above the break the AGN are nearing the Eddington limit, where the strong radiation pressure may act to regulate black-hole growth through powerful outflows that evacuate gas from the central regions \citep[e.g.][]{Hopkins2006, Hopkins2008, Fabian2012}. In fact, \citet{Aird2013} show that such a steep break in the modelled $p(\log \lambda_{\rm SAR} | M_*, z)$ distribution is necessary to reproduce the X-ray luminosity function (XLF) well. Obscuration alone cannot explain this break as the obscured AGN fraction is found to decrease steeply with an increasing Eddington ratio \citep[e.g.][]{Ricci2017, Ananna2022b, Ricci2022}. 

Interestingly, we observe the break to shift to lower $\lambda_{\rm SAR}$ values for higher stellar masses. This means that massive galaxies are less likely to reach high accretion rates than lower-mass galaxies, possibly due to such galaxies being more quenched and gas-poor, with fewer cold gas inflows able to fuel the AGN \citep[e.g.][]{Hopkins2008, Saintonge2011, Saintonge2017, Tacconi2018}. Below the break, we observe the normalisation of $p(\log \lambda_{\rm SAR} | M_*, z)$ to decrease for decreasing stellar mass, diverging from the \citet{Aird2012} curve.

\begin{figure*}[ht!]
    \centering
    \includegraphics[width=0.33\linewidth]{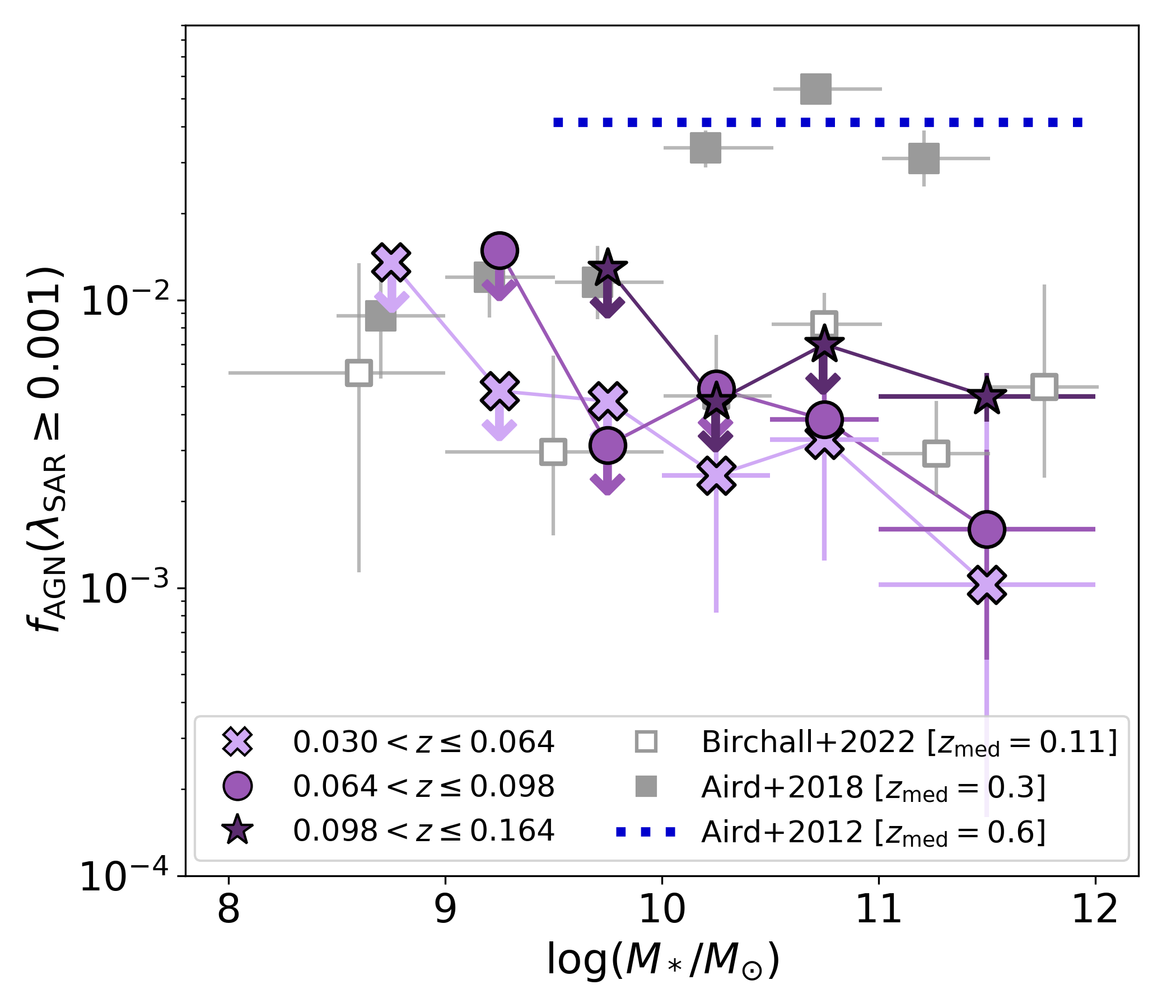}
    \includegraphics[width=0.33\linewidth]{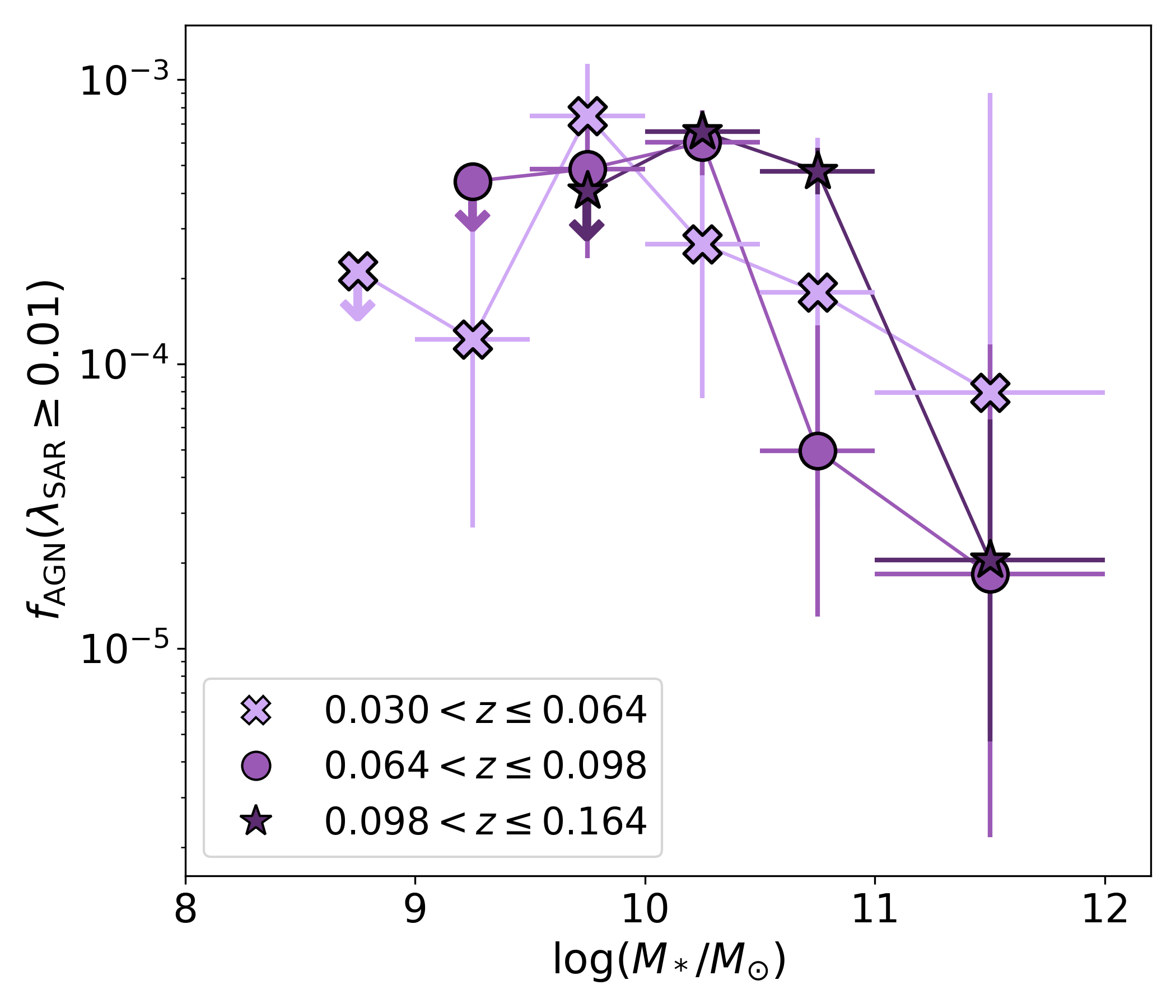}
    \includegraphics[width=0.33\linewidth]{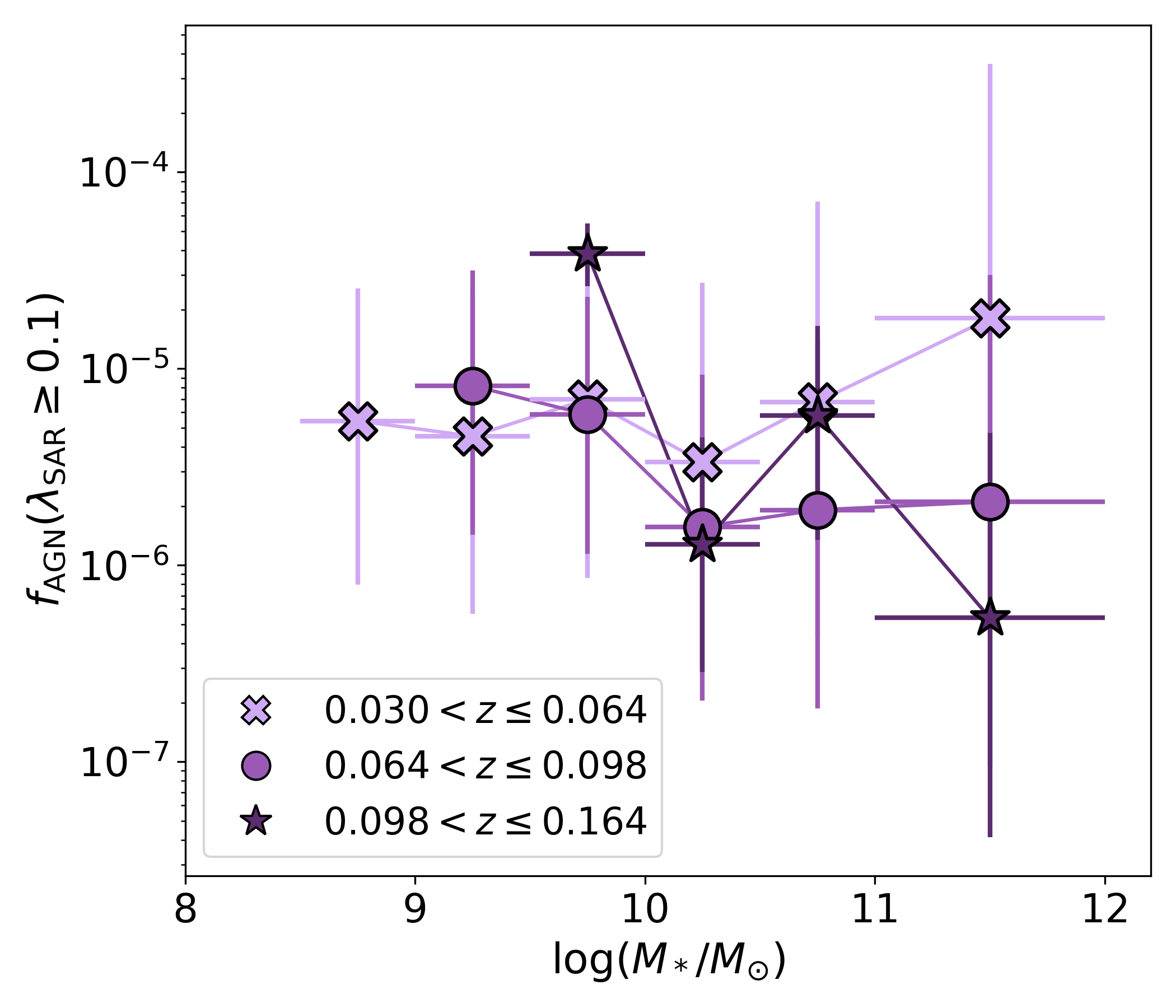}
    \caption{Cumulative AGN fraction for $\lambda_{\rm SAR}\geq10^{-3}$ (left), $\lambda_{\rm SAR}\geq10^{-2}$ (middle) and $\lambda_{\rm SAR}\geq10^{-1}$ (right) as a function of stellar mass, in different redshift bins (see legend). Results from previous works that selected AGN in the hard X-ray band at various redshifts are also shown as a comparison \citep{Aird2012, Aird2018, Birchall2022}.}
    \label{fig:agnfrac}
\end{figure*}

In the low-$\lambda_{\rm SAR}$ regime, we show that spurious contamination among the X-ray `undetected' sample strongly affects our ability to constrain $p(\log \lambda_{\rm SAR} | M_*, z)$. This is most prominently visible in the $0.098\leq z \leq 0.164$ and $9.5 < \log M_*/M_{\odot} \leq 10$ bin (yellow curve), where there is a steep rise below $\log \lambda_{\rm SAR}<-2$. As shown by the grey histogram, this happens in a region of parameter space where the observed X-ray detections are dwindling due to the flux limit of the survey (vertical dashed red lines). We note that the high-mass sources are less affected, as expected given the low ($24\%$) spurious contamination. Upon quantifying this effect in a statistical manner using shifted apertures as described in Sect. \ref{sec:plambda_methods}, we find that we cannot reliably constrain $p(\log \lambda_{\rm SAR} | M_*, z)$ below the approximate eRASS:4 flux limit, even when using the X-ray information from all galaxies in our Bayesian method.

As the additional contamination from unresolved stellar remnants (XRBs) is expected to be important at much lower $\lambda_{\rm SAR}$ than the eRASS:4 flux limit (see Fig.~\ref{fig:xrb_hot_gas_removal_IMBH} and the vertical black dash-dotted lines in Fig.~\ref{fig:plambda_curves}), we do not discuss this in detail here. We do note, however, that past studies \citep[e.g.][]{Gilfanov2004, Brorby2014, Lehmer2019, Kouroumpatzakis2020, Kyritsis2025} have found enhanced X-ray emission for lower-metallicity, lower-mass, higher-SFR galaxies and an increased scatter in the determination of $L_{\rm X,G}$, owed to high variability in the XRB population. Therefore, it may be that Eq.~\ref{eq:aird_lx_gal} is currently underestimating the galactic X-ray emission. Future studies of eRASS X-ray emission from more nearby galaxies may help address these uncertainties.

Phenomenologically, a low-$\lambda_{\rm SAR}$ turnover may be expected given that AGN transition from a `radiative' to `kinetic' mode of accretion towards lower specific accretion rates \citep[e.g.][]{MerloniHeinz2008, HeckmanandBest2014, Hardcastle&Croston2020, Harrison2024}. AGN at these low-$\lambda_{\rm SAR}$ values would no longer emit dominantly at X-ray wavelengths (as the accretion flow can no longer energetically maintain a hot X-ray-emitting corona), but instead become detectable in the radio bands through their jet kinetic emission \citep[see e.g.][]{Kondapally2022, Igo2024, Igo2025}. Mathematically, this turnover is also required such that the $p(\log \lambda_{\rm SAR} | M_*, z)$ probability distribution is bounded and integrates to one (or lower, if the BHOF is $<$100\%). However, deeper X-ray and optical data are required to reliably constrain this low-$\lambda_{\rm SAR}$ turnover.

\subsection{The cumulative AGN fraction}
\label{sec:duty_cycle}

One way to summarise the information contained in Figs. \ref{fig:plambda_curves} and \ref{fig:plambda_zpanels} is to compute the cumulative AGN fraction. We define this in a similar way to \citet{Aird2018}, where the cumulative AGN fraction in a given stellar mass and redshift bin is the fraction of galaxies hosting an X-ray AGN with $\lambda_{\rm SAR}\geq\lambda_{\rm thresh}$: 
\begin{equation}
f_{\rm AGN}(\lambda_{\rm SAR} \geq \lambda_{\rm thresh}) = \int_{\log\lambda_{\rm thresh}}^\infty p(\log \lambda_{\rm SAR} | M_*,z) \; d\log \lambda_{\rm SAR}. 
\label{eq:dutycycle}
\end{equation}
The cumulative AGN fraction can be interpreted as an AGN duty cycle, meaning the fraction of time that a central massive black hole spends in an `active' state.  

Figure \ref{fig:agnfrac} shows the cumulative AGN fraction for $\lambda_{\rm SAR}\geq10^{-3}$, $\lambda_{\rm SAR}\geq10^{-2}$ and $\lambda_{\rm SAR}\geq10^{-1}$ as a function of stellar mass, in the different redshift bins (see legend). Stellar mass bins affected by spurious contamination at low $\lambda_{\rm thresh}$ values are plotted as upper limits at the 95th percentile of Model A; otherwise, 1$\sigma$ uncertainties are propagated from the full $p(\log \lambda_{\rm SAR} | M_*, z)$ posterior.

Figure \ref{fig:agnfrac} (left) shows that we can constrain the overall $f_{\rm AGN}(\lambda_{\rm SAR} \geq 10^{-3})$ to be $\lesssim1\%$ across the mass scale at $0.03<z\leq0.164$. There is also a slight increase in $f_{\rm AGN}(\lambda_{\rm SAR} \geq 10^{-3})$ with redshift, but we do not discuss the cosmic evolution further as we probe only a very local volume \citep[see e.g.][for samples over larger redshift ranges]{Aird2018, Mezcua2018, Zou2023, Zou2024,Cho2024, Guetzoyan2025}.

The middle and right panels in Figure \ref{fig:agnfrac} show that, for these very low redshifts, the cumulative AGN fraction for $\lambda_{\rm SAR}\geq10^{-2}$ and $\lambda_{\rm SAR}\geq10^{-1}$ is $\lesssim0.1$\% and $\lesssim0.01$\%, respectively, in the range of $8.5 \leq \log M_*/M_{\odot} \leq 10$. This highlights the rare, but non-zero, presence of highly accreting AGN in low-mass galaxies. 

Interestingly, we observe the cumulative AGN fraction to vary as a function of stellar mass. Most notably, \mbox{$f_{\rm AGN}(\lambda_{\rm SAR}\geq10^{-2})$} is found to peak around $\log M_*/M_{\odot}\sim 10-10.5$, meaning that the cumulative fraction of AGN at lower and higher masses is suppressed, alluding to varying fuelling efficiencies in these regimes. For the case of $f_{\rm AGN}(\lambda_{\rm SAR}\geq10^{-3})$, there is a tentative increase as a function of decreasing stellar mass, although spurious contamination at the lowest masses prevents clear conclusions from being drawn. There is no observable trend with stellar mass for the case of $f_{\rm AGN}(\lambda_{\rm SAR}\geq10^{-1})$.

It is important to note that by virtue of our soft selection using eRASS:4, our $p(\log \lambda_{\rm SAR} | M_*,z)$ distributions are not sensitive to obscured X-ray sources. According to \citet{Ricci2017}, the obscured AGN fraction depends primarily on the Eddington ratio and is mass-independent, decreasing dramatically from $\sim70$\% at $\lambda_{\rm SAR}\sim10^{-2}$ to $\sim 20\%$ at $\lambda_{\rm SAR}\sim10^{-1}$. If we assume that, to first order, obscuration preferentially removes sources from our parent sample instead of attenuating their luminosity and that $\lambda_{\rm SAR}$ is a proxy for the Eddington ratio, we can estimate a value of the cumulative AGN fraction including both obscured and unobscured sources. This assumption implies that the shape of the $p(\log \lambda_{\rm SAR} | M_*,z)$ at high specific accretion rates would be similar for obscured and unobscured selections to first order. Thus, by convolving the cumulative AGN fraction calculation as in Fig.~\ref{fig:agnfrac} with the unobscured AGN fraction ($1 -$obscured AGN fraction) as a function of $\lambda_{\rm SAR}$, we estimate an increase of $f_{\rm AGN}(\lambda_{\rm SAR} \geq 10^{-3})$, $f_{\rm AGN}(\lambda_{\rm SAR} \geq 10^{-2})$ and $f_{\rm AGN}(\lambda_{\rm SAR} \geq 10^{-1})$ by a factor of around 4.64, 2.81 and 1.07, respectively (averaged across all mass and redshift bins).

\section{Discussion}
\label{sec:imbh_discussion}

Using our well-defined, complete and rigorously cleaned sample of X-ray AGN in low- and high-mass galaxies in the local Universe, we are able to place tight constraints on the
$p(\log \lambda_{\rm SAR} | M_*, z)$ distribution and on the cumulative AGN fraction. In this section, we compare our findings to existing literature samples and previous constraints on the specific accretion rate distribution, along with discussing the physical interpretation of its shape in the context of black hole formation and growth across the mass scale.

\subsection{Comparison to previous literature}
\label{discussion:lit_comparison}

Firstly, almost all of the host galaxies of the X-ray AGN found in this study are catalogued in the \textit{NASA/IPAC Extragalactic Database} (NED; using a 5\arcsec of the optical host galaxy co-ordinates, at the time of writing), but only 14/874 and 104/12,618 are classed as `X-ray Sources' from the low- and high-mass sample, respectively. Therefore, our work provides the largest sample of X-ray AGN in low-mass galaxies to date, enabled by eROSITA, revealing their previously unrecognised X-ray nature. At the same time, it enlarges our database of X-ray AGN in high-mass galaxies, offering a control sample for comparison studies.

Detailed X-ray spectroscopic and multi-wavelength follow-up of our low-mass X-ray sample may be key to test if black hole accretion mechanisms scale universally across the mass scale or there are some fundamental differences at the low-mass regime (Igo et al., in prep.). This is motivated by studies at both low- and high-redshift that find differences in the expected X-ray emission from such sources, in comparison to their emission in other wavebands. For example, recent \textit{JWST}-discovered high-redshift black holes, including the `Little Red Dots', have been found to be X-ray undetected, even after stacking \citep[e.g.][]{Maiolino2025chandra, Yue2024}. Similarly, at low redshift, studies on stacked samples of optical and IR variability-selected, X-ray-undetected MBHs in low-mass galaxies indicate that the observed X-ray emission originates from galactic processes rather than from a central accreting AGN, potentially alluding to a lack of X-ray corona in these sources \citep{Arcodia2024}.

\begin{figure*}[t!]
    \centering
    \includegraphics[width=0.8\linewidth]{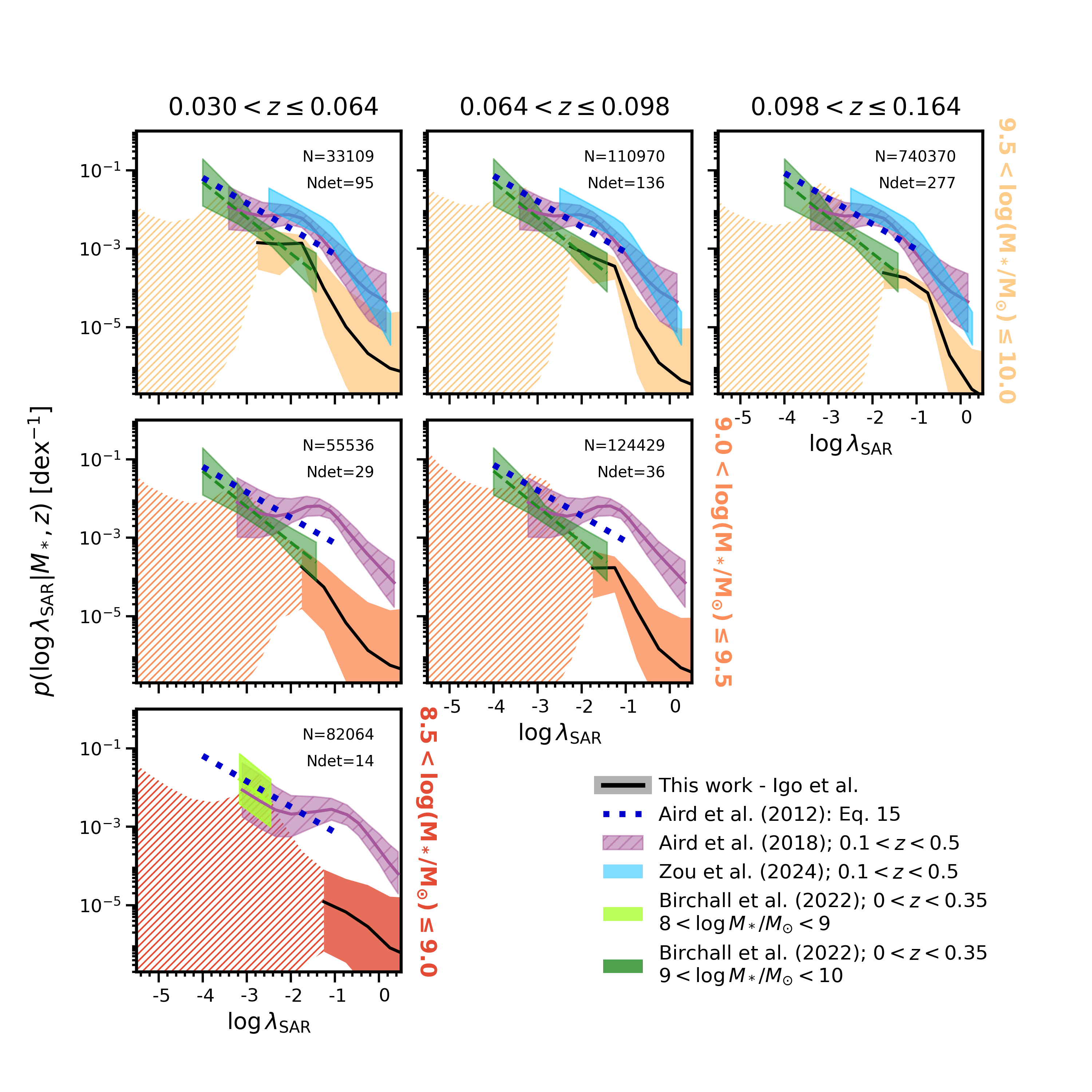}
    \caption{Comparison of the X-ray AGN incidence as a function of $\lambda_{\rm SAR}$ to other key works in the low-mass regime: \citealt{Aird2018} (purple, hatched), \citealt{Birchall2022} (light and dark green for different mass bins, see legend) and \citealt{Zou2024} (blue). Differently to Fig.~\ref{fig:plambda_curves}, the shaded intervals mark 90th percentile confidence intervals to be comparable with those of \citet{Aird2018, Zou2024} and the hatched region begins when the difference between the upper and lower envelopes of Model A and B, respectively, drops below 2~dex. The confidence intervals on the curves from \citet{Birchall2022} are still at the 1$\sigma$ level.}
    \label{fig:literature_comparison_plot}
\end{figure*}

As mentioned in the introduction, past work has lacked large statistical samples to probe the X-ray AGN incidence in the low-mass regime. Figure \ref{fig:literature_comparison_plot} directly compares our $p(\log \lambda_{\rm SAR} | M_*, z)$ distributions (with the shaded curves now representing the 90th percentile confidence intervals and same colour scheme as Fig.~\ref{fig:plambda_curves}) to the work of \citealt{Aird2012, Birchall2022, Zou2024}. The results of \citet{Birchall2020} are not shown as they are in agreement with \citet{Birchall2022} and probe a very limited redshift range, $z<0.06$, with only 28 low-mass sources.

Fig.~\ref{fig:literature_comparison_plot} shows that our work sets the tightest constraints on $p(\log \lambda_{\rm SAR} | M_*, z)$ for $-3\lesssim \log \lambda_{\rm SAR} \lesssim 0$, at a low redshift of $0.03<z<0.164$, for low-mass galaxies in the range $\log M_*/M_{\odot}=8.5-10$. \citet{Birchall2022}, who identify AGN in the hard X-ray band with \textit{XMM-Newton} from a parent galaxy sample defined by the MPA–JHU catalogue, recover a simple power-law form for the X-ray AGN incidence at moderate specific accretion rates over a redshift range comparable to ours. Importantly, their distributions agree with our results in that they show a deviation from the mass-invariant results of \citealt{Aird2012} (dotted blue curves; extrapolated to the median redshift range of each bin).

\citet{Aird2018} select X-ray AGN using deep \textit{Chandra} observations of the Cosmic Assembly Near-Infrared Deep Extragalactic Legacy Survey (CANDELS) and UltraVISTA surveys. They constrain $p(\log \lambda_{\rm SAR} | M_*, z)$ to a similar confidence level, but probe a higher redshift range, $0.1<z<0.5$ than our sample. Their relatively large statistical sample also highlights second-order mass-dependent effects in the specific accretion rate distribution, in agreement with our work. In the context of cosmic evolution, they find an increase in the X-ray AGN incidence in the high-$\lambda_{\rm SAR}$ regime, compared to our work, especially towards lower masses ($\log M_*/M_{\odot}<9.5$). Future studies extending the redshift range of X-ray AGN samples in low-mass galaxies \citep[e.g.][]{Mezcua2018} will be vital to further understand this evolution in the low-mass regime. 

\citet{Zou2024} compiles a sample of X-ray AGN among host-galaxies found in the CANDELS fields, four of the LSST deep-drilling fields (DDFs), and the eFEDS field, also at $0.1<z<0.5$, sampling down to $\log M_*/M_{\odot}=9.5$. Their large sample size and in-depth statistical modelling considerably tighten the constraints upon
$p(\log \lambda_{\rm SAR} | M_*, z)$, relative to previous work \citep[e.g.][]{Aird2018}.

As shown in Figure \ref{fig:agnfrac} (left), our cumulative AGN fraction results for $\lambda_{\rm SAR} \geq 10^{-3}$ are in agreement with \citet{Birchall2020, Birchall2022} (unfilled grey squares). Meanwhile, \citet{Aird2018} find an increasing $f_{\rm AGN}(\lambda_{\rm SAR} \geq 10^{-3})$ (filled grey squares) with higher galaxy stellar mass, which further increases as a function of redshift. Figure \ref{fig:agnfrac} (left) also features the extrapolated results of \citet{Aird2012}, who find a mass-independent $p(\log \lambda_{\rm SAR} | M_*, z)$ at their median redshift of $z=0.6$, corresponding approximately to $f_{\rm AGN}(\lambda_{\rm SAR} \geq 10^{-3}) \sim 4\%$. Integrating Eq.~15 of \citet{Aird2012} at our median redshift range of 0.13 gives $\sim1.3\%$, slightly higher than our (unobscured-only) results (as is already clear from Figure \ref{fig:plambda_curves}). In terms of $f_{\rm AGN}(\lambda_{\rm SAR} \geq 10^{-2})$, \citet{Aird2018, Zou2023} find a value of $\sim0.1-1\%$ in comparison to our $\lesssim0.1$\%, which is consistent given that their studies probe a higher median redshift. 

Regarding the shape of the specific accretion rate distribution, or similarly, the Eddington ratio distribution function (ERDF), there is a range of observational literature discussing the mass dependence (or not), the presence (or not) of a high-$\lambda_{\rm SAR}$ break and the physical meaning of the shape of the distribution in relation to the black hole mass and luminosity functions. For example, \citet{Ananna2022} use the BAT AGN Spectroscopic Survey (BASS) DR2 to compute the ERDF and find a break around a similar value of $\log \lambda_{\rm SAR} \sim -1.5$. However, they do not find any mass dependence in the shape of the ERDF \citep[see also e.g.][]{Caplar2015, Weigel2017}. On the other hand, \citet{Bongiorno2016}, probing a higher redshift range $0.3<z<2.5$ from the XMM-COSMOS survey, find a comparable mass dependence in the high-$\lambda_{\rm SAR}$ break as our results. They attribute this to the `AGN downsizing' phenomenon in the AGN luminosity function, whereby the space density of high-luminosity AGN peaks at earlier cosmic epochs than that of low-luminosity AGN \citep[e.g.][]{Ueda2003, Hasinger2005,Aird2015} and state that this is a consequence of the (weak) mass-dependent evolution of the host-galaxy mass function and the stronger mass-dependent evolution of the specific accretion rate distribution function. Similarly, \citet{Georgakakis2017} use a combination of deep and shallow \textit{Chandra} and \textit{XMM-Newton} surveys to show that, within a high-mass galaxy sample, more massive systems tend to avoid high specific accretion rates, consistent with the trends observed in Fig.~\ref{fig:plambda_zpanels}. Lastly, \citet{Schulze2010}, who optically select Type 1 AGN at $z<0.1$, find a decline in $f_{\rm AGN}(\lambda_{\rm SAR} \geq 10^{-1})$ as a function of black hole mass for their high-mass galaxy study, a further evidence of the same anti-hierarchical growth AGN phenomenon \citep[e.g.][]{Merloni2004}. Overall, our work provides a robust statistical characterisation of the $p(\log \lambda_{\rm SAR} | M_*, z)$ distribution, enabled by an unprecedented sample size across $\lambda_{\rm SAR}$ and stellar mass, and will be key to disentangling the interplay between black hole growth, galaxy evolution, and the underlying seeding and fuelling mechanisms.

\subsection{Towards understanding early black hole seeding: interpretation of cumulative AGN fractions}
\label{disc:agn_fractions}

In Section \ref{sec:duty_cycle}, we showed that $f_{\rm AGN}(\lambda_{\rm SAR} \geq 10^{-3})\lesssim1\%$ for low-mass galaxies at $0.03<z\leq0.164$. Obviously, this is only a strict lower limit on the BHOF as there exists a strong degeneracy between low occupation and high active fractions versus high occupation and low active fractions: we cannot currently distinguish between $\sim 1\%$ of low-mass galaxies hosting a central massive black hole, all of which are accreting, or $\gg1\%$ hosting a central massive black hole, but only $\sim 1\%$ of which accreting. 

If one assumes a universal $p(\log \lambda_{\rm SAR} | M_*, z)$ distribution shape across the mass scale, one could make a statement on the BHOF by looking at the relative normalisations of the curves as a function of mass, anchoring the BHOF at $\sim100$\% at the highest masses \citep[see e.g.][]{Miller2015, Burke2025, Zou2025}. Yet, we show in this work that the X-ray AGN incidence has a more complex mass-, redshift- and $\lambda_{\rm SAR}$-dependent form that cautions against such a simplifying assumption. 

This is further complicated by the fact that different multi-wavelength selection techniques find different cumulative AGN fractions (see e.g. \citealt{Wasleske2024} for a systematic selection of dwarf galaxies across various wavelengths or e.g. \citealt{Menzel2016} for the case of higher-mass systems). Additionally, mergers or other secular instabilities can produce off-centre `wandering' black holes \citep[e.g.][]{Bellovary2021, Ricarte2021, DiMatteo2023, Erostegui2025}, which are harder to detect (and may even become dormant), potentially leading to an underestimation of the BHOF. Although \citet{Wu2025} predict that such off-centre wandering black holes could produce micro-lensing effects on quasars detectable in upcoming surveys by LSST \citep{Ivezic2019}.

Past studies on high-mass galaxies have attributed the increase in cumulative AGN fractions towards lower masses to AGN downsizing. However, our work shows a subsequent decline in $f_{\rm AGN}(\lambda_{\rm SAR}>10^{-2})$ for low-mass galaxies with a peak around $\log M_*/M_{\odot}\sim 10-10.5$ (while $f_{\rm AGN}(\lambda_{\rm SAR}>10^{-1})$ shows no conclusive trends as a function of stellar mass). If the AGN downsizing persists to such low-mass galaxies, our observed decrease could allude to reduced BHOF in this regime, potentially explaining the lower normalisation of the $p(\log \lambda_{\rm SAR} | M_*, z)$ curves below the break. 

Nevertheless, we highlight that the cumulative AGN fractions derived in this work show a more nuanced view of AGN fuelling, in particular its efficiency (as shown by Fig.~\ref{fig:agnfrac}, middle), which extends beyond the early mass-invariant results. There may even be an indication of varying trends with stellar mass for cumulative AGN fractions above different $\lambda_{\rm SAR}$ thresholds. These effects should be incorporated into future multi-scale, multi-phase models of AGN fuelling to disentangle the mass- and $\lambda_{\rm SAR}$-dependent properties of accretion, including deciphering the relative importance between feedback-limited versus gas supply-limited cases \citep[e.g.][]{Hopkins2006, Hopkins2008}.
 
From the simulations perspective, SAMs have repeatedly shown that distinguishing different black hole seeding models is very difficult with current data \citep[e.g.][]{Ricarte2018, Burke2023}. For example, \citet{Chadayammuri2023} find that, in order to do so, we would need to detect (or model) all AGN with $L_{X}>10^{37}$~erg~s$^{-1}$ in galaxies of $\log M_*/M_{\odot} \sim 8-10$. This is orders of magnitude below the flux limit of eRASS:4 (see Fig.~\ref{fig:xrb_hot_gas_removal_IMBH}) and still out of reach of the deepest X-ray survey fields \citep[e.g.][]{Aird2018}. Even detecting a low BHOF at low masses is degenerate to slowly accreting heavy seeds or light seeds on fast growth channels \citep{Chadayammuri2023}. Although there are still uncertainties on the simulated BHOF itself that arise from the strong dependence on black hole and galaxy sub-grid prescriptions, as well as the challenges of accounting for obscured AGN in simulations \citep{Haidar2022, Alonso-Tetilla2026}.

\subsection{AGN growth and feedback in the low-mass regime: can local low-mass galaxies be considered high-redshift primordial galaxy analogues?}
\label{disc:agn_growth}

We detect a significant population of X-ray emitting low-mass galaxies, with $L_{\rm X} > 10^{42}$~erg~s$^{-1}$ (see Figs.~\ref{fig:xrb_hot_gas_removal_IMBH} and \ref{fig:agnfrac}, right). This means that not only are AGN present in this low-mass regime -- a long-standing uncertainty only clarified in the past few years -- but they may also be growing and exerting a significant energetic influence on their surroundings.

For example, AGN feedback in low-mass galaxies is increasingly thought to play an important role in observations and simulations, on par with or potentially even exceeding, stellar feedback from supernovae \citep[e.g.][]{Mezcua2015, Koudmani2019, Koudmani2021, Koudmani2022, Gim2024, RodriguezMorales2025, Salehirad2025}. This may weaken the link between the black holes in low-mass galaxies and their original seed masses, thereby complicating attempts to test seeding models using local-Universe analogues \citep{Mezcua2019}.

From a theoretical point of view, semi-analytical models and hydrodynamical simulations struggle to create such highly X-ray luminous AGN powered by low-mass black holes as observed in this work \citep[e.g.][]{Bellovary2019, Beckmann2023}. This is because their accretion prescriptions (often following the simplistic `Bondi-Hoyle-Lyttleton' model, \citealt{Hoyle1939, Bondi1944, Bondi1952}), have a strong dependence on the black hole mass as $\dot{m} \propto M_{\rm BH}^2$, where $\dot{m}$ is the physical accretion rate, which render the fuelling of low-mass black holes more difficult. In light of our new findings \citep[see also e.g.][]{Mezcua2018}, such accretion prescriptions may need to be refined. Recent work by \citet{Ortame2026} tests alternatives to Bondi-based models, such as sink-particle methods, and shows that efficient black hole growth and AGN feedback can occur in simulated dwarf galaxies. Overall, ensuring a high degree of consistency between simulations and observations in the local Universe is crucial for testing the extent to which local dwarf galaxies can serve as meaningful analogues of high-redshift galaxies.

\section{Summary}
\label{sec:imbh_conclusion}

In this work, we compile the largest X-ray selected sample of AGN in low-mass ($\log M_*/M_{\odot}\leq 10$) galaxies in the local Universe using the deep four-pass eROSITA-DE all-sky survey. By combining this sample with a complementary set of high-mass galaxies ($\log M_*/M_{\odot}>10$), we identify X-ray-detected AGN across the full mass range and investigate the distribution of their specific accretion rates, $p(\log \lambda_{\rm SAR} | M_*, z)$, through a Bayesian hierarchical inference framework. From this, we estimate the cumulative AGN fraction at varying thresholds in $\lambda_{\rm SAR}$.

Our parent galaxy sample consists of $\sim 5.35$ million galaxies from LS10 that have good-quality optical photometry, are brighter than a $z$-band magnitude of 20, and are in the redshift range $0.03<z<0.2$. We carefully derived galaxy properties through SED fitting with an AGN component (when needed) and computed X-ray fluxes for all sources using X-ray aperture photometry. We performed extensive validation and cleaning to ensure the reliability of X-ray detections and their optical host galaxy associations. This proves to be an essential step, as 79\% of the initial X-ray detections in the low-mass galaxy sample are found to be spurious X-ray sources or are the most probable counterpart of a nearby or background high-mass AGN. In comparison, the high-mass sample suffers much less from contamination (only 24\%).

The $2-10$~keV luminosity range of the X-ray AGN is between $10^{40}-10^{44}$~erg~s$^{-1}$, with several low-mass sources also exceeding $L_{\rm X} \sim 10^{43}$~erg~s$^{-1}$. This observational finding sets an important constraint on the seeding and subsequent growth of black holes in low-mass galaxies in simulations.

Our Bayesian framework takes into account the detected X-ray counts (and background) from all galaxies, and thereby allows us to place tight constraints on the $p(\log \lambda_{\rm SAR} | M_*, z)$ distribution from $\lambda_{\rm SAR}\sim10^{-4}-10^{-3}$ all the way to the Eddington limit, $\lambda_{\rm SAR}=1$. This range is wider than any past study at the same stellar mass and redshift range and allows us to reveal second-order mass-dependent properties of $p(\log \lambda_{\rm SAR} | M_*, z)$. We find a steep break in the distribution at high $\lambda_{\rm SAR}\gtrsim 10^{-2}-10^{-1}$, which could indicate Eddington-limited, self-regulated black hole growth.

Using our $p(\log \lambda_{\rm SAR} | M_*, z)$ distributions, we derive a cumulative AGN fraction ($f_{\rm AGN}$) as a function of stellar mass for $\lambda_{\rm SAR}\geq10^{-3}$, $\lambda_{\rm SAR}\geq10^{-2}$ and $\lambda_{\rm SAR}\geq10^{-1}$. We find that $f_{\rm AGN}(\lambda_{\rm SAR} \geq 10^{-3})\lesssim1\%$ for galaxies with $\log M_*/M_{\odot}<10$ at $0.03<z\leq0.164$, setting a firm lower limit on the BHOF in the low-mass regime. Interestingly, we find varying trends of the cumulative AGN fractions as a function of stellar mass, for different thresholds of $\lambda_{\rm SAR}$. For example, we find a suppression in the efficiency of fuelling AGN beyond $\lambda_{\rm SAR} \geq 10^{-2}$ at both low and high masses, in comparison to those living in galaxies with $\log M_*/M_{\odot}\sim 10-10.5$. This highlights a more nuanced view of AGN fuelling that must be taken into account in future modelling.

Our cumulative AGN fraction results clearly indicate that AGN do exist even in low-mass galaxies, and a non-negligible fraction of them are highly accreting. If AGN feedback is active in these AGN in low mass galaxies, as recent studies are finding, then this could challenge the long-standing view that stellar feedback alone drives low-mass galaxy evolution, revealing AGN as an equally, if not more, critical factor \citep[e.g.][]{Mezcua2019, Koudmani2019, Koudmani2021, Koudmani2022}.

Overall, the eRASS:4 X-ray survey is an unparalleled dataset for such studies, due to the combination of its depth and coverage of over $13,000$~deg$^{2}$, allowing our work to present the largest statistical sample of X-ray emitting low-mass galaxies at low redshift to date. Future wide-area surveys covering a few $100$~deg$^{2}$ by \textit{NewAthena} \citep{Nandra2013, Cruise2025} will push studies of the incidence of AGN in low-mass galaxies to higher redshifts, which is currently only possible in the deepest \textit{Chandra} fields and with much lower statistics. Meanwhile, LSST \citep{Ivezic2019} and SPHEREx \citep{Dore2018,Crill2020} will open up new channels to detect even larger samples of AGN in low-mass galaxies, through variability and spectroscopic data in the infrared, respectively. Finally, such studies help pave the way for next-generation observatories, such as \textit{LISA} \citep{AmaroSeoane2023}, which will transform our understanding of black hole growth by directly tracing the mergers of low-mass black hole seeds through their gravitational wave signatures, thereby opening an unprecedented window onto these extreme events.

\bibpunct{(}{)}{;}{a}{}{,} % to follow the A&A style

% for the bibliography, at the end
\bibliographystyle{aa} % style aa.bst
\bibliography{bib.bib} % your references Yourfile.bib

\begin{appendix}

\section{Extragalactic redshift compilation and validation of photometric redshifts}
\label{sec:exgal_specz_compil}

We construct an extragalactic ($z>0.002$) spectroscopic redshift compilation, aiming towards high completeness, including most of the largest catalogues in the literature\footnote{The public version of the extragalactic spectroscopic redshift catalogue developed for this paper is available upon request.}, along with the latest redshifts from SDSS-V \citep{Bowen1973, Gunn2006, Smee2013, Kollmeier2026} and 1229 unpublished redshifts from Balzer et al. (in prep.) using VIRUS on the Hobby-Eberly Telescope \citep{Ramsey1998, Hill2021}. Our compilation includes galaxies, AGN and QSOs that have good quality redshifts, as defined by the cuts described in Table \ref{table:speczcompil}. It builds on the compilation of galaxy-only redshifts assembled for the work by \citet{Kluge2024} (see their Appendix D and Table D.1 for further information and all the references). The quality cuts shown in Table \ref{table:speczcompil} are especially important for AGN and QSO spec-zs as standard pipelines attribute erroneous redshifts to $\sim5\%$ of such sources \citep[e.g.][]{Aydar2025}, whereas a higher accuracy is achieved for galaxy dominated spectra, due to precise absorption features (e.g. Calcium II H and K at 3969.59\AA\ and 3934.78\AA, respectively).

\begin{figure}
    \centering
    \includegraphics[width=\linewidth]{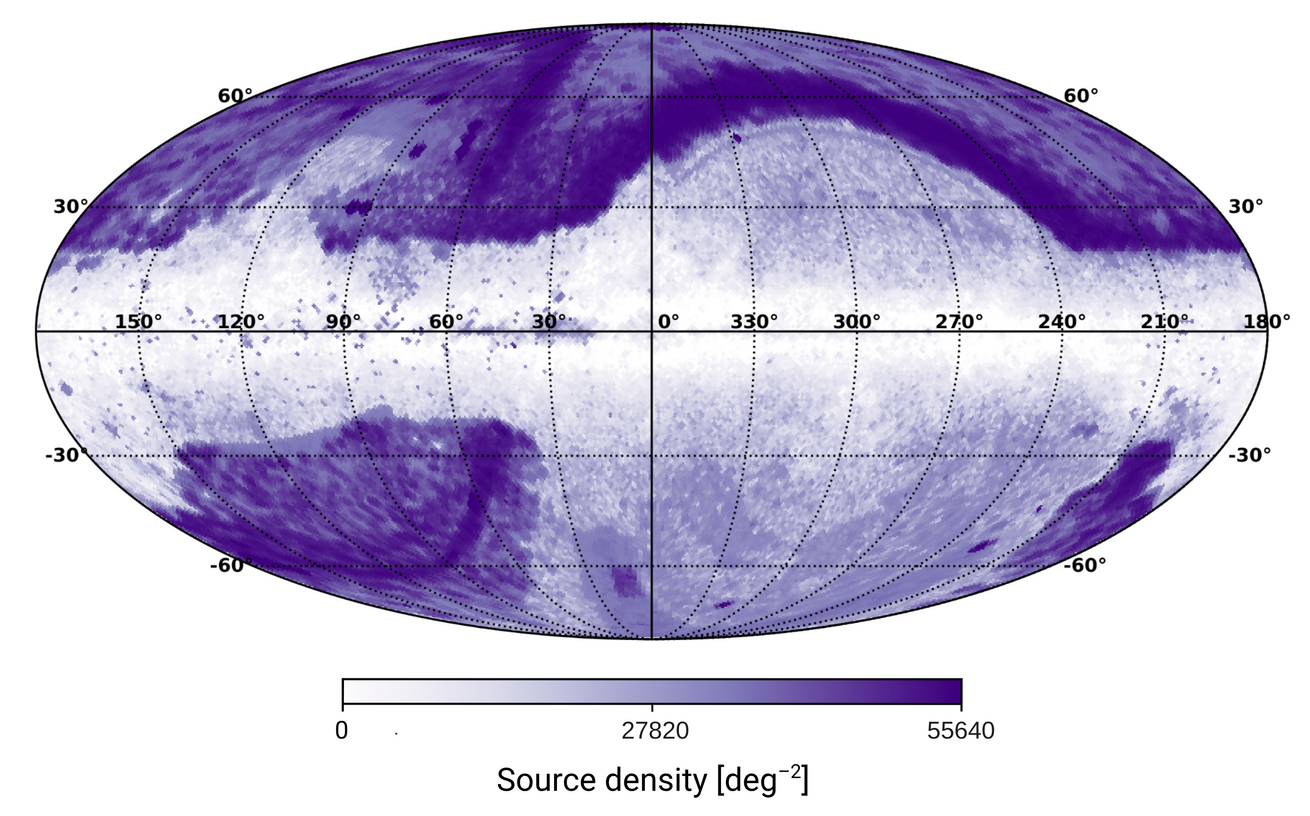}
    \caption{Sky map showing the source density of the unique extragalactic redshift compilation in Galactic coordinates and Mollweide projection.}
    \label{fig:specz_skyplot}
\end{figure}

\begin{figure}
    \centering
    \includegraphics[width=0.85\linewidth]{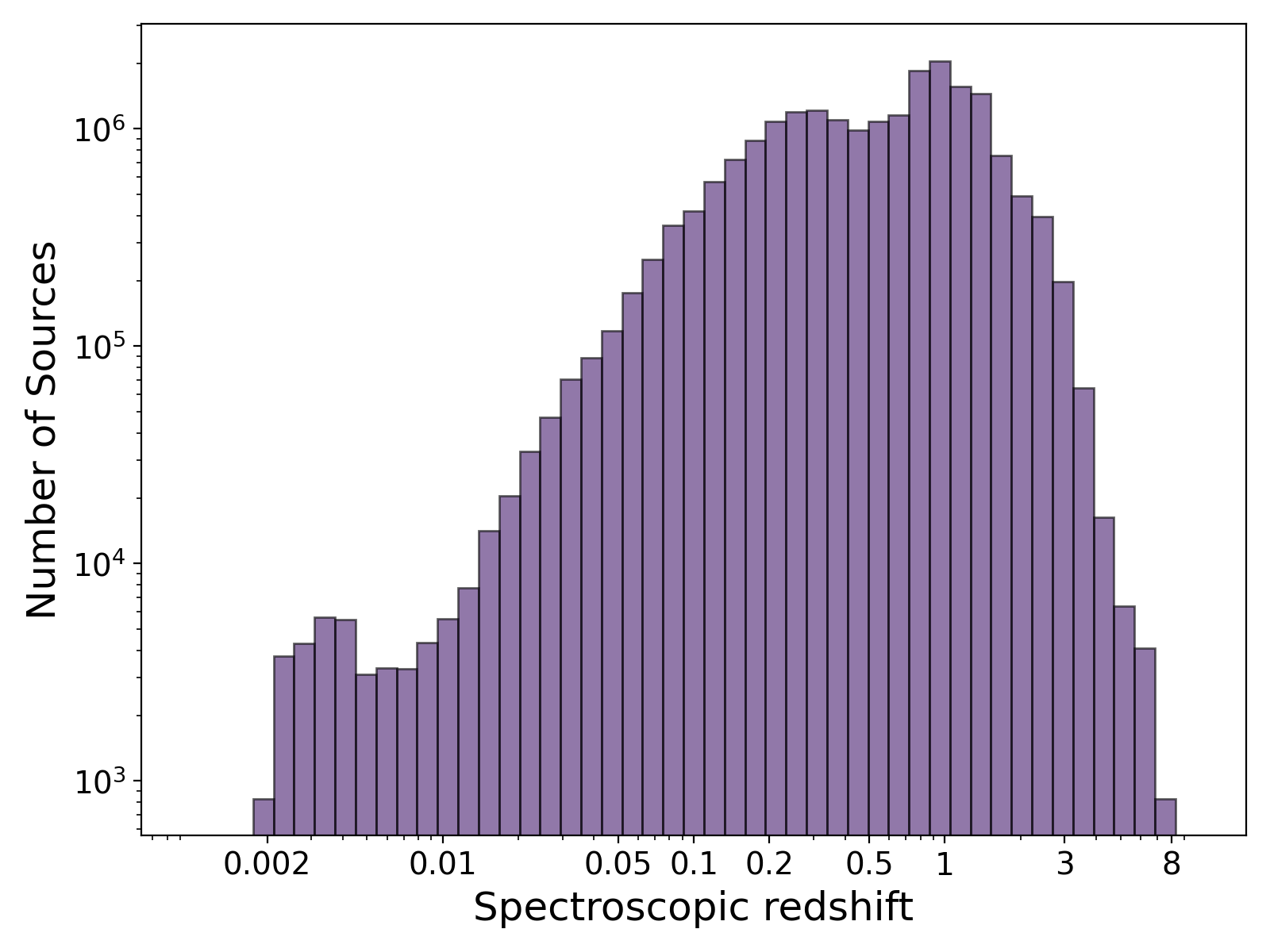}
    \caption{Histogram showing the redshift distribution of the unique extragalactic redshift compilation.}
    \label{fig:specz_hist}
\end{figure}

We first compile all available spec-zs and then create a unique table, listing only the `best' spec-z entry per source, which is subsequently used to update the parent sample as explained above. The flagging algorithm is as follows: 
\begin{enumerate}
    \item Individually sort large catalogues with repeated target observations by S/N (e.g. SDSS, LAMOST) and take only the highest quality redshift per target.
    \item Compile all catalogues together to form the complete extragalactic spectroscopic redshift catalogue (with duplicates). 
    \item Make an internal match within 1\arcsec\ and sort duplicate sources into groups with a unique \texttt{GroupID} and \texttt{GroupSize}.
    \item Compute the range in redshift (max-min) per group identified with \texttt{GroupID}. 
    \item Flag groups with large redshift ranges $>0.01$: \texttt{z\_discrepant\_range\_0pt01=1} (otherwise \texttt{0}).
    \item If \texttt{GroupSize==2}:
    \begin{enumerate}[label=\alph*)]
        \item And group contains a spec-z from DESI DR1 (whether group redshifts are discrepant or not), keep DESI DR1 redshift.
        \item If not (a) and group contains a spec-z from SDSS in the priority order of latest release, that is SDSS-V, then IV, then III and earlier (whether group redshifts are discrepant or not), keep SDSS redshift.
        \item If not (a) nor (b) and \texttt{z\_discrepant\_range\_0pt01=0}, keep first redshift of the group. Otherwise, if \texttt{z\_discrepant\_range\_0pt01=1}, discard all redshift in the group from the unique compilation. 
    \end{enumerate}
    \item If \texttt{GroupSize>2} and  \texttt{z\_discrepant\_range\_0pt01=0}, keep entries as 6(a)-(c) above.
    \item If \texttt{GroupSize>2} and  \texttt{z\_discrepant\_range\_0pt01=1}, do `majority agreement' flagging: for each entry in a given \texttt{GroupID}, check how many other entries have $|\Delta z|<0.01$ (i.e. are in agreement). Find the subset with the maximum number of redshifts in agreement and keep first entry. Mark the other sources not in the agreeing subset with \texttt{z\_outlier=1}. If no majority agreement is found, all sources in the group are flagged as outliers and are discarded from the unique compilation.
\end{enumerate}
The flagging algorithm results in 20,527,574 unique spec-z sources, with the top 20 contributors, as well as additional AGN and QSO catalogues added to the compilation by \citet{Kluge2024}, described in Table \ref{table:speczcompil}. Note that this number refers to the number of unique sources over the full sky (not just the eROSITA-DE footprint), as shown by the sky plot in Fig.~\ref{fig:specz_skyplot}. Additionally, the range of redshifts covered by this compilation is shown in Figure~\ref{fig:specz_hist}.

\begin{figure}
    \centering
    \includegraphics[width=\linewidth]{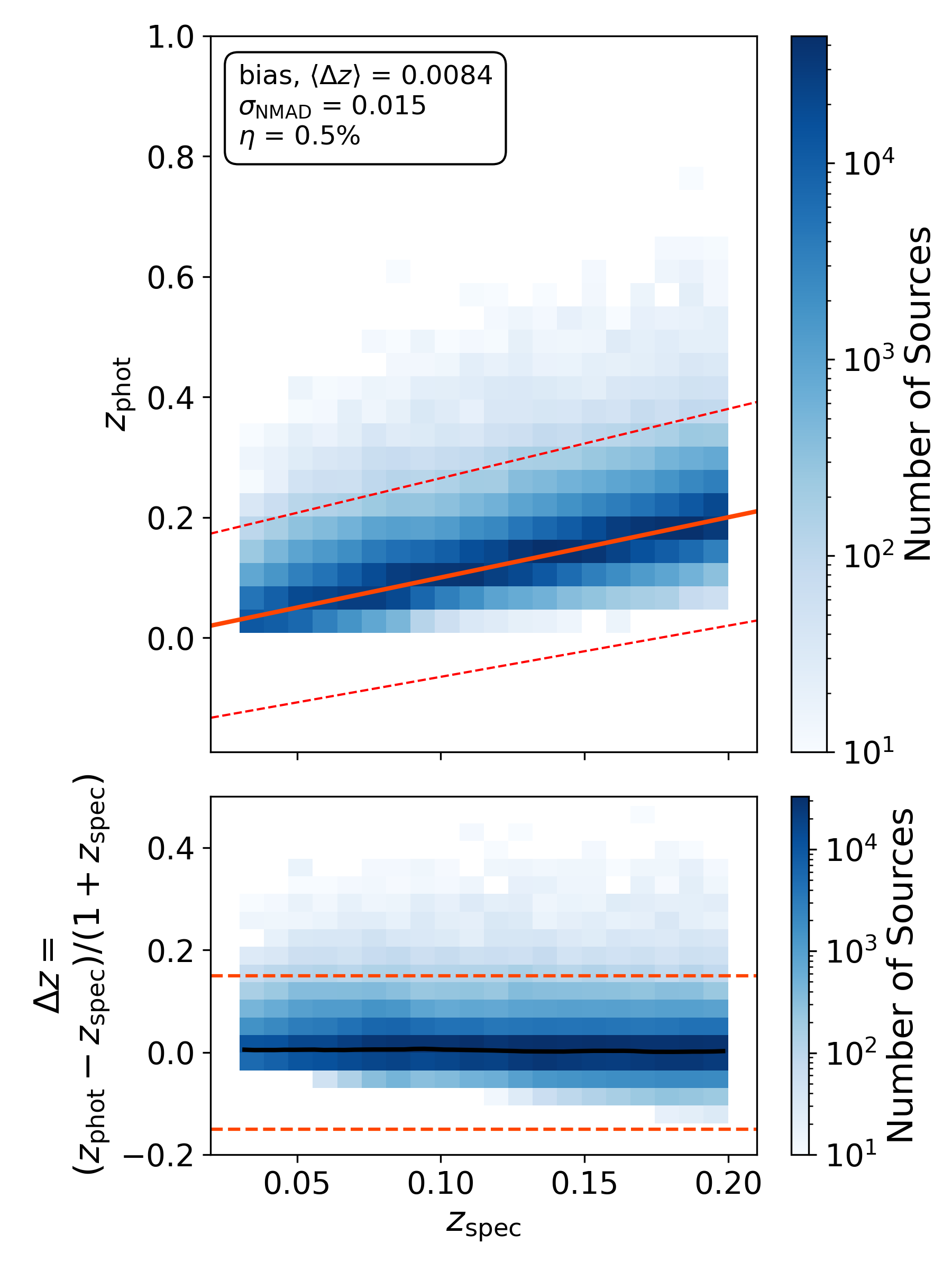}
    \caption{Validation of the subset of photometric redshifts from the parent sample of galaxies with spectroscopic redshift from the extragalactic redshift compilation. Minimal outlier fraction of 0.5\% and almost negligible bias of 0.0084 show the excellent quality of the photo-zs. The solid red line marks the 1:1 relation and the dashed red lines are used to define $\eta$ (see text for details). Darker blue colours indicate higher number of sources, as shown by the colour bar.}
    \label{fig:photoz-specz}
\end{figure}

We append LS10 catalogue columns to the sources within the LS10 footprint and apply the same selection criteria as described in Sect.~\ref{sec:building_parent_sample} to the compilation, except for the cut on the redshift. This results in around 3.57 million unique sources which can then be used to improve upon the photo-zs in the parent galaxy sample. Figure \ref{fig:photoz-specz} shows the comparison of the photo-zs versus spec-zs, using now the same redshift cut as the parent sample, $0.03<z\leq0.2$, leaving around 1.03 million sources. Following the standard statistical metrics to assess the quality of the photo-zs  \citep{Ilbert2006}, we find a bias of 0.0084 (defined as the mean of the normalised residuals: $\langle \Delta z \rangle  = \frac{(z_{\mathrm{spec}} - z_{\mathrm{phot}})}{(1+z_{\mathrm{spec}})}$), a standard deviation from the normalised median absolute deviation of 0.015 (defined as $\sigma_{\mathrm{NMAD}} = 1.4826 \times \mathrm{Median}  \frac{|z_{\mathrm{spec}} - z_{\mathrm{phot}}|}{(1+z_{\mathrm{spec}})}$) and a fraction of outliers of 0.5\% (defined as $\eta$ = $ \frac{|z_{\mathrm{spec}} - z_{\mathrm{phot}}|}{(1+z_{\mathrm{spec}})} > 0.15$). The fraction of catastrophic outliers is $<0.03\%$, defined arbitrarily as $\frac{|z_{\mathrm{spec}} - z_{\mathrm{phot}}|}{(1+z_{\mathrm{spec}})} > 1$. These excellent metrics thereby validate the use of the photo-zs for this work.

\clearpage
\onecolumn

\begin{table*}[h!]
\centering
\begin{tabular}{p{0.15cm} p{2.95cm} p{3.6cm} p{1.2cm} p{8.7cm}}
\toprule
\toprule
\textbf{\#} & \textbf{Survey} & \textbf{Reference} & \textbf{Number$^{a}$} & \textbf{Quality Flag} \\
1  & DESI DR1 & \citet{DESICollab2026} & 15679799 & \texttt{Z > 0.002 \& ZWARN == 0 \& COADD\_FIBERSTATUS == 0} \\
2  & SDSS-IV DR17 & \citet{Abdurrouf2022} & 1492476 & \texttt{Z > 0.002 \& ZWARNING == 0 \& SN\_MEDIAN\_ALL > 2 \& SPECPRIMARY == 1} \\
3  & Quaia (G<20.5) & \citet{Storey-Fisher2024} & 693386 & \texttt{redshift\_quaia > 0.002} \\
4  & SDSS-IV DR16 & \citet{Ahumada2020} & 681464 & \texttt{zwarning == 0 \& z > z\_err \& z\_err > 0} \\
5  & SDSS-V DR20 & (priv. comm.) & 212918 & \texttt{MJD <= 60708 \& Z > 0.002 \& ZWARNING == 0 \& SN\_MEDIAN\_ALL > 2 \& SPECPRIMARY == 1} \\
6  & Compilation$^{b}$ & \citet{Zou2019} & 181775 & \texttt{z > 0.002} \\
7  & SDSS-III DR11-12 & \citet{Alam2015} & 177324 & \texttt{z > 0.002 \& zWarning == 0} \\
8  & PRIMUS & \citet{Coil2011,Cool2013} & 160862 & \texttt{z > 0.002 \& ZQUALITY >= 3} and not a star \\
9  & 2dF Galaxy Redshift Survey & \citet{Colless2001} & 159491 & \texttt{z > 0.002 \& q\_z >= 3} \\
10 & 3D-HST Survey$^{c}$ & \citet{Momcheva2016} & 153681 & \texttt{z > 0.002 \& in\_DESComplilation \& des\_flags >= 3} \\
11 & WiggleZ Dark Energy Survey & \citet{Drinkwater2018} & 125702 & \texttt{z > 0.002 \& q\_z >= 3} \\
12 & 6dFGS DR3 & \citet{Jones2009} & 93866 & \texttt{z > 0.002 \& q\_z == 4} \\
13 & HETVIPS & \citet{Zeimann2024} & 68118 & \texttt{z > 0.002 \& classification == GALAXY} \\
14 & Milliquas v8 & \citet{Flesch2023} & 60975 & \texttt{Z > 0.002} \textnormal{ (not photo-z, i.e. 1 decimal place redshift)} \\
15 & LAMOST DR10 & (see caption$^{d}$) & 54422 & \texttt{z > 0.002 \& z\_err < z \& z\_err > 0} \\
%removed \citet{Luo2022} ref bc that is DR7
16 & GAMA DR4 & \citet{Driver2022} & 49977 & \texttt{Z > 0.002 \& NQ > 2} \\
17 & VERONCAT (corrected) & \citet{VERONCAT2010,Flesch2013} & 47565 & \texttt{z > 0.002} \\
18 & 2dFLenS & \citet{Blake2016} & 42756 & \texttt{z > 0.002 \& in\_DESComplilation \& des\_flags >= 3} \\
19 & VIPERS DR1 & \citet{Garilli2014} & 41775 & \textnormal{2.6 < flag < 4.6; also 22.6–24.6 (secondary), 12.6–14.6 (AGN), 212.6–214.6 (AGN, secondary)} \\
20 & Las Campanas Redshift Survey & \citet{Shectman1996} & 21907 & \texttt{z > 0.002} \\
 & $\vdots$ & $\vdots$ & $\vdots$ & $\vdots$ \\
 & SDSS-V DR19 & (priv. comm.) & 10810  & \texttt{Z > 0.002 \& ZWARNING == 0 \& SN\_MEDIAN\_ALL > 2} \\
  & VVDS DR2 & \citet{LeFevre2013} & 10584 & \texttt{z>0.002 \& f\_z>1 \& z!=9.99} \\
 & OzDES DR2 & \citet{Lidman2020} & 10276 & \texttt{z>0.002 \& qop>2 \& qop!=6} \\
 & AGES & \citet{Kochanek2012} & 8704 & \texttt{z>0.002} \\
& COSMOS compilation & \citet{Khostovan2026} & 8552 & \texttt{specz>0.002 \& Confidence\_level>=80} \\
 & 2SLAQ & \citet{Cannon2006, Croom2009} &  3207 & \texttt{z>0.002 \& q\_z>2, z2S>0.002 \& q\_z2S==1} \\
 & 2Qz & \citet{Croom2004} & 411 & \texttt{z1>0.002 \& (q\_z1==11 or q\_z1==21)} \\
 & VANDELS DR4 & \citet{Garilli2021} & 381  & \texttt{zsp>0.002 \& q\_zsp!=0 \& q\_zsp!=1 \& q\_zsp!=10 \& q\_zsp!=11 \& q\_zsp!=20 \& q\_zsp!=21 \& q\_zsp!=210 \& q\_zsp!=211 \& q\_zsp!=220} \\
& SDSS-IV DR16Q  & \citet{Lyke2020} & 227 & \texttt{Z>0.002 \& ZWARNING==0 \& SN\_MEDIAN\_ALL>2} \\
 & VUDS DR1 & \citet{Tasca2018} & 204 & \texttt{zspec>0.002 \& zflags>1 \& zflags!=11 \& zflags!=21 \& zflags!=31 \& zflags!=41 \& zflags!=32 \& zflags!=33 \& zflags!=34}
 \\
 & SDSS-V DR18 eFEDS & \citet{Aydar2025} & 84 & \texttt{Z > 0.002 \& ZWARNING == 0 \& SN\_MEDIAN\_ALL > 2 \& SPECPRIMARY == 1}  \\
& QUBRICS & \citet{Boutsia2020}  & 14 & \texttt{zspec>0.002} \\
 & z>5.6 QSOs & \citet{Fan2023} & 406 &  \\
\bottomrule
\bottomrule
\end{tabular}
\caption{Top 20 contributors to the extragalactic redshift compilation used for this work, with columns for the survey name, literature reference, number of sources and quality flag used for selection. Entries appearing after the triple dots are additional AGN and QSO catalogues added in this work that are not documented in \citet{Kluge2024}. \textit{a}: Number of sources refers to the unique entries that made it into the final compilation after the flagging algorithm (see text). \textit{b}: Note that the \cite{Zou2019} entries come from a compilation of several spectroscopic surveys: 2dFGRS \citep{Colless2001}, 2SLAQ \citep{Cannon2006}, 6dFGS \citep{Jones2004, Jones2009}, CFRS \citep{Lilly1995}, CNOC2 \citep{Yee2000}, DEEP2 \citep{Davis2003, Newman2013}, SDSS DR14 \citep{Abolfathi2018}, VIPERS \citep{Garilli2014, Guzzo2014}, VVDS \citep{LeFevre2005,Garilli2008}, WiggleZ \citep{Drinkwater2010,Parkinson2012}, and zCOSMOS \citep{Lilly2007}. \textit{c}: The cuts \texttt{in\_DESComplilation \& des\_flags >= 3} refer to good-photometric quality sources present in the compilation from \citet{Gschwend2018}. \textit{d}: Documentation for the LAMOST DR10 catalogue can be found here: \url{https://www.lamost.org/dr10/v2.0/doc/lr-data-production-description}.}
\label{table:speczcompil}
\end{table*}

\twocolumn

Using this well-defined sample of 1.03 million sources with the same selection as the parent galaxies, we also quantify the potential systematic effects of sources scattering in and out of our redshift range. We find that 8.3\% of sources with known spec-z between $0.03<z\leq0.2$ would be scattered out of the sample because the $z_{\mathrm{phot}}>0.2$, and 2.3\% with known $z_{\mathrm{spec}}>0.2$ would be scattered in because their $0.03<z_{\mathrm{phot}}\leq0.2$. This will be improved with future spectroscopic surveys such as LSST, 4MOST and future SDSS releases.

\section{Training our XGBClassifier to find outlier LePHARE-derived stellar masses in need of SED fitting using \texttt{GRAHSP}}
\label{appendix:details_ml_grahsp}

\begin{figure*}
    \centering
    \includegraphics[width=\linewidth]{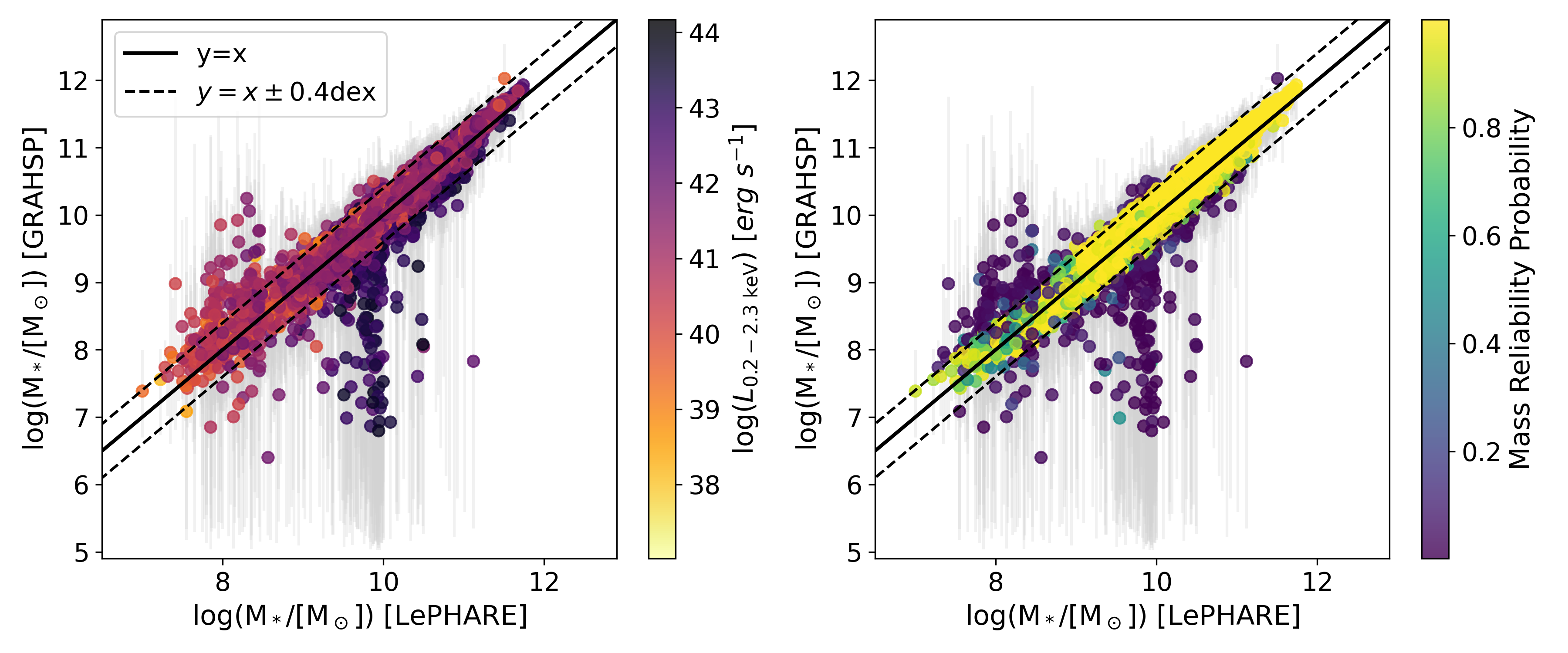}
    \caption{A comparison of the median stellar mass derived via LePHARE (x axis) and \texttt{GRAHSP} (y axis) SED fitting for the training sample of 3027 galaxies with solid and dashed lines indicating $y=x$ and $y=x\pm0.4$~dex. The left and right panel are colour-coded as a function of observed $0.2-2.3$~keV X-ray luminosity and mass reliability probability (as computed by the XGBClassifier), respectively. 1$\sigma$ uncertainties on LePHARE-derived stellar masses are often too small to be seen, whereas the 2$\sigma$ uncertainties on \texttt{GRAHSP}-derived stellar masses are clearly visible (see text for more details on the handling of such large confidence intervals).} 
    \label{fig:mass_outliers_training}
\end{figure*}

In this section, we describe in detail how we trained an eXtreme Gradient Boosting machine learning classifier \citep[XGBClassifier;][]{Chen2016} to distinguish galaxies for which their optical photometry and LePHARE fit results indicate a biased stellar-mass measurement, such that we can re-compute it with a more sophisticated fully-Bayesian SED fitting code, \texttt{GRAHSP} \citep{Buchner2024_grahsp}. XGBClassifier is a gradient-boosted decision tree algorithm, meaning that it builds an ensemble of decision trees sequentially, where each new tree is trained to correct the errors (residuals) of the previous ones using gradient descent on a given loss function. We train our XGBClassifier to assign a low `mass reliability probability' to galaxies whose LePHARE-derived stellar masses differ from the \texttt{GRAHSP}-derived values by more than $\pm 0.4$ dex. This threshold is chosen to be smaller than the stellar mass bin width used in computing $p(\log \lambda_{\rm SAR} | M_*, z)$, while still being sufficiently large for the XGBClassifier to robustly learn the properties of outlier sources (see Fig.~\ref{fig:mass_outliers_training}).

As a first step, we need to build a training sample on which we run SED fitting using \texttt{GRAHSP}. We do so by random sampling our parent galaxy sample in stellar mass and redshift space. We also note that for a classifier to perform well, it needs to learn from enough variety, meaning that there need to be enough outliers. Previous studies \citep[e.g.][and references therein]{Buchner2024_grahsp} have shown that when an AGN is present, but not included in the SED fitting, the stellar mass tends to be overestimated, since the multi-wavelength emission of the AGN is incorrectly attributed to the stellar component. An indication of an AGN being present can be given by nuclear X-ray emission, but several cleaning and validation procedures must be undertaken before attributing any X-ray detection by eROSITA to an AGN, as well as properly assigning it to its most likely host galaxy counterpart (a step especially difficult for the as-of-yet under-explored low-mass galaxy regime). This is why we have to make an iterative step in our analysis to first determine real X-ray emitting parent sample sources, as well as the correct associations to their host galaxy counterparts, such that we can feed our XGBClassifier with a well-understood sample of (potential) outliers. In Section \ref{sec:xrayvalidation_catalogues}, we describe these cleaning and validation steps that we perform on the parent sample galaxies using only their LePHARE-derived stellar masses. We find 892 secure X-ray detected sources among the low-mass galaxies ($\log M_{*,\rm LePHARE}/M_{\odot}\leq10$), which are all added to the training sample. We note that this X-ray selection will, by definition, not provide a complete AGN sample across all wavelengths, but the goal of this step is to boost the potential outlier class in the training sample. Then we sample (up to) 50 random galaxies in each $M_*-z$ bin for both the high-mass ($\log M_{*,\rm LePHARE}/M_{\odot}>10$) X-ray detected galaxies (totalling to 583) and the full non-X-ray detected parent galaxies (totalling to 1552). Overall, our training sample is made up of 3,027 galaxies sampled over the entire $M_*-z$ distribution to ensure coverage of the entire parameter space. We note that the final sample numbers presented in Section \ref{sec:xrayvalidation_catalogues} refer to the second iteration of these cleaning and validation procedures, using the LePHARE- and \texttt{GRAHSP}-derived stellar masses, as an outcome of the analysis in this section.

We proceed to re-compute the stellar masses of these 3,027 galaxies with \texttt{GRAHSP}. We use the same SED set-up as for the LePHARE runs, except for the following key differences (for which full details are given in \citealt{Buchner2024_grahsp}). We account for continuum and line emission from ionised gas with the \texttt{nebular} module \citep{Boquien2013, Boquien2019}, which contribute increasingly for galaxies with recently formed stars. We use the \texttt{biattenuation} module to account for dust attenuation from a SMC attenuation curve \citep{Prevot1984}, which shows a steep rise with $\lambda^{-1}$ with no strong 2175\AA\, feature, indicating smaller dust grains compared to the galactic ISM average, and is parametrised using \texttt{E(B-V)}. As this attenuated optical light is then reprocessed and re-emitted (conserving energy) at IR wavelengths, we model this using the \citet{Dale2014} templates (\texttt{galdale2014}). Within the \texttt{biattenuation} module we also separately account for the effect of attenuation on the AGN models, as they are affected by both the galactic and nuclear attenuation. Importantly, we include several AGN components, including a big blue bump at optical/UV wavelengths using an empirically motivated smooth bending power-law parametrisation (\texttt{activatepl}), broad and narrow emission lines (\texttt{activatelines}) and torus emission associated to the reprocessed optical/UV emission by hot and cold dust emitting in the NIR and MIR, respectively (\texttt{activatetorus}). Photometric redshift errors (with or without \textit{i}-band, depending on availability, see Appendix \ref{sec:exgal_specz_compil}) are also ingested as priors in the SED fitting. \texttt{GRAHSP} efficiently samples this complex parameter space with a nested sampling Monte Carlo algorithm called \textit{Ultranest} \citep{Buchner2021}, which additionally allows for finer sampling than traditional grid-based SED-fitting codes. 
The same \textit{g,r,i,z,W1,W2} photometry, derived from the best-fit model fluxes in the LS10 catalogue, is used as in the LePHARE SED fitting for consistency.

The results are shown in Figure \ref{fig:mass_outliers_training}. There are 344 sources with $|\Delta \log M_*|>0.4$~dex (11\%), confirming the conclusions of Sect.~\ref{sec:lephare} that for most cases the simpler and faster SED fitting method with LePHARE produces unbiased results. However, it is also clear that the stellar mass can be catastrophically wrong and off by several orders of magnitude for the outlier cases. Interestingly, the majority of outliers below the 1:1 line are highly X-ray luminous in the $0.2-2.3$~keV band and have overestimated stellar mass from LePHARE, as predicted above. However, being X-ray luminous does not necessarily mean that the stellar mass will be over-estimated, as shown by the numerous such sources lying within the scatter of the 1:1 line. It is important to correct the stellar masses of these X-ray luminous outlier objects as they fall exactly in our statistical sample (see Sect.~\ref{sec:mstar_completeness_paper3}) and boost the incidence of X-ray AGN in low-mass galaxies. At lower stellar masses, $\log M_{*,~\rm{LePHARE}}/M_{\odot}<9$ there are also a cloud of outliers above the 1:1 line, which are sources where \texttt{GRAHSP} fits a steeply rising MIR emission from hot galactic dust. As this model was not included in the LePHARE SED fitting run, the stellar masses of these objects are under-estimated. However, this dust-obscured star-forming galaxy fit is also found to be degenerate with a heavily obscured AGN solution with high nuclear $E_{\rm (B-V),~AGN}$ attenuation and high AGN luminosity, as traced by 5100\AA\, emission. The overall impact on the stellar mass estimation of this degeneracy is non-trivial as stellar age is a confounding factor: a galaxy could have a lot of dust-obscured star formation now, but it might be young overall and not yet have built up much stellar mass. Conversely, a galaxy could be old and massive but currently forming fewer stars or hosting a dust-obscured AGN. We defer in-depth analysis of this issue to future work where we use more photometric bands that can help disentangle such degeneracies.

The advantage of machine-learning-based classifiers is that they can learn from a combination of several different features, referring in this case to the optical photometric properties and LePHARE fit statistics of the training sample galaxies (see below), allowing them to detect outliers with high completeness. In contrast, if we simply used a cut in X-ray luminosity, we would be re-computing stellar masses for sources which do not need it and simultaneously missing many outliers above the 1:1 line, as shown in Figure \ref{fig:mass_outliers_training}. In particular, the advantage of an XGBClassifier, in comparison to Logisitic Regression or Random Forests, is that it can handle non-linear feature interactions naturally, it is more efficient (thanks to the boosting, meaning that fewer trees are required), it includes regularisation to control over-fitting and handles imbalanced data well \citep{Chen2016}.

The ultimate goal of any classifier is to maximise the precision\footnote{Precision is defined as the ratio of the true positives over the sum of the true positives and false positives ($\frac{TP}{TP + FP}$), i.e. of all the instances the model predicted as positive, how many were actually positive?} and recall\footnote{Recall is defined as the ratio of the true positives over the sum of the true positives and false negatives ($\frac{TP}{TP + FN}$), i.e. of all the actual positive instances, how many did the model correctly identify?}, simultaneously. This is encoded in the F1 Score, the harmonic mean between the two\footnote{The F1 Score is defined as:  $\rm{F1~Score = \frac{2}{(1/Precision)+(1/Recall)}=2 \cdot \frac{Precision \cdot Recall}{Precision + Recall}}$.}. For our scientific goal, where we define `positive' detections (\texttt{class==1}) as outliers, we must prioritise maximising recall because false negatives (i.e. not identifying an outlier) are costly. Accordingly, the methodological choices outlined below are designed with this priority in mind. We train the model on the following features, which we found to provide strong discriminatory power: 

\begin{enumerate}
    \item Colour indicators of the presence of an AGN or MIR dust emission component: \textit{W1-W2}, \textit{r-W2} \citep{Andonie2025}, \textit{g-r} magnitudes.
    \item Stellar mass proxies: absolute $\textit{z}$-band magnitude (\texttt{abs\_Mag\_z}), galaxy half-light radius for the best fitting galaxy type from LS10 (\texttt{SHAPE\_R} in units of arcseconds).
    \item  Fit quality indicators: uncertainty in the redshift (with or without $i$-band and zero for spectroscopic redshifts: \texttt{redshift\_err}),  LePHARE $\chi^2$ fit statistic (\texttt{log\_LPH\_CHI\_BEST}). 
    \item Potential X-ray AGN indicator: $0.2-2.3$~keV luminosity (\texttt{log\_Lx\_soft} in units of erg~s$^{-1}$).

\end{enumerate}

For the 77 sources with missing \textit{W2} flux, we replace the \textit{W1-W2} and \textit{r-W2} feature values with $-999$, such that the model learns the impact of the missing photometric point in such a key part of the SED, given the above findings. As is commonly done, we split the 3,027 sources into training (80\%) and test (20\%) sets, and downsample 30\% of the inliers to improve class balance and help the model learn that outliers are not so rare that they will never be assigned. After tuning the hyper-parameters via grid search, we settle on 300 estimators (weak learners), a maximum tree depth of 6, a learning rate of 0.05, a subsample fraction of 0.8, and a column subsampling rate of 0.7. We also apply a class-weighting factor (\texttt{scale\_pos\_weight}: the ratio of inliers to outliers) so that each outlier contributes \texttt{scale\_pos\_weight} times more to the loss function, thereby penalising false negatives more strongly and increasing recall. 
Lastly, we choose our evaluation metric (or loss function) to optimise the area under the precision–recall curve (AUCPR). After training, we choose an optimal decision threshold of the mass reliability probability by maximising recall whilst keeping the precision$>=60\%$ via AUCPR (see Fig.~\ref{fig:confusion_matrix_and_PRcurve}, right). We find the optimal decision threshold to be $<0.38$, which we use to flag outlier sources. 

Applying the model on the unseen test sample, using the optimal decision threshold above, 537 sources are classed as inliers and 69 as outliers, in the ratios shown by the confusion matrix in Figure \ref{fig:confusion_matrix_and_PRcurve} (left). Therefore, the overall accuracy\footnote{Accuracy = (Number of correct predictions) / (Total number of predictions) = (TP + TN) / (TP + TN + FP + FN)} of our classifier is (497+59)/606 $\sim$ 92\%. It has an area under the Receiver Operating Characteristic Curve (ROC AUC) of 0.937, meaning the model can distinguish inliers from outliers very well, and has high recall for both inliers (93\%) and outliers (86\%). Meanwhile an AUCPR of 0.672 highlights that, though the model captures outliers well, the precision suffers, meaning that there are some false positives (inliers marked as outlier) included. However, as mentioned above, it is more important to catch all outliers than to potentially spend extra computational time re-computing inliers that have been erroneously marked as outliers. Figure \ref{fig:mass_outliers_training} shows the stellar mass derived by LePHARE and \texttt{GRAHSP} colour-coded by the probability of being an inlier, in other words the `mass reliability probability', highlighting the excellent performance of the XGBClassifier. We find that the top three most important features to distinguish outliers are the $g-r$ and $r-W2$ magnitudes, as well as the $\chi^2$ from the LePHARE fit; the $0.2-2.3$~keV luminosity is the second-to-last in the ranking.

\begin{figure}
    \centering
    \includegraphics[width=\linewidth]{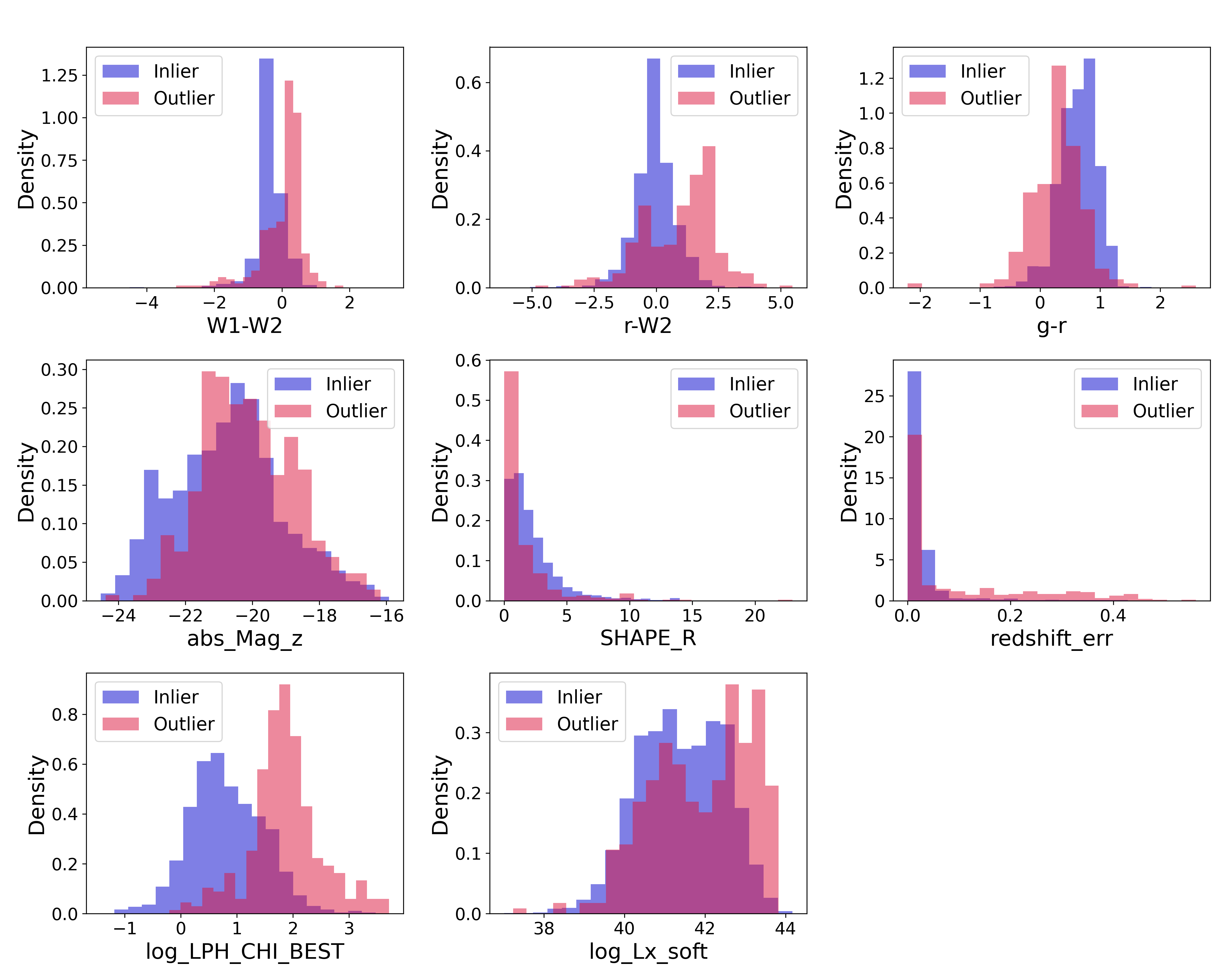}
    \caption{Histograms showing the difference in feature-space between the inlier ($|\Delta \log M_*|\leq0.4$~dex) and outlier ($|\Delta \log M_*|>0.4$~dex) sources in the training sample for the XGBClassifier (see text for more details of the meaning of the features and their units).}
    \label{fig:feature_hist}
\end{figure}

\begin{figure}
    \centering
    \includegraphics[width=0.42\linewidth]{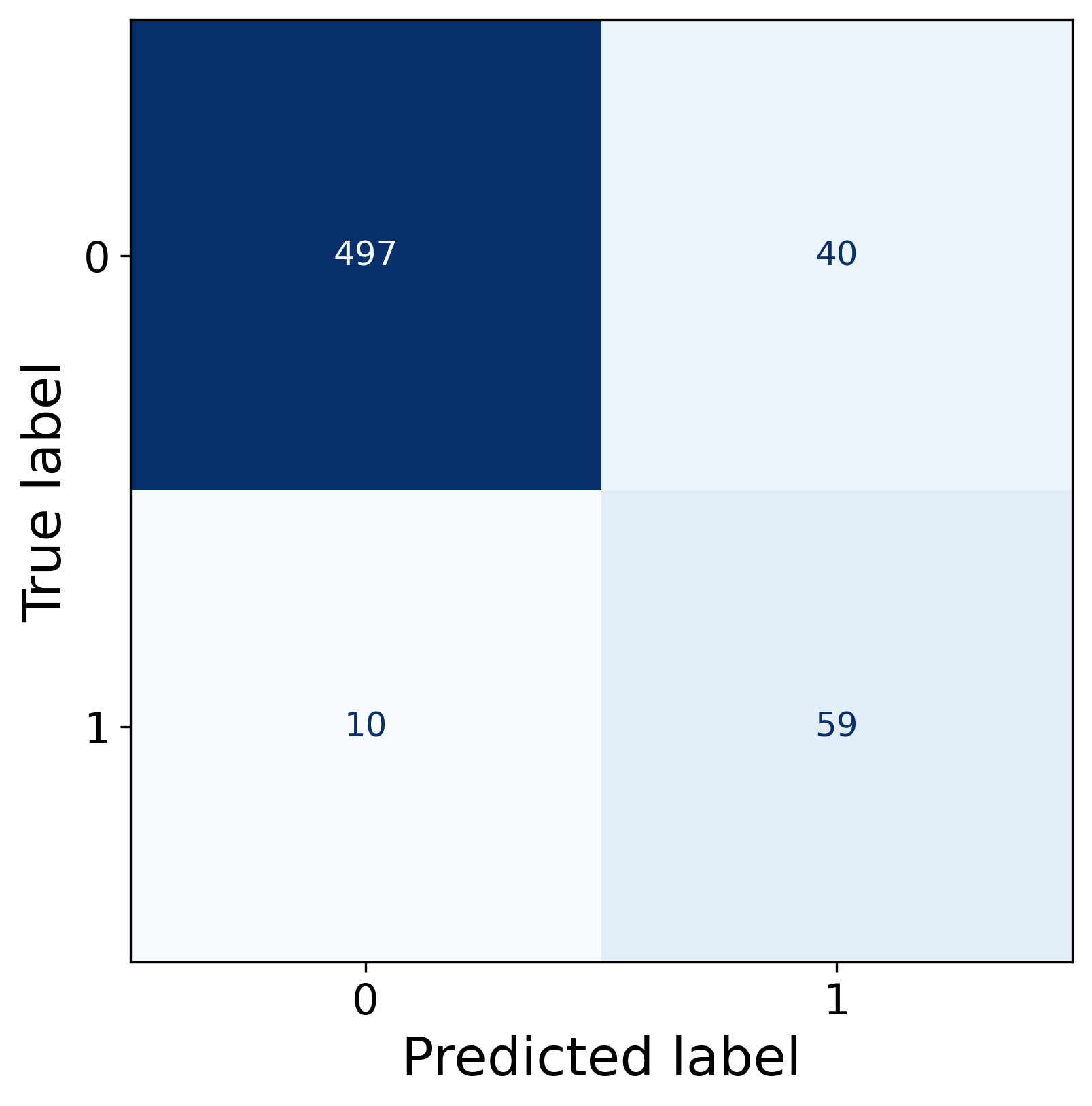}
    \includegraphics[width=0.55\linewidth]{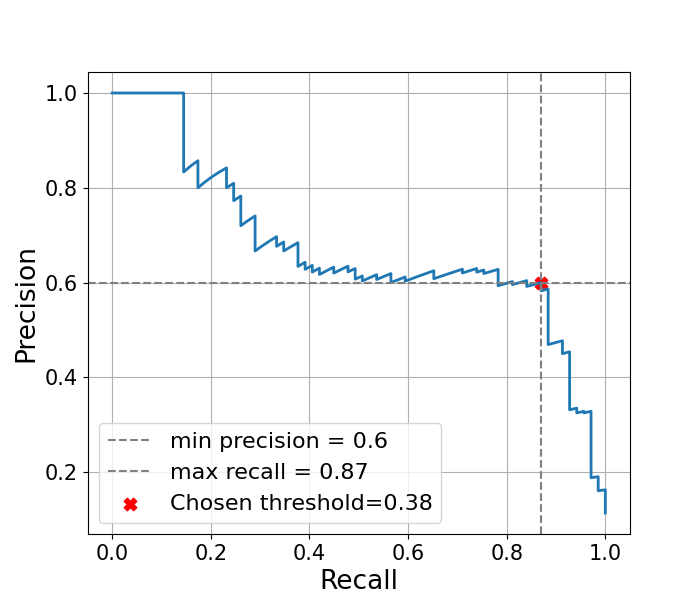}
    \caption{Left: Confusion matrix showing the performance of the XGBClassifier on the test sample. The total accuracy of the model is 92\% and there is a high recall for both inliers (93\%; \texttt{class==0}) and outliers (86\%; \texttt{class==1}). Right: Precision-recall curve showing the optimal threshold chosen for selecting outliers from final galaxy sample.}
    \label{fig:confusion_matrix_and_PRcurve}
\end{figure}

\begin{figure*}[ht!]
    \centering
    \includegraphics[width=0.87\linewidth]{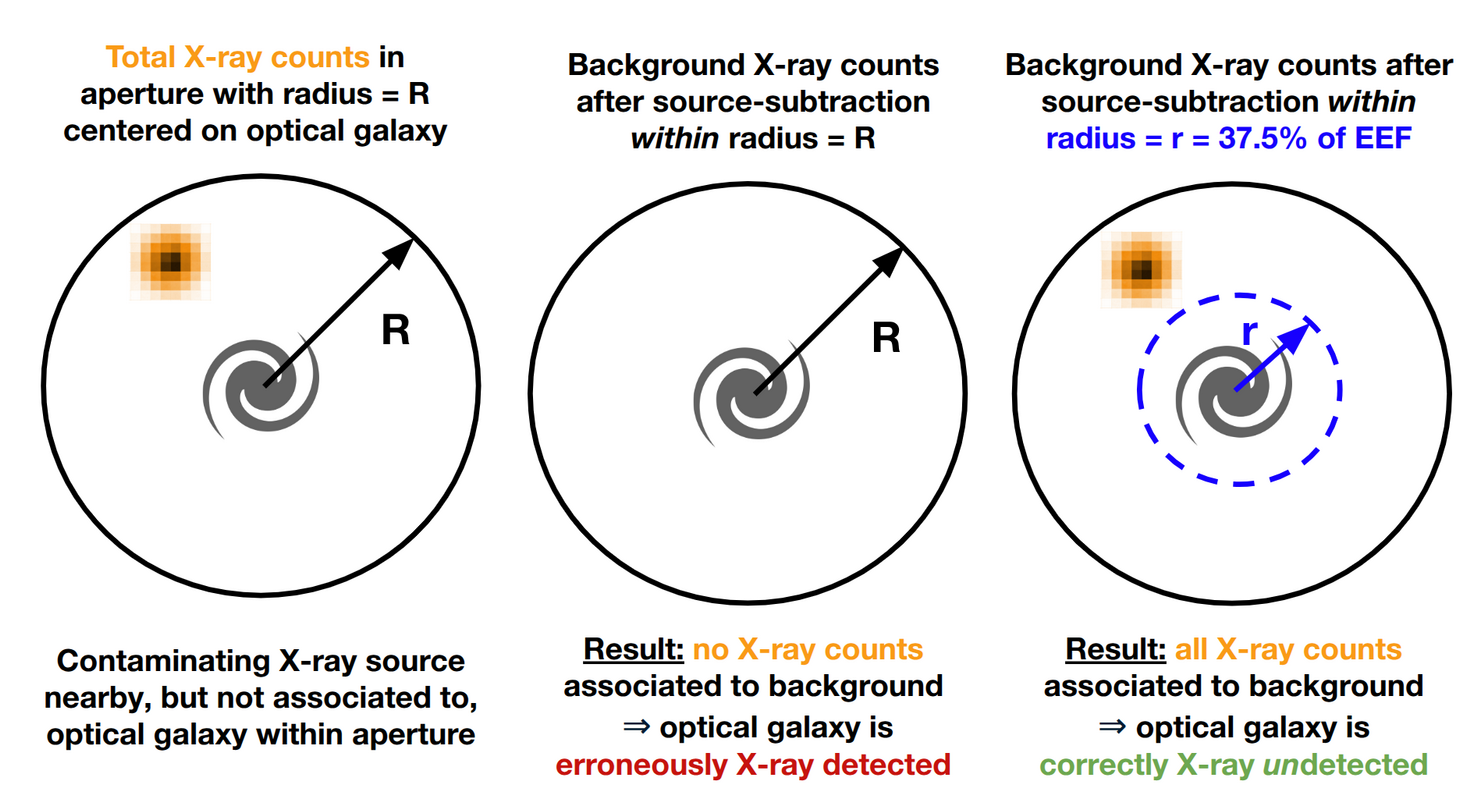}
    \caption{Schematic showing the set-up of the \texttt{apetool} parameters to extract meaningful X-ray information for the parent galaxy sample by implementing an inner source-extraction region.}
    \label{fig:apetool_explanation}
\end{figure*}

Finally, we can apply the trained model on the full parent sample of galaxies and find 32,548 with mass reliability probability $<0.38$ to be re-computed with \texttt{GRAHSP}. \citet{Buchner2024_grahsp} extensively validate the galaxy properties obtained by \texttt{GRAHSP} by comparing to a benchmark photometric dataset (`Chimera') where non-AGN pure galaxies are paired with optically-selected pure quasars at the same redshift. They show that the galaxy properties (e.g. $M_*$, SFR) obtained by \texttt{GRAHSP} on the Chimera sample show no measurable bias, even for galaxies with dominant AGN component or sources with few photometric bands, and very low outlier fraction (defined as when the estimated error bars lie completely outside a 1 dex wide band centred around the true value) of 5\%. \texttt{GRAHSP} is also shown to estimate more realistic uncertainties as the fully Bayesian fit includes uncertainties in the model and the data, making the inference highly robust \citep{Buchner2024_grahsp}. However, one consequence of this is that often the uncertainties on the galaxy properties are very large, especially for sources with few photometric bands, as is the case with our sample. For example, 9,423 out of 32,548 sources have stellar masses that are uncertain by more than 2.5~dex. We deem these sources unconstrained and set them as $2\sigma$ upper limits. Importantly, we note that this is a small minority of the overall parent sample consisting of over 5 million galaxies, so there is limited impact on the statistical incidence results presented in this work. Future studies will improve on this aspect by using more photometric bands and the full stellar mass posterior, to better account for the uncertainties in this parameter for all parent sample galaxies.

\begin{figure*}[ht!]
    \centering
    \includegraphics[width=\linewidth]{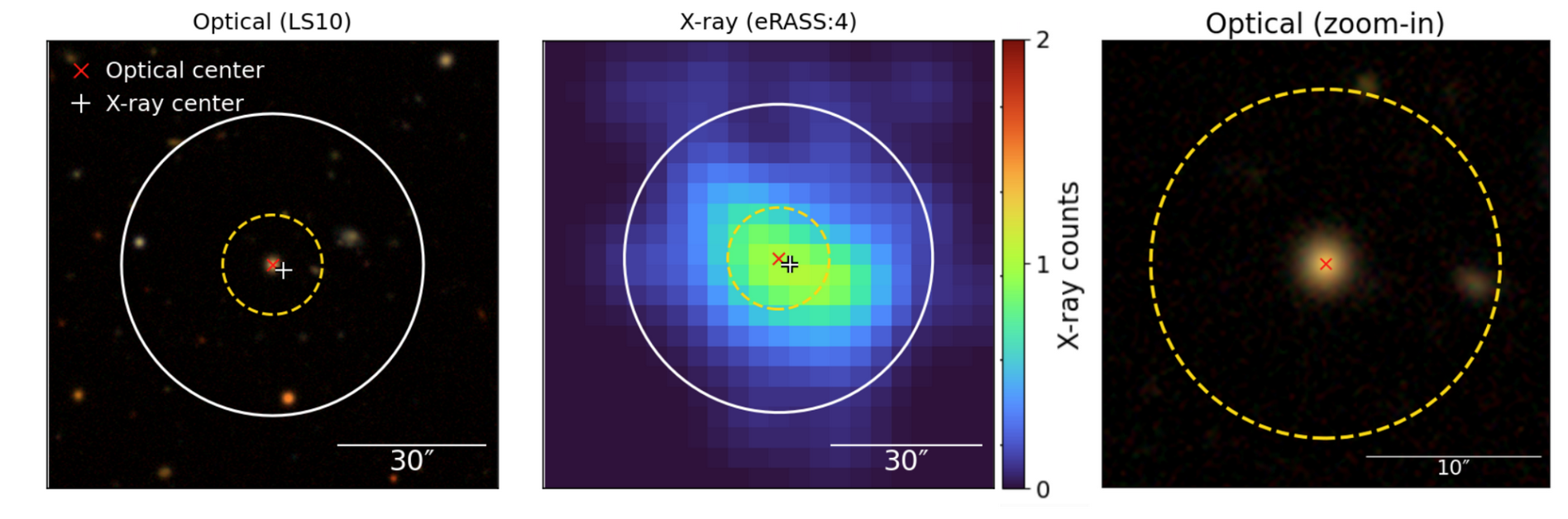}
    \caption{Example optical LS10 (left panel; right panel, zoom-in) and smoothed X-ray eRASS:4 (middle panel) images of an X-ray detected low-mass galaxy ($\log M_*/M_{\odot}\sim 9.55$ and spec-$z$ of 0.18). The optical and X-ray centres are marked with a red and white cross, respectively. A colour bar indicates the number of X-ray counts. The white solid circle denotes the aperture used to extract X-ray photometry, it has a radius of $\sim30$\arcsec. The yellow dashed circle with radius $\sim10$\arcsec\ marks the region outside which eRASS:4 sources contribute to the background counts within the total aperture.} 
    \label{fig:cutout}
\end{figure*}

\section{Details of \texttt{apetool} set-up}

In Fig.~\ref{fig:apetool_explanation}, we show an example case of an aperture of radius R centred on an optical galaxy with a nearby contaminating X-ray source that is not associated to the galaxy (left panel). Let us assume that the only counts in the aperture originate from this source (i.e. neglect the fluctuating base background level). The contaminating X-ray source is detected by eROSITA source-detection algorithms and since its centroid lies within the aperture, it will be subtracted from the eROSITA source map before computing the background (middle panel). However, this results in no X-ray counts being attributed to the background, thus reducing $P_{\rm thresh}$ and erroneously classifying this optical galaxy as a secure X-ray detection. On the contrary, if the radius within which such source-subtraction takes place is set to r$<$R, as in the right panel, the contaminating source can correctly be associated to the local background emission around the optical galaxy, meaning that it will not be significantly X-ray detected. We choose to set this radius, r, to 37.5\% of the EEF, which corresponds to roughly three times the 50th percentile of the eRASS:4 positional error (i.e. $\sim$ 10\arcsec). Overall, this means that X-ray sources with centroids at radii between r$=$37.5\% and R$=$75\% of the EEF are not being removed from the source map and thus they will contribute to an increased local background level.

\section{Details of cleaning spurious associations}
\label{appendix:spur_flowchart}

The flowchart in Figure \ref{fig:flowchart} describes the steps taken to validate the X-ray detections found via \texttt{apetool} and their optical host associations. We explain the procedure in detail below, focusing on low-mass galaxies (high-mass galaxies are treated in the same way, except we do not visually inspect any sources). We note that identifying sources with unreliable LePHARE-derived stellar masses using our machine-learning classifier required prior knowledge of X-ray emission from a cleaned sample of galaxies (recall Section \ref{sec:grahsp_mstar} and Appendix \ref{appendix:details_ml_grahsp}). Therefore, an iteration of this flowchart was first completed using only the LePHARE-derived stellar masses. Figure \ref{fig:flowchart} shows the second iteration using the final LePHARE and \texttt{GRAHSP} stellar masses for which some sources changed from low- to high-mass subsets (or vice versa). 

The first step is to see whether the parent sample galaxy has a match within 30\arcsec\ to a source in the entire eRASS:4 X-ray catalogue (multiple parent sample galaxies could match to the same eROSITA source at this point). This results in 3,979 matches from the low-mass galaxy sample. The 142 unmatched sources are not necessarily spurious as \texttt{apetool} may be able to detect sources below the detection threshold of the catalogue and there is a small minority of real (bright) X-ray detections that are not present in the eRASS:4 catalogue. Therefore, all 142 unmatched sources are visually inspected to recover real X-ray sources associated with the target low-mass galaxies. Visual inspection is done by two authors of this work and the answer to the following four questions must be positive for the source to be considered an X-ray-detected low-mass galaxy: (i) is there a visually discernable agglomeration of X-ray photons indicating an X-ray source (i.e. the counts in the aperture are not scattered due to high local background levels)?; (ii) is the X-ray emission centred on the low-mass galaxy?; (iii) Are there no other possible contaminating X-ray sources that could have erroneously lowered the $P_{\rm thresh}$ value making it a false detection?; (iv) is the optical photometry of good quality, meaning that there are no artifacts, nearby bright objects such as stars or fragmentation? Figure \ref{fig:cutout} shows an example optical LS10 and X-ray eRASS:4 cutout of an X-ray detected low-mass galaxy. 

The result of visually inspecting the 142 sources in this subset shows that in many cases these sources are in regions of high X-ray background; are located near very bright X-ray sources which leak photons into the aperture; or simply the \texttt{ERMLDET} algorithm has failed to catalogue nearby X-ray sources properly, leading to them not being subtracted from the source map and thereby erroneously lowering the $P_{\rm thresh}$ value. Visual inspection helps save 24 X-ray sources passing the criteria defined above, although the large majority of these source lie close to the $P_{\rm thresh}$ cut-off, meaning that they are faint.

The second step of the flowchart involves associating the 3,979 low-mass galaxies with matches in the eRASS:4 catalogue to the LS10-eRASS:4 counterpart catalogue (Salvato, priv. comm.). This step makes use of the Bayesian cross-matching algorithm called NWAY \citep{maraNWAY2018} which not only uses astrometric (distance) information, but also multi-wavelength priors learned from known X-ray sources and their host galaxy counterparts\footnote{If a source was heavily influenced by these additional priors the \texttt{bias\_LS10\_Xray\_proba} will be $>>1$. Additional columns returned by NWAY are the best-match flag (\texttt{match\_flag = 1}), a probability for the match being the correct one (\texttt{p\_i}) and a probability of the source in question having any counterpart at all in the search region (\texttt{p\_any}).} \citep{Mara_ctp_efeds, Salvato2025}. Out of 34 sources with no matches within 30\arcsec\ to the eRASS:4 counterpart catalogue, 9 sources remain as X-ray emitting low-mass galaxy candidates as per the visual inspection criteria defined above. These are typically faint, diffuse and isolated low-mass galaxies with clear X-ray detections, that were likely under-represented in the training sample used to derive the X-ray prior or in some cases lie just outside the declination cut used in the eRASS:4-CTP catalogue.

\begin{figure}[t!]
    \centering
    \includegraphics[width=0.95\linewidth]{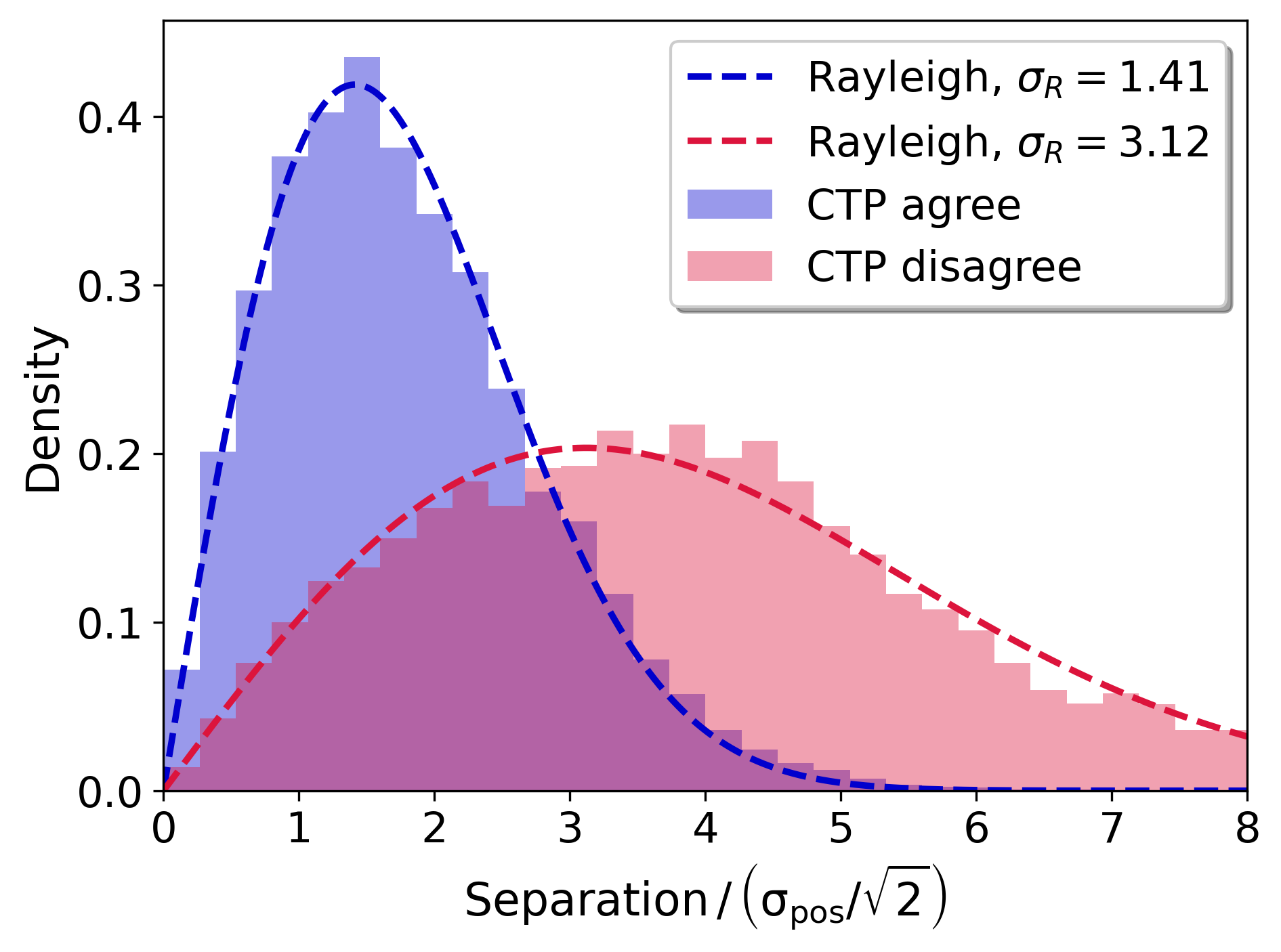}
    \caption{Rayleigh curve fit to the histogram of the separation between the parent galaxy and the nearest eRASS:4 source normalised by the one-dimensional eRASS:4 positional error for the sources in agreement (blue) and disagreement (red) with the eRASS:4 CTP catalogue (see Figure \ref{fig:flowchart}).}
    \label{fig:rayleigh_dwarf}
\end{figure}

For the other 3,945 sources with a match, we split the sample into unique triplets of parent low-mass galaxy -- eRASS:4 X-ray source -- LS10 counterpart (making sure the match is to the same X-ray source from step 1 and 2) and duplicated match triplets, where multiple low-mass galaxies match to the same X-ray source or LS10 counterpart galaxy. For both of these subsets, we then check if the optical co-ordinates of the low-mass parent galaxy are in agreement with the LS10 optical counterpart assigned by the catalogue. If so, we keep these sources as X-ray detected low-mass galaxy candidates and if not, they are discarded from the sample. In the case of duplicated match triplets, this step also efficiently identifies the most favoured match, leaving only a unique optical-host-X-ray pair where the counterparts are in agreement. For the unique triplets we also make a sanity check to compare the total aperture counts derived using our method and those quoted in the eRASS:4 X-ray catalogue (using the same aperture radius) and they are in good agreement, as expected. 

Given the high probability of spurious matches, as shown by the shifted aperture analysis of Section~\ref{sec:xrayvalidation_catalogues}, it makes sense that 3,104 (3,018+86; see ``No'' branches from Q4 in Fig.~\ref{fig:flowchart}) sources are discarded this way. Simply put, if there is a nearby galaxy which is behaving exactly like a known X-ray emitter, it is the most likely origin of the detected X-ray emission; the target low-mass galaxy is at-most constrained by an upper limit in X-ray emission. However, it is clear statistically that there is a high spurious association as: (i) 82\% of best-match LS10 eRASS:4 counterparts are closer in position to the X-ray source than the low-mass galaxy; (ii) 97\% of them are within three times the positional error (\texttt{POS\_ERR}) of the X-ray source; (iii) 90\% of them have \texttt{p\_i} $>0.9$; (iv) 92\% of them are significantly influenced by the X-ray prior, meaning that their LS10 catalogue properties match the host galaxy properties of known X-ray emitters; and (v) the LS10 eRASS:4 counterparts lie in typical regions where bluer, more massive quasars are in the $g-r$ versus $z-W1$ colour-colour plot \citep[see Fig. 18 in][]{Mara_ctp_efeds}, along with the majority of these having best-fit LS10 galaxy \texttt{TYPE=PSF}. Figure \ref{fig:rayleigh_dwarf} shows the histogram of the separation between the parent galaxy and the nearest eRASS:4 source normalised by the one-dimensional eRASS:4 positional error for the sources in agreement (blue) and disagreement (red) with the eRASS:4 CTP catalogue. Sources in agreement can be well fitted by a Rayleigh distribution with $\sigma_{R} \sim 1$, as would be expected from cross-matching catalogues with Gaussian distributed astrometric errors \citep{Pineau2017, Mara_ctp_efeds}. Sources in disagreement show a very broad distribution, indicating the unreliability of the counterpart association, confirming our approach in Figure \ref{fig:flowchart}. Additionally, examining the NWAY information of the low-mass target sample in these cases reveals very low counterpart probabilities, meaning that they are not even close secondary counterparts (which would otherwise justify treating them as X-ray upper limits). All sources labelled as `discarded' in Figure \ref{fig:flowchart} are masked out from the parent sample and no longer used in future analysis.  
\end{appendix}

\begin{acknowledgements}
    The authors thank the anonymous referee for their careful reading of the paper and their constructive comments. 
    ZI acknowledges the support by the Excellence Cluster ORIGINS which is funded by the Deutsche Forschungsgemeinschaft (DFG, German Research Foundation) under Germany´s Excellence Strategy – EXC-2094 – 390783311 and support through the European Space Agency (ESA) Research Fellowship in Space Science. BT also acknowledges support by the Excellence Cluster ORIGINS and from the European Research Council (ERC) under the European Union’s Horizon 2020 research and innovation program (grant agreement No. 950533). RS is supported by Swiss National Science Foundation project grant 200021\_21257.

    This work is based on data from eROSITA, the soft X-ray instrument aboard SRG, a joint Russian-German science mission supported by the Russian Space Agency (Roskosmos), in the interests of the Russian Academy of Sciences represented by its Space Research Institute (IKI), and the Deutsches Zentrum für Luft- und Raumfahrt (DLR). The SRG spacecraft was built by Lavochkin Association (NPOL) and its subcontractors, and is operated by NPOL with support from the Max Planck Institute for Extraterrestrial Physics (MPE). The development and construction of the eROSITA X-ray instrument was led by MPE, with contributions from the Dr. Karl Remeis Observatory Bamberg \& ECAP (FAU Erlangen-Nuernberg), the University of Hamburg Observatory, the Leibniz Institute for Astrophysics Potsdam (AIP), and the Institute for Astronomy and Astrophysics of the University of Tübingen, with the support of DLR and the Max Planck Society. The Argelander Institute for Astronomy of the University of Bonn and the Ludwig Maximilians Universität Munich also participated in the science preparation for eROSITA. The eROSITA data shown here were processed using the eSASS software system developed by the German eROSITA consortium.
    
    The DESI Legacy Imaging Surveys consist of three individual and complementary projects: the Dark Energy Camera Legacy Survey (DECaLS), the Beijing-Arizona Sky Survey (BASS), and the Mayall z-band Legacy Survey (MzLS). DECaLS, BASS and MzLS together include data obtained, respectively, at the Blanco telescope, Cerro Tololo Inter-American Observatory, NSF’s NOIRLab; the Bok telescope, Steward Observatory, University of Arizona; and the Mayall telescope, Kitt Peak National Observatory, NOIRLab. NOIRLab is operated by the Association of Universities for Research in Astronomy (AURA) under a cooperative agreement with the National Science Foundation. Pipeline processing and analyses of the data were supported by NOIRLab and the Lawrence Berkeley National Laboratory (LBNL). Legacy Surveys also uses data products from the Near-Earth Object Wide-field Infrared Survey Explorer (NEOWISE), a project of the Jet Propulsion Laboratory/California Institute of Technology, funded by the National Aeronautics and Space Administration. Legacy Surveys was supported by: the Director, Office of Science, Office of High Energy Physics of the U.S. Department of Energy; the National Energy Research Scientific Computing Center, a DOE Office of Science User Facility; the U.S. National Science Foundation, Division of Astronomical Sciences; the National Astronomical Observatories of China, the Chinese Academy of Sciences and the Chinese National Natural Science Foundation. LBNL is managed by the Regents of the University of California under contract to the U.S. Department of Energy. The complete acknowledgments can be found at https://www.legacysurvey.org/acknowledgment/.

    The Photometric Redshifts for the Legacy Surveys (PRLS) catalog used in this paper was produced thanks to funding from the U.S. Department of Energy Office of Science, Office of High Energy Physics via grant DE-SC0007914.

    Funding for the Sloan Digital Sky Survey V has been provided by the Alfred P. Sloan Foundation, the Heising-Simons Foundation, the National Science Foundation, and the Participating Institutions. SDSS acknowledges support and resources from the Center for High-Performance Computing at the University of Utah. The SDSS web site is www.sdss.org.
    
    SDSS is managed by the Astrophysical Research Consortium for the Participating Institutions of the SDSS Collaboration, including the Carnegie Institution for Science, Chilean National Time Allocation Committee (CNTAC) ratified researchers, the Gotham Participation Group, Harvard University, Heidelberg University, The Johns Hopkins University, L’Ecole polytechnique fédérale de Lausanne (EPFL), Leibniz-Institut für Astrophysik Potsdam (AIP), Max-Planck-Institut für Astronomie (MPIA Heidelberg), Max-Planck-Institut für Extraterrestrische Physik (MPE), Nanjing University, National Astronomical Observatories of China (NAOC), New Mexico State University, The Ohio State University, Pennsylvania State University, Smithsonian Astrophysical Observatory, Space Telescope Science Institute (STScI), the Stellar Astrophysics Participation Group, Universidad Nacional Autónoma de México, University of Arizona, University of Colorado Boulder, University of Illinois at Urbana-Champaign, University of Toronto, University of Utah, University of Virginia, and Yale University.

\end{acknowledgements}

\end{document}